\documentclass[a4paper,fleqn,usenatbib,useAMS]{mnras}

\usepackage{amsmath}	
\usepackage{amssymb}	
\usepackage{multicol}        
\usepackage{bm}		
\usepackage{pdflscape}	
\usepackage[T1]{fontenc}
\usepackage{ae,aecompl}
\usepackage{txfonts}
\usepackage{longtable}
\usepackage{ulem}
\usepackage[pdftex]{graphicx}

\newcommand{\kms}{\,km\,s$^{-1}$} 

\def\H2O       {H$_2$O }

\def\h       {\ifmmode{^{\rm h}}\else{$^{\rm h}$}\fi}
\def\m       {\ifmmode{^{\rm m}}\else{$^{\rm m}$}\fi}
\def\s       {\ifmmode{^{\rm s}}\else{$^{\rm s}$}\fi}
\def\deg     {\ifmmode{^{\circ}}\else{$^{\circ}$}\fi}

\def\decdeg  {\ifmmode{{\rlap.}^{\circ}} \else ${\rlap.}^{\circ}$\fi}
\def\decs    {\ifmmode{{\rlap.}^{\rm s}} \else ${\rlap.}^{\rm s}$\fi}
\def\decas   {\ifmmode{{\rlap.}{''}}\else{${\rlap.}{''}$}\fi}

\def\arcsec  {$^{\prime}$}
\def\arcsec  {$^{\prime\prime}$}

\def\Ta        {$T^\ast_A$}

\title{\textit{SiO Maser Survey towards off-plane O-rich AGBs around the orbital plane of the Sagittarius Stellar Stream}}
\author[Y. W. Wu]{Y. W. Wu$^{1,2}$\thanks{\href{yuanwei.wu@ntsc.ac.cn}{yuanwei.wu@ntsc.ac.cn}}, Noriyuki Matsunaga$^{3}$, Ross A. Burns$^{4}$, B. Zhang$^{5}$
\\
$^{1}$Mizusawa VLBI Observatory, National Astronomical Observatory of Japan
Mitaka, Tokyo 181-8588, Japan\\
$^{2}$National Time Service Center, Chinese Academy of Sciences, Xi'an 710600, China\\
$^{3}$Department of Astronomy, School of Science the University of Tokyo
7-3-1 Hongo, Bunkyo-ku, Tokyo 113-0033, Japan\\
$^{4}$Joint Institute for VLBI ERIC, Postbus 2, 7990 AA Dwingeloo, The Netherlands\\
$^{5}$Shanghai Astronomical Observatory, Chinese Academy of Sciences, Shanghai 200030, China}
\date{Last updated 2017 April 15}
\pubyear{2017}
\begin{document}
\label{firstpage}
\pagerange{\pageref{firstpage}--\pageref{lastpage}}
\maketitle

\begin{abstract}
We conducted an SiO maser survey towards 221 O-rich AGB stars with the aim of
identifying maser emission associated with the Sagittarius stellar stream. In
this survey, maser emission was detected in 44 targets, of which 35 were new
detections. All of these masers are within 5~kpc of the Sun. We also compiled a
Galactic SiO maser catalogue including $\sim$2300 SiO masers from the
literature.  The distribution of these SiO masers give a scale height of 0.40
kpc, while 42 sources deviate from the Galactic plane by more than 1.2~kpc,
half of which were found in this survey.  Regarding SiO masers in the disc, we
found both the rotational speeds and the velocity dispersions vary with the
Galactic plane distance. Assuming Galactic rotational speed $\Theta_0$~=~240
\kms, we derived the velocity lags are 15 \kms\ and 55 \kms\ for disc and
off-plane SiO masers respectively.  Moreover, we identified three groups with
significant peculiar motions (with 70\% confidence). The most significant group
is in the thick disc that might trace stream/peculiar motion of the Perseus
arm. The other two groups are mainly made up of off-plane sources. The northern
and southern off-plane sources were found to be moving at $\sim$ 33 \kms\ and
54 \kms\ away from the Galactic plane, respectively. Causes of these peculiar
motions are still unclear.  For the two off-plane groups, we suspect they are
thick disc stars whose kinematics affected by the Sgr stellar stream or very
old Sgr stream debris.\end{abstract}

\begin{keywords}
Galaxy: structure --- masers --- radio lines: star --- stars: AGB and post-AGB
\end{keywords}


\section{Introduction} \label{sec:intro}

Large scale optical and infrared surveys have proved that the Milky Way halo
contains a number of accretion-derived stellar features. These long-lived,
tail-like features are produced by encounters with satellite dwarf galaxies.
Contrary to disc and bulge stars, stream stars are located in the halo, with
Galactocentric distances ranging from $\sim$ 10 kpc to more than 100 kpc and are thus very
valuable targets for constraining the shape of the dark matter halo. 

The most prominent and well studied stream is the Sagittarius tidal stream
(hereafter Sgr stream), which was produced by the interaction of the Milky Way with
its nearest satellite, the Sagittarius Dwarf Spheroidal Galaxy (Sgr
dSph). The existence of the Sgr stream was originally anticipated by
\cite{1995MNRAS.275..429L} by investigating the positions and proper motions of
global clusters. \citet{2001ApJ...551..294I} firstly identified this structure
from carbon stars, while \citet{2003ApJ...599.1082M} found that M giants can better trace
its detailed structures and identified the ``southern arc'' and ``northern arm''. More
recently, a variety of tracers were used to characterize the stream, including
RR Lyrae \citep{2006AJ....132..714V, 2013ApJ...765..154D}, horizontal branch
stars \citep{2011ApJ...731..119R, 2012ApJ...751..130S}, red clump stars
\citep{2010ApJ...721..329C, 2012AJ....144...18C}, Carbon stars
\citep{2001ApJ...551..294I, 2015MNRAS.453.2653H}, upper main-sequence and
main-sequence turn-off stars \citep{2006ApJ...642L.137B, 2012ApJ...750...80K,
2013ApJ...762....6S}.

SiO, \H2O and OH masers have been found in the envelopes of
O-rich($[$C$/$O$]<$~1) asymtotic giant branch (AGB) stars, in $\sim$3000
sources \citep{1990AJ.....99.1173L, 2007IAUS..242..200D, 2012JKAS...45..139K}.  Most circumstellar maser sources are distributed in the disc of the
Galaxy, while they also have been detected in globular clusters
\citep{2005PASJ...57L...1M}.  \cite{2007PASJ...59..559D} found one SiO maser,
J1923554$-$1302029, which may be associated with the Sgr stellar tidal stream,
while \cite{2010PASJ...62..525D} studied radial velocities of stellar SiO
masers away from the Galactic plane and found some SiO masers with peculiar non
circular motions larger than 100 \kms, indicating a possible signature of
streaming motions. 

In order to extend our investigations from the disc to the halo region, we
conducted an SiO maser survey towards the Sgr stream region. Sample and
observations are presented in Section \ref{sec:obs}. In Section
\ref{sec:results} we present the result of this survey and compile a Galactic SiO maser catalogue. In Section \ref{sec:discuss} we discuss the Galactic
locations and kinematics of SiO masers. A summary is given
in Section \ref{sec:summary}.

\section{Sample and Observation} \label{sec:obs}
\subsection{Sample}  \label{subsec:sample}

The infrared colour-colour diagram is an effective tool to select AGB stars and
has been explored intensively since the all-sky IRAS photometry became
available in the 1980s \citep{1988A+A...194..125V}. Many SiO maser surveys have
been performed towards cold luminous IRAS sources \citep{1990MNRAS.243..480H,
1994A+AS..103..107H, 1994ApJ...437..419I, 1995ApJS...98..271I,
1995PASJ...47..815J, 2001A+A...376..112I}.  Compared with the IRAS catalogue, the
sensitivity of the recently released Wide-filed Infrared Survey Explorer (WISE)
all-sky catalogue is much better \citep{2010AJ....140.1868W}. In fact, the number
of point sources in the catalogues of IRAS, AKARI and WISE are 245889, 877091 and
563921584 \citep{1988iras....7.....H, 2010A+A...514A...1I,
2012yCat.2311....0C}.  In order to detect more distant and fainter sources, we
selected our maser survey sample from the WISE all-sky point source catalogue.

\begin{figure*}
 \includegraphics[width=16cm]{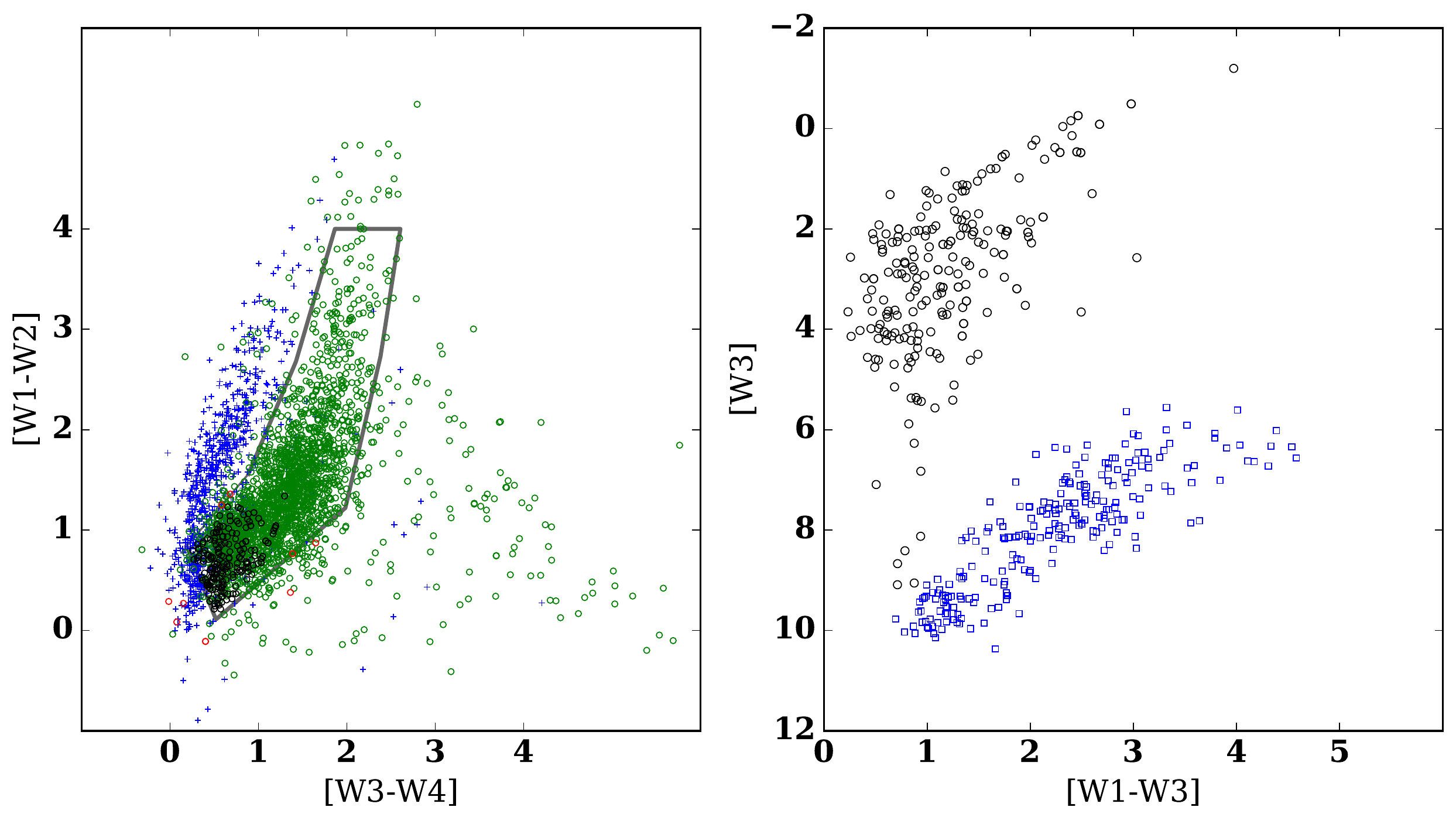}
 \caption{\textit{Left panel:} WISE colour-colour diagram ([$W3$-$W4$] versus [$W1$-$W2$]),
  blue crosses and green circles denote verified O-rich and C-rich AGB catalogue of
  \citet{2009JKAS...42...81S, 2011MNRAS.417.3047S}. An empirical boundary ,the
 polygon, is used to select O-righ AGBs candidates. Black and red circles are
 our target sources. 
 Red circles denote 9 Miras added at the late stage of the
 observational sessions.
 \textit{Right panel}: WISE colour-magnitude
 diagram([$W1$-$W3$] versus $W3$) diagram. Blue squares are O-rich AGBs
 in the SMC selected based on the same colour-colour selection
 criteria. Black circles are 221 sources that have been observed by this survey.
 \label{fig-1}}
\end{figure*}

\cite{2014MNRAS.442.3361N} and \cite{2014A+A...564A..84L} showed that the WISE
colour-colour diagram ($W1$-$W2$ versus $W3$-$W4$) is effective at identifying
O-rich and C-rich AGB stars. We therefore used this colour-colour diagram to
select targets for this SiO maser survey. Firstly, we used the verified C-rich
and O-rich AGBs catalogue of \cite{2009JKAS...42...81S, 2011MNRAS.417.3047S},
shown as blue pluses and green circles in the colour-colour diagram to define
an empirical boundary (polygon in the colour-colour diagram) of O-righ AGBs
(see the left panel of Fig.~\ref{fig-1}).  Secondly, we used a colour-magnitude
([$W3$] versus [$W1$-$W3$]) diagram to remove contaminations, which are mainly
extragalactic sources. In the right panel of Fig.~\ref{fig-1}, we show our
observed target sources and O-rich AGB stars in the Small Magellanic Cloud
(SMC). Using the colour-magnitude diagram, fainter extragalactic sources with
locations below SMC AGBs are easily identified and excluded. Thirdly, we
considered the following three constraints on the sky position,
DEC.~$>$~$-$25$^\circ$ to be observable from the Nobeyma 45m telescope,
Galactic latitude higher than 30$^\circ$ to exclude Galactic plane
contaminations, and angular separation $<$~20$^\circ$ from the Sgr stream
orbital plane which follows a great circle with the normal vector towards
\textit{l}~=~5.6$^\circ$ \textit{b}~=~$-$14.2$^\circ$
\citep{2003ApJ...599.1082M}. We further checked the spectral type and stellar
classifications with SIMBAD\footnote{\url{http://simbad.u-strasbg.fr/}}
database to exclude tens of known C stars and galaxies. These criteria define a
sample of 274 sources. 

\begin{figure*}
\includegraphics[width=16cm]{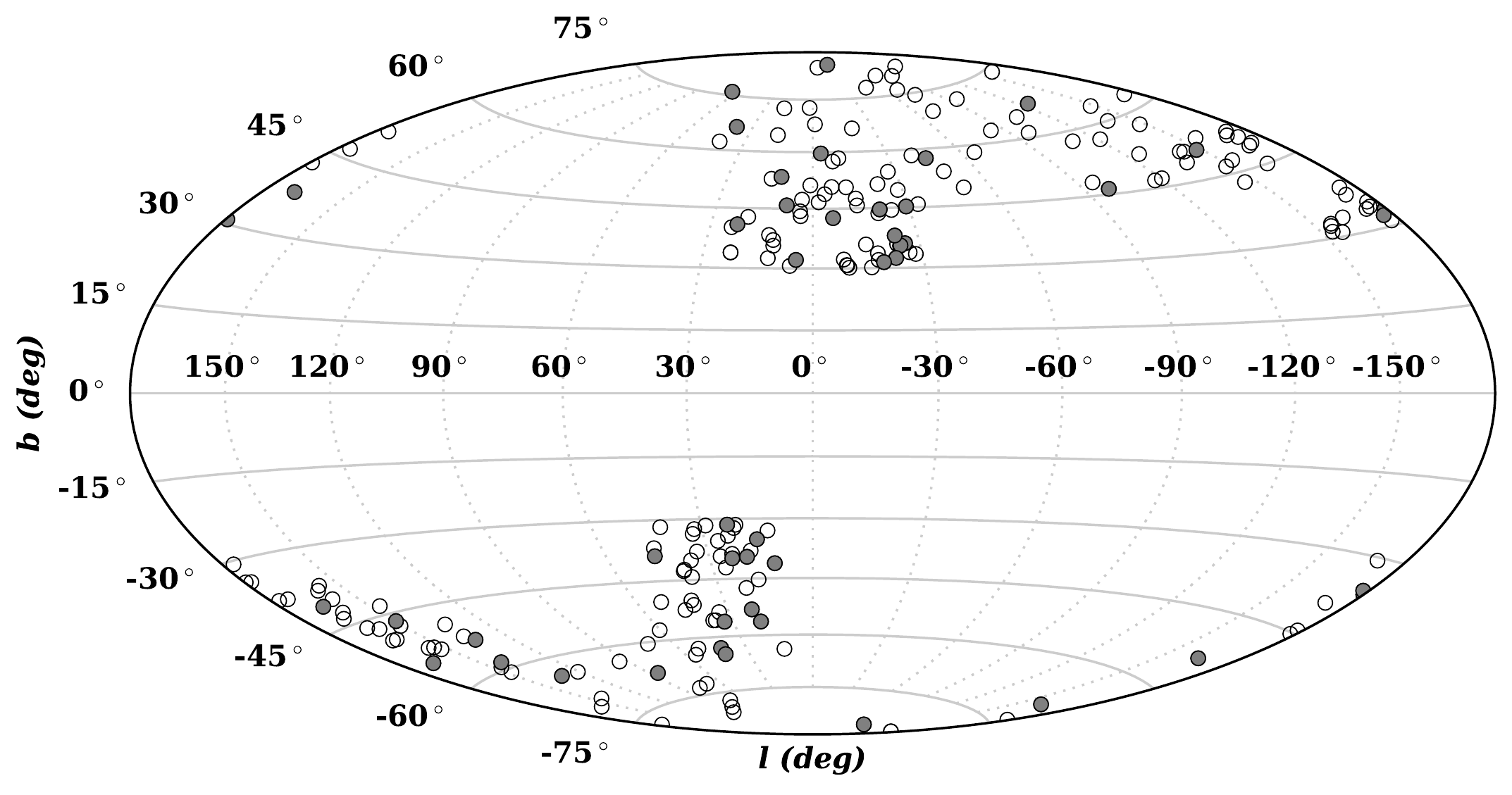}
\includegraphics[width=16cm]{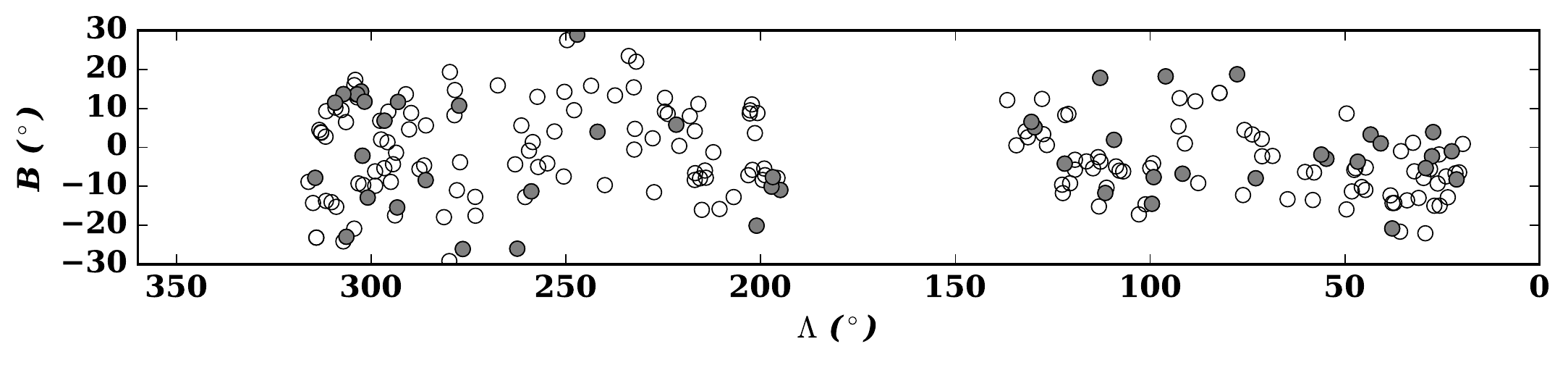}
\caption{\textit{Top panel}: Galactic distribution of observed targets, filled grey circles
indicates sources with detection of SiO masers. 
\textit{Bottom panel}: Same as top panel but in the Stellar Stream coordinate
system.  
\label{fig-2}} 
\end{figure*}

Around 30 sources from our sample were not observed due to time constraints and
conflict with the Solar position. In addition, considering the
high detection rate of SiO masers in Miras, in the last stage of our
observation sessions, we observed nine additional Miras which were just
outside of our original selection polygon (red circles in Fig.~\ref{fig-1}). In total, we searched for SiO maser emission in
221 sources, including 108 (49\%) Miras, 37 (17\%) semi-regular stars, 23
(10\%) long period variables, 42 (19\%) stars and 11 (5\%) other types,
including 5 variable stars , 2 Infra-red sources, 2 Mira candidates, 1 S star
and 1 long period Variable candidate. In Fig.~\ref{fig-2}, we show the
locations of these sources in Galactic coordinates and Sgr stellar stream
coordinates, where filled grey circles indicate sources with detections of
maser emissions. The defination of the Sgr stellar stream coordinates can be found in Section 5.2.3 of \citet{2003ApJ...599.1082M}, with the latitudes of Sgr stellar stream coordinates, $B$, defined
by the Sgr debris projected on the sky as viewed from the Sun. The normal vector of the Sgr orbital plane is 
toward the direction of ($l$, $b$) = (273$^\circ$.8, $-13^\circ$.5) and with $A$~=~0$^\circ$ towards the center of the Sgr
dSph. Equations to convert between the equatorial and the Sgr stream coordinate systems can be found in the appendix of \citet{2014MNRAS.437..116B}.
Source details, i.e., source name (Galactic coordinate
notation), equatorial coordinates, star type and spectral type are given in Table
\ref{tab-8}.  For variable stars, Bayer designation names are also
given, where star type and spectral type were queried from the SIMBAD database. 

\subsection{Observation}\label{subsec:obs} 

We carried out observations of the SiO v=1, 2, J=1-0 (43.122 and 42.820~GHz)
transitions for 221 sources with the Nobeyama 45-m telescope\footnote{The 45-m
radio telescope is operated by Nobeyama Radio Observatory, a branch of National
Astronomical Observatory of Japan.} in April and May of 2016. During
observation sessions, pointing was checked every 2 hours using known sources of strong
SiO maser emission. The half power beam width was 38.7\arcsec\ at 43 GHz, with
an aperture efficiency of 0.53.  The backend was set with bandwidth of 125 MHz
and channel spacing of 30.52 kHz, covering a velocity range of
$\pm$430~\kms\ with a velocity spacing of 0.21 \kms.

The data were calibrated using the chopper wheel method, which corrected for
atmospheric attenuation and antenna gain variations to yield an antenna
temperature \Ta. System temperatures  were within 150 to 230 K. The integration
time per source was 10-30 minutes giving a 1$\sigma$ level of 0.03-0.05K. The
conversion factor from the antenna temperature, \Ta\, in units of K, to
flux density in units of Jy, was about 2.71Jy~K$^{-1}$, which was
estimated using an aperture efficiency of 0.53 and forward efficiency of 0.65.

\section{Survey Results and SiO catalogue} \label{sec:results}

\subsection{Survey Results}\label{subsec:results}

\begin{table*}
\caption{List of detections of SiO maser \label{tab-1}}
\begin{tabular}{rcccccccc}
\hline
ID & Source & Other  & R.A.(J2000) & DEC.(J2000) & $V_{\rm LSR}$ & Period &  PLR Distance & WISE Distance \\
 & Name  & Name & (h:m:s) & (d:m:s) & (km s$^{-1}$) & (day)  & (kpc)  & (kpc)  \\
\hline
1  &G004.482$+$32.104 & BD Oph      & 16 05 46.29 & $-$06 42 27.8 &   -8.8 &    340.440    &   2.003&  2.791  \\
2  &G008.104$+$45.840 & MW Ser      & 15 28 43.67 & $+$03 49 43.6 &   41.0 &               &        &  1.659  \\
3  &G011.025$+$53.268 & FV Boo      & 15 08 25.77 & $+$09 36 18.4 &    4.5 &    313.500    &   1.990&  1.821  \\
4  &G011.159$-$41.196 & X Mic       & 21 04 36.85 & $-$33 16 47.3 &   17.8 &    239.720    &   1.463&  1.528  \\
5  &G015.405$-$35.139 & R Mic       & 20 40 02.99 & $-$28 47 31.2 &   19.9 &    138.620    &   0.765&  1.212  \\
6  &G019.002$-$39.495 & RR Cap      & 21 02 20.78 & $-$27 05 14.9 &  -51.8 &    277.540    &   2.071&  1.135  \\
7  &G019.509$-$56.308 & R PsA       & 22 18 00.24 & $-$29 36 13.8 &  -29.0 &    294.950    &   1.692&  1.691  \\
8  &G021.513$-$53.023 & S PsA       & 22 03 45.83 & $-$28 03 04.2 & -100.8 &    271.350    &   1.908&  2.467  \\
9  &G022.158$+$40.858 & U Ser       & 16 07 17.66 & $+$09 55 52.5 &  -15.0 &    238.250    &   1.707&  1.611  \\
10 &G022.943$-$31.448 & RU Cap      & 20 32 34.16 & $-$21 41 26.5 &    9.8 &    347.370    &   1.444&  1.706  \\
11 &G023.376$-$39.816 & V Cap       & 21 07 36.63 & $-$23 55 13.4 &  -20.9 &    276.260    &   1.458&  1.572  \\
12 &G038.070$+$66.469 & R Boo       & 14 37 11.58 & $+$26 44 11.7 &  -42.4 &    223.255    &   0.576&  1.091  \\
13 &G045.734$-$38.770 & HY Aqr      & 21 31 06.50 & $-$07 34 20.4 &   26.3 &    311.350    &   3.026&  3.718  \\
14 &G064.549$+$76.014 & RT CVn      & 13 48 44.69 & $+$33 43 34.2 &   26.3 &    253.600    &   3.651&  4.738  \\
15 &G085.581$-$67.859 & V Cet       & 23 57 54.07 & $-$08 57 31.2 &   50.9 &    258.910    &   1.865&  2.153  \\
16 &G131.720$-$64.091 & Z Cet       & 01 06 45.20 & $-$01 28 51.8 &    4.0 &    184.405    &   1.053&  1.270  \\
17 &G133.797$-$53.388 & S Psc       & 01 17 34.56 & $+$08 55 52.0 &    4.3 &    404.620    &   0.941&  1.246  \\
18 &G141.940$-$58.536 & R Psc       & 01 30 38.35 & $+$02 52 52.5 &  -57.0 &    345.250    &   0.958&  1.484  \\
19 &G149.396$-$46.550 & S Ari       & 02 04 37.67 & $+$12 31 36.9 &   9.5  &    291.000    &   2.602&  2.335  \\
20 &G165.616$-$40.899 & YZ Ari      & 02 57 27.52 & $+$11 18 05.3 &   13.0 &    447.000    &   2.691&  2.635  \\
21 &G166.965$-$54.751 & R Cet       & 02 26 02.31 & $-$00 10 42.0 &   35.2 &    166.240    &   0.623&  1.376  \\
22 &G168.980$+$37.738 & X UMa       & 08 40 49.50 & $+$50 08 11.9 &  -82.9 &    249.040    &   2.329&  3.397  \\
23 &G179.379$+$30.743 & ---         & 08 05 03.70 & $+$40 59 08.1 &  -10.6 &               &        &  3.344  \\
24 &G180.069$-$36.185 & V1083 Tau   & 03 43 43.89 & $+$06 55 30.5 &   58.0 &    343.000    &   3.322&  4.451  \\
25 &G180.829$+$32.784 & W Lyn       & 08 16 46.88 & $+$40 07 53.3 &  -24.8 &    295.200    &   2.106&  2.368  \\
26 &G182.006$-$35.653 & V1191 Tau   & 03 49 27.68 & $+$06 04 40.4 &   61.0 &    338.000    &   2.373&  1.654  \\
27 &G183.614$+$31.966 & RT Lyn      & 08 14 50.64 & $+$37 40 11.7 &   27.5 &    394.600    &        &  1.616  \\
28 &G195.025$-$53.735 & SS Eri      & 03 11 53.14 & $-$11 52 32.4 &   33.2 &    316.700    &   3.065&  3.334  \\
29 &G198.593$-$69.596 & RY Cet      & 02 16 00.08 & $-$20 31 10.5 &   -2.8 &    369.000    &   1.270&  1.433  \\
30 &G211.919$+$50.661 & V Leo       & 10 00 01.99 & $+$21 15 43.9 &  -25.5 &    273.350    &   1.530&  1.002  \\
31 &G235.246$+$67.258 & TZ Leo      & 11 23 40.03 & $+$16 51 07.0 &   12.9 &    327.750    &   0.795&  1.122  \\
32 &G248.071$-$84.665 & U Scl       & 01 11 36.38 & $-$30 06 29.4 &  -10.3 &    333.730    &   1.771&  2.436  \\
33 &G261.694$+$46.256 & RT Crt      & 11 01 55.14 & $-$07 39 41.8 &   32.9 &    263.100    &   2.032&  3.177  \\
34 &G315.566$+$57.522 & VY Vir      & 13 18 30.52 & $-$04 41 03.2 &   72.0 &    278.520    &   1.579&  2.133  \\
35 &G325.570$+$85.690 & T Com       & 12 58 38.90 & $+$23 08 21.0 &   26.1 &    406.000    &   1.838&  1.875  \\
36 &G330.757$+$45.262 & Z Vir       & 14 10 22.10 & $-$13 18 11.8 &   66.8 &               &        &  3.022  \\
37 &G334.109$+$36.043 & LY Lib      & 14 37 29.14 & $-$20 19 41.2 &  -25.4 &    283.500    &   3.379&  4.111  \\
38 &G335.504$+$35.524 & SX Lib      & 14 42 46.26 & $-$20 12 36.1 &  -31.2 &    331.450    &   1.952&  1.842  \\
39 &G336.532$+$38.006 & V Lib       & 14 40 22.18 & $-$17 39 26.9 &   18.1 &    255.650    &   3.047&  2.714  \\
40 &G337.373$+$32.451 & EG Lib      & 14 55 21.62 & $-$22 00 19.6 &   -7.8 &    386.000    &   1.312&  1.805  \\
41 &G339.224$+$44.663 & KS Lib      & 14 32 59.87 & $-$10 56 03.2 &   70.7 &    375.500    &   3.788&  3.076  \\
42 &G340.829$+$31.460 & YY Lib      & 15 08 10.66 & $-$21 10 00.3 &   -3.9 &    229.865    &   3.406&  3.465  \\
43 &G353.826$+$42.588 & Y Lib       & 15 11 41.26 & $-$06 00 41.2 &   14.9 &    276.350    &   1.285&  1.875  \\
44 &G356.642$+$59.618 & AP Vir      & 14 28 30.27 & $+$07 17 37.1 &   37.2 &    306.500    &   2.295&  1.961  \\
\hline
\multicolumn{9}{l}{{{\bf Note:} Column 1 are ID of sources; Column 2 are Galactic coordinate notated source names; column 3 are Bayer designation}}\\
\multicolumn{9}{l}{{names of variables; column 4 and 5 are equatorial coordinates;column 6 are $V_{\rm LSR}$; column 7 are periods of }}\\
\multicolumn{9}{l}{{variables; column 8 and 9 are PLR and WISE distances.}}\\
\end{tabular}
\end{table*}

In Table \ref{tab-1}, we list the SiO masers detected in this survey. Of the 44
objects, 35 have no previous report of SiO masers. Apart from 1 infrared source
(G179.370+30.743) and 2 Mira candidates (G011.025+53.268 and G353.826+42.588),
all others are known as Miras. The detection rate of this SiO maser survey was 20\%
 or 39\% considering a Mira-only subsample. Table
\ref{tab-9} details the observational results of the 44 SiO maser
sources. 

\begin{figure*}
\includegraphics[width=16cm]{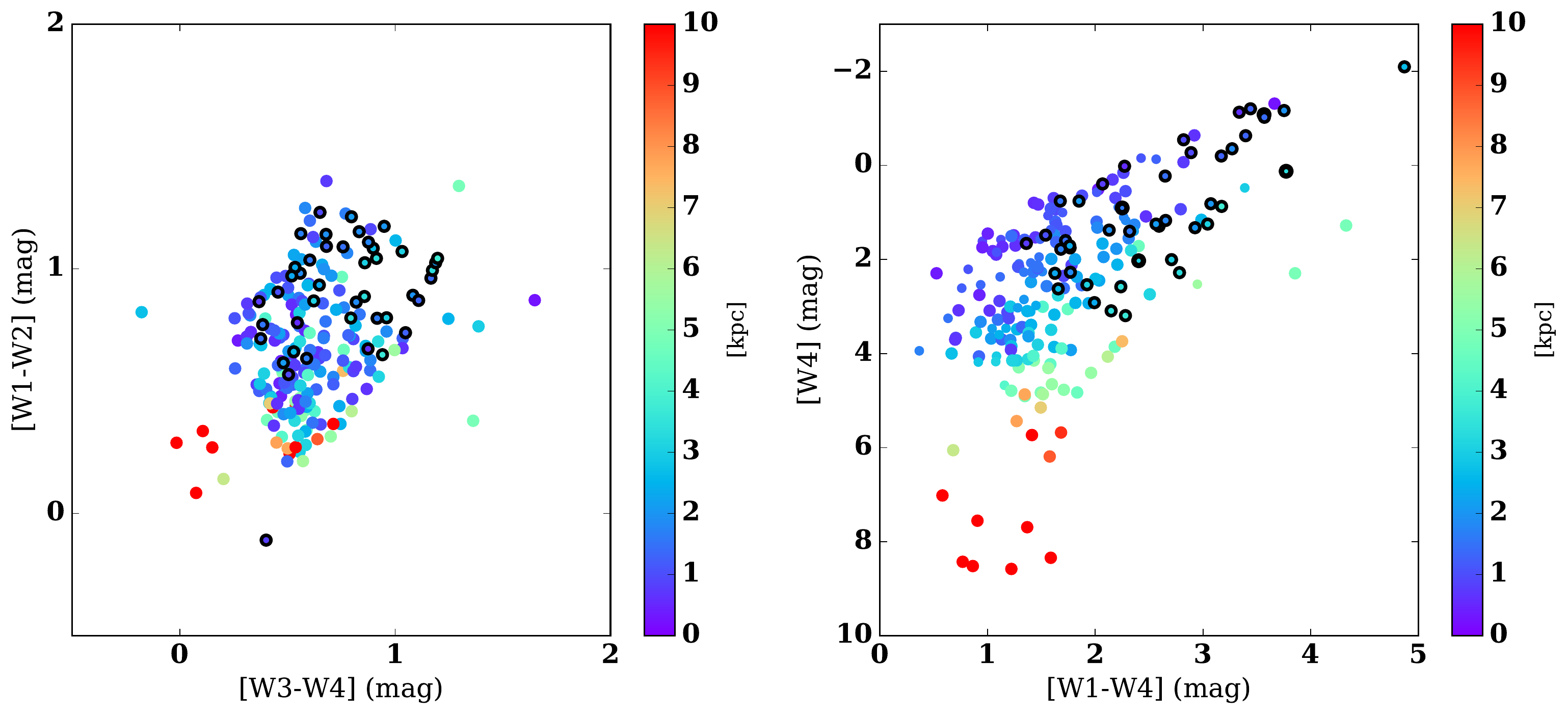} 

\caption{\textit{Left panel}: WISE colour-colour ([$W3$-$W4$]V.S.[$W1$-$W2$]) diagram.
Colors denote distances. With/without black edge circles denote our targets
with/without SiO maser detections.  \textit{Right panel}: Symbols are same as
left panel but for WISE colour-magnitude ([$W1$-$W4$]V.S.[$W4$]) diagram.
\label{fig-3}}
\end{figure*}

\begin{figure}
\includegraphics[width=8cm]{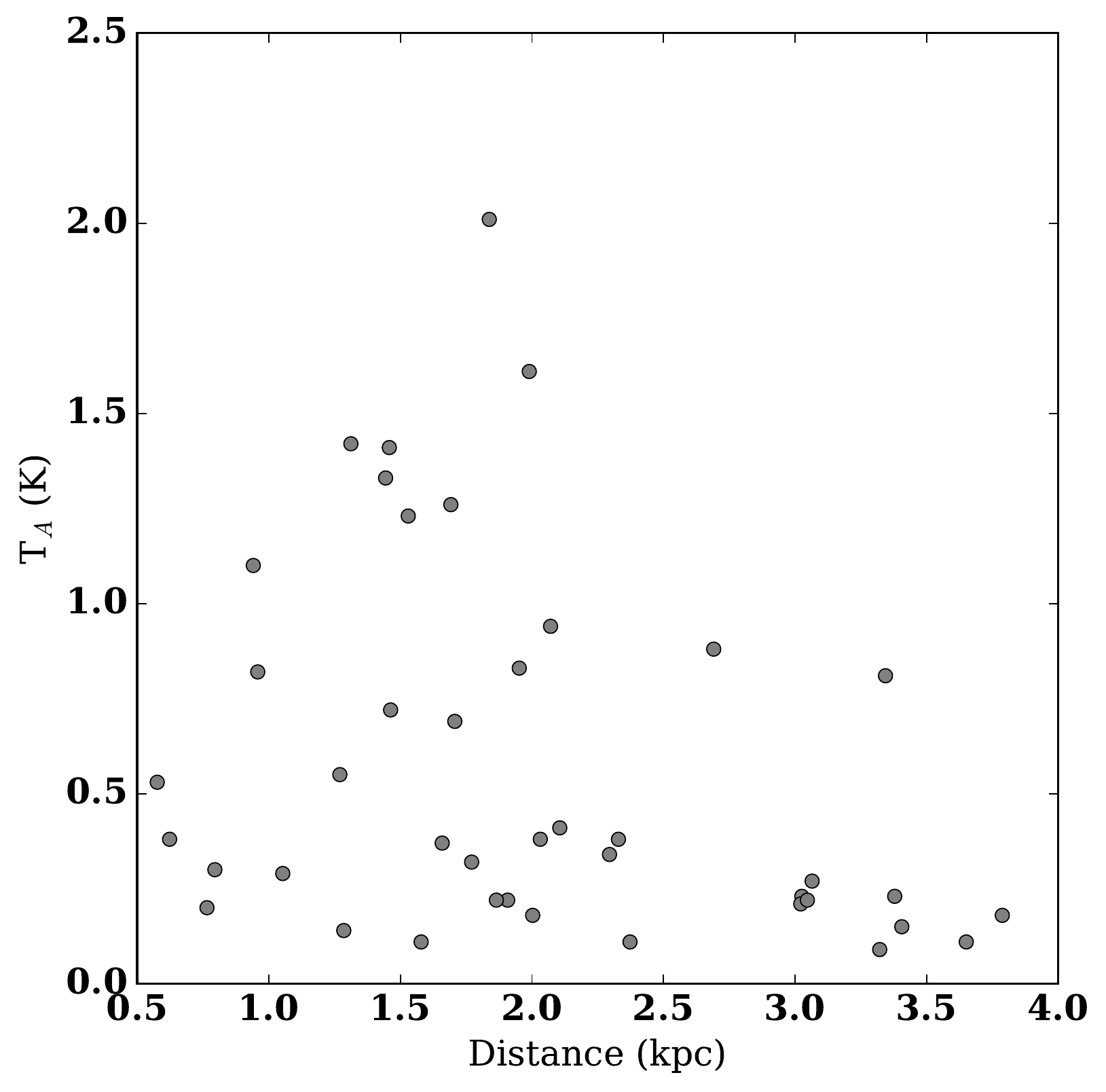} 
\caption{Distance versus SiO maser flux density in unit of
antenna temperature. \label{fig-4}}
\end{figure}

Fig.~\ref{fig-3} shows the colour-colour and colour-magnitude diagrams of our
sample with/without maser detections. We found that the $W1$-$W4$ colour is
higher than 1.3 for all SiO maser sources, thus presenting a suitable selection
criteria for future AGB maser surveys. As indicated by colours of dots in
Fig.~\ref{fig-3}, all the SiO masers detected by this survey are within 5~kpc.
Methods used to derive distances are presented in Section
\ref{subsec:distance}. 

In Fig.~\ref{fig-4}, we plot distances versus antenna temperatures of SiO
v=1 J =1-0 maser emission. There is a trend that SiO maser line emission at
near distances tend to be brighter than those of sources at further distances.


\begin{table*}
\caption{Galactic SiO maser catalogue}
\label{tab-2}
\begin{tabular}{rcccccccccccccc}
\hline
ID & Source & R.A.(2000) &DEC.(2000) & $V_{\rm LSR}$ &J &H &Ks &$W1$ &$W2$ &$W3$ &$W4$ & Period & Distance  & Ref.\\
 &Name & (h:m:s) & (d:m:s) & (km s$^{-1}$) & (mag)  & (mag)  & (mag) & (mag) & (mag) & (mag)  & (mag) & (day)& (kpc) & \\
\hline
1&G108.713$-$35.564& 00 00 06.6  &25 53 11.3 & -29.0&2.225& 1.317& 0.915& 2.838& 1.890&-0.615&-0.712& 327.400& 0.53/---  &  17 \\
2&G116.144$-$06.556& 00 03 21.3  &55 40 50.0 & -17.5&2.770& 1.810& 1.135& 1.211& 1.012&-0.958&-2.165& 413.480& 0.70/---  &  43 \\
3&G116.145$-$06.557& 00 03 21.6  &55 40 48.0 &   2.0&2.770& 1.810& 1.135& 1.211& 1.012&-0.958&-2.165& 413.480&  ---/---  &  6  \\
4&G113.251$-$21.875& 00 04 20.5  &40 06 36.0 & -91.0&3.570& 2.592& 2.084& 2.315& 1.504& 0.560& 0.171& 313.000&  ---/1.07 &  6  \\
5&G039.912$-$80.045& 00 07 36.2  &25 29 40.0 &  22.9&4.254& 3.199& 2.652& 2.444& 1.501& 0.114&-0.606& 411.000& 1.39/1.43 &  43 \\
6&G116.964$-$07.516& 00 10 09.1  &54 52 34.3 & -35.2&4.822& 3.747& 3.043& 2.683& 1.762& 1.425& 0.745& 396.000& 1.63/1.20 &  46 \\
\hline
\multicolumn{15}{l}{{{\bf Note:} Column 1  are ID of sources; Column 2 are Galactic coordinate notated source names;; 
column 3 and 4 are equatorial coordinates; column 5 are $V_{\rm LSR}$;}}\\
\multicolumn{15}{l}{{olumn 6, 7, 8 are 2MASS J, H, Ks magnitudes; column 9,10, 11, 12 are WISE 4 bands magnitudes;  column 13 are periods of 
variables; column 14 are PLR and}}\\
\multicolumn{15}{l}{{WISE distances; last column gives references. The full catalogue together with 48 references is present in online materials.}} \\
\end{tabular}
\end{table*}

\subsection{Galactic SiO maser survey catalogue}\label{subsec:sio-catalogue}

In order to identify potential SiO maser sources in the Sgr stellar stream, in
addition to our observations, we compiled a Galactic SiO maser catalogue that
includes $\sim$2300 sources from 48 papers published before 2017 (Table
\ref{tab-2}). These SiO masers were observed at transitions of v=0,1 J=1-0
(43/42 GHz) or v=0,J=2-1 (86 GHz). For sources which were reported in more
than one papers, the mean $V_{\rm LSR}$ value was calculated. 

The positional accuracy of the WISE point source catalogue is 0.5 \arcsec.
Regarding SiO masers surveys, aside from several blind surveys towards the
Bugle and Galactic centre region \citep{1997ApJ...478..206S,
1998ApJ...494L..89I}, and interferometric (JVLA \&ATCA) surveys toward the
Galactic centre region \citep{2002A+A...391..967S, 2010ApJ...720L..56L}, nearly
all other surveys have samples selected from IRAS, 2MASS, MSX or variable star
catalogues. Despite the majority of maser surveys being conducted with single
dish observations with beam sizes $\sim$ 40 \arcsec, the positional accuracy of
most infrared selected sources is as good as 3 \arcsec.  When matching with the
WISE point source catalogue, we used a search radius of 5 \arcsec. We further
removed extended sources and spurious sources, such as diffraction spikes and
halos in nearby bright sources. Most sources without 2MASS or WISE conterparts
are sources towards the Bulge and Galactic centre where high stellar densities
result in source confusion. In summary, we have compiled a catalogue including
$\sim$2300 SiO masers, and derived distances for $\sim$1000 of them. The full
catalogue is included in the online material.

\section{Discussions} \label{sec:discuss}

\begin{figure*}
\includegraphics[width=16cm]{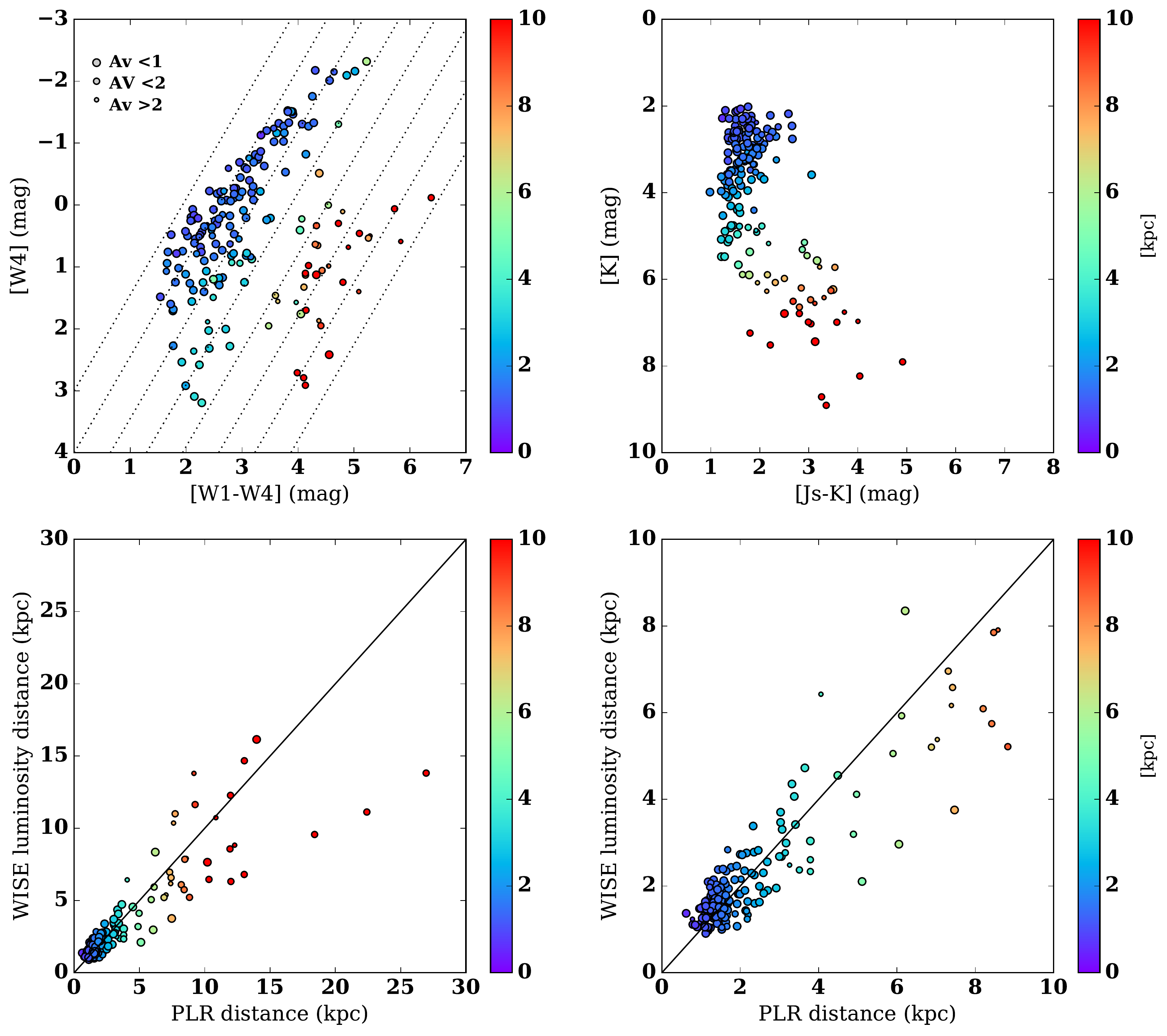}
\caption{
\textit{Top Left panel}: WISE colour-magnitude ([$W1$-$W4$] VS [$W4$]) diagram, with
colour of symbols denotes distances and size of symbols denotes interstellar
extinction.\textit{Top Right panel}: 2MASS colour-magnitude ([Js]-[K] VS [K]).
\textit{Bottom Left panel}: PLR distances versus WISE luminosity distances.
\textit{Bottom Right panel}: sample as bottom left panel, but with a zoom on
inner 10~kpc range. \label{fig-5}}
\end{figure*}

\subsection{Distance and Galactic distribution}\label{subsec:distance}
Distance is a key parameter in investigations of the distribution of Galactic objects \citep{2012PASJ...64..136H, 2014ApJ...783..130R, 2016ARA+A..54..529B}. For SiO
masers in Miras with known periods and Ks magnitudes, their distances can be
calculated via the period luminosity relation (PLR). Formulas used
to derive the PLR distances are given in Appendix \ref{subsec:A1}. 

In the top panel of Fig.~\ref{fig-5}, we present the WISE and 2MASS
colour-magnitude diagram of O-rich Miras in our combined catalogue.
In the WISE [$W1$-$W4$] versus
[$W4$] diagram, a clear gradient of distances from the top left to the bottom
right direction can be seen. Redder sources tend to be brighter at $W4$ band
due to reddening by the circumstellar dusty envelope.
As described in  Appendix \ref{subsec:A2}, 
by using Miras with known PLR distances, we derived an empirical relation
to derive the distance based on the WISE [$W1$-$W4$] colour and [$W4$] magnitude for
O-rich Miras (Equation \ref{eq-A9}). The details on the derivation and
calibration of this relationship is given in appendix \ref{subsec:A2}. This
relation can be used to estimate WISE luminosity distances of O-rich Miras
without known periods or Ks magnitudes. In the lower panels of
Fig.~\ref{fig-5}, we plot the WISE distances against the PLR distances for Miras with
both distances. The mean and standard deviation of
$D_{WISE}$-$D_{PLR}$ is $-$0.26 and 1.6~kpc. Generally, distances estimated by
these two methods agree within 2 kpc. Most large deviations are sources towards
the Bulge and Galactocentrc region where interstellar extinction is high (Av~$>$~2). In
summary, we obtained PLR distances for 587 sources and WISE distances for 749
sources. 

\begin{figure*}
\includegraphics[width=16cm]{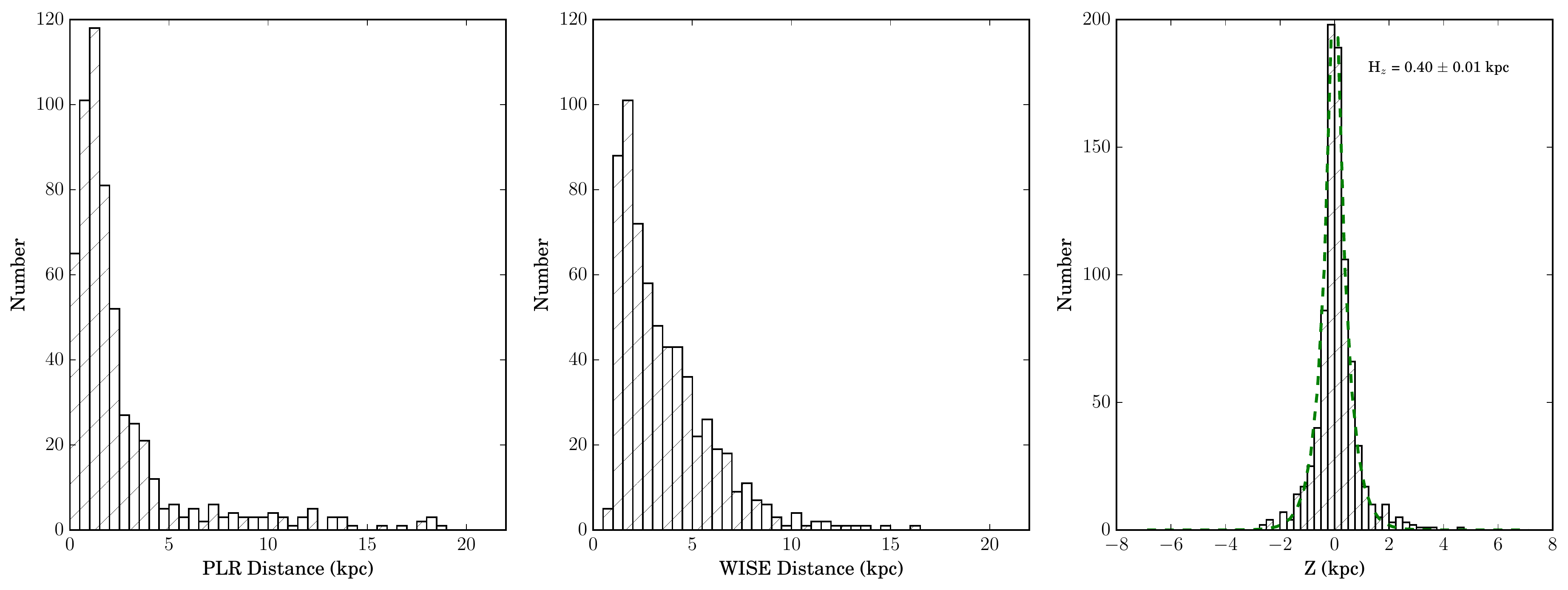} 
\caption{\textit{Left and middle panels} are histograms of PLR
distances and WISE distances. \textit{Right panel} is the histogram of Galactic
plane distances.  The green dash line denotes the fitted exponential density profile, with a scale height of 0.40~kpc. \label{fig-6}}
\end{figure*}

In Fig.~\ref{fig-6}, we show histograms of heliocentric distances and distances
from the Galactic plane. The majority of PLR distances are within 5 kpc. This
is due to high extinction at further distances. In contrast, there are only a
few sources with the WISE distances within 2~kpc. This is due to the saturation
of WISE phototmetry for nearby AGBs. 

\begin{figure*}
\centering
\includegraphics[width=16cm]{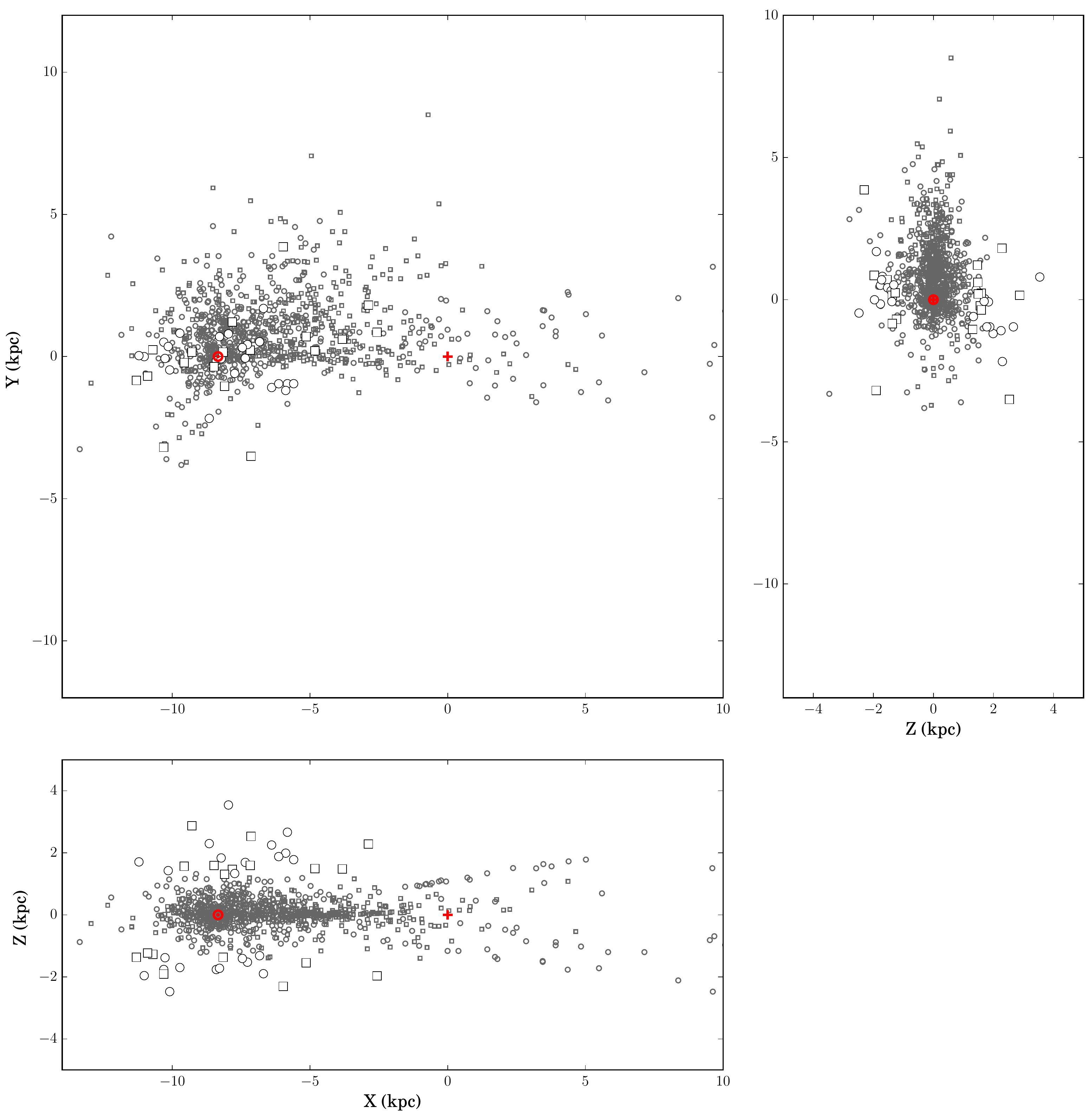} 

\caption{3D locations of SiO masers in the Galactocentric Cartesian
coordinates, with the Galactic centre at the zero point and the Sun at [$-$8.34,
0, 0] kpc. Squares denote sources with PLR distances and circles denote sources
with WISE luminosity distances. In $XZ$ plot, large symbols denote off-plane SiO
masers. Circles denote sources detected by this survey, Squares denote sources
from published references. \label{fig-7}} 
\end{figure*}

With these distances, we estimated a Galactic plane scale height of 0.40~kpc
for Miras with SiO masers, by fitting the histogram of Galactic plane distances
with an exponential density profile(right panel of Fig.~\ref{fig-6}). When
estimating the scale height, only sources with distances smaller than 7~kpc
were used, given that distance uncertainties of sources towards the Bulge and
Galactocentric region tend to be large (Fig. \ref{fig-5} and
Fig.~\ref{fig-7}). It should be mentioned that more than half of the SiO masers are
located within 3~kpc of the Sun, as can be seen in Fig.~\ref{fig-6} and
Fig.~\ref{fig-7}, so that the scale height of 0.40~kpc we estimated here can
be an average scale height of O-rich AGBs around 3~kpc of the sun.

In Fig.~\ref{fig-7} we present the locations of these SiO masers in the Galaxy.
We identified 42 off-plane sources with Galactic plane distances higher
than 1.2~kpc, which are listed in Table \ref{tab-3} and highlighted in
Fig.~\ref{fig-7}. Around 50\% of these off-plane SiO maser were found by this
survey. 

\begin{table*}
\caption{Off plane SiO masers}
\label{tab-3}
\begin{tabular}{rccccccccl}
\hline
ID & Source & R.A.(2000) &DEC.(2000) & $V_{\rm LSR}$ & Period & PLR Distances& WISE Distances & $Z$ & Ref.\\
 &Name & (h:m:s) & (d:m:s) & (km s$^{-1}$) &  (day) &(kpc) &(kpc) & (kpc)  & \\
\hline
1 &G039.912$-$80.045& 00 07 36.2 &$-$25 29 40.0&  22.9&  411.000& 1.39&  1.43&-1.37&8          \\
2 &G248.071$-$84.665& 01 11 36.4 &$-$30 06 29.4& -10.3&  333.730& 1.77&  2.44&-1.76&1          \\
3 &G149.396$-$46.550& 02 04 37.7 &$ $12 31 36.9&   9.5&  291.000&     &  2.34&-1.70&1          \\
4 &G165.616$-$40.899& 02 57 27.5 &$ $11 18 05.2&  14.2&  447.000& 2.69&  2.63&-1.76&1, 2       \\
5 &G195.025$-$53.735& 03 11 53.1 &$-$11 52 32.4&  33.2&  316.700& 3.07&  3.33&-2.48&1          \\
6 &G180.069$-$36.185& 03 43 43.9 &$ $06 55 30.5&  58.0&  343.000& 3.32&  4.45&-1.96&1          \\
7 &G182.006$-$35.653& 03 49 27.7 &$+$06 04 40.4&  61.0&  338.000& 2.37&  1.65&-1.38&1          \\
8 &G174.233$-$28.113& 03 54 23.6 &$+$16 01 01.9&  45.5&  334.500& 2.70&  1.96&-1.27&2          \\
9 &G195.134$-$24.895& 04 50 57.3 &$+$03 08 32.3& -13.2&  358.500& 2.92&  1.97&-1.23&2          \\
10&G195.853$-$23.965& 04 55 30.3 &$+$03 04 28.1&  47.0&  396.000& 3.37&  3.45&-1.37&2          \\
11&G238.258$-$26.921& 05 47 58.7 &$-$33 05 10.9& 110.8&  199.300& 4.21&  7.27&-1.91&3          \\
12&G179.379$+$30.743& 08 05 03.7 &$ $40 59 08.1&  -9.8&         &     &  3.34& 1.71&1, 4       \\
13&G168.980$+$37.738& 08 40 49.6 &$ $50 08 12.1& -82.9&  249.040& 2.33&  3.40& 1.43&1          \\
14&G190.023$+$51.458& 09 53 43.5 &$ $34 55 32.0& -11.0&  233.830&     &  2.00& 1.56&5          \\
15&G261.694$+$46.256& 11 01 55.1 &$-$07 39 41.9&  32.9&  263.100&     &  3.18& 2.30&1          \\
16&G170.681$+$71.517& 11 41 40.3 &$ $38 28 29.3& -56.3&  252.460& 3.03&  3.48& 2.87&3          \\
17&G288.878$+$34.294& 11 59 19.1 &$-$27 09 03.5&  72.5&  301.000& 4.49&  4.59& 2.53&4          \\
18&G282.845$+$50.684& 12 00 20.8 &$-$10 11 04.9&  -4.3&  294.765& 1.69&  1.70& 1.31&2          \\
19&G248.033$+$76.318& 12 04 15.5 &$ $18 47 00.0&  -4.0&         &     &  1.64& 1.59&5          \\
20&G325.570$+$85.690& 12 58 38.9 &$ $23 08 21.1&  27.0&  406.000& 1.84&  1.88& 1.83&1, 6       \\
21&G315.565$+$57.522& 13 18 30.4 &$-$04 41 05.1&  72.0&  278.520& 1.58&  2.13& 1.33&1          \\
22&G064.549$+$76.014& 13 48 44.7 &$ $33 43 34.3&  26.3&  253.600& 3.65&  4.74& 3.54&1          \\
23&G330.755$+$45.262& 14 10 21.4 &$-$13 18 14.8&  66.8&  304.105& 3.17&  3.02& 2.25&1          \\
24&G356.642$+$59.618& 14 28 30.3 &$ $07 17 37.1&  37.2&  306.500&     &  1.96& 1.69&1          \\
25&G339.224$+$44.663& 14 32 59.9 &$-$10 56 03.4&  69.2&  375.500& 3.79&  3.08& 2.66&1, 4       \\
26&G334.109$+$36.043& 14 37 29.1 &$-$20 19 41.2& -25.4&  283.500& 3.38&  4.11& 1.99&1          \\
27&G336.532$+$38.006& 14 40 22.2 &$-$17 39 26.9&  18.1&  255.650& 3.05&  2.71& 1.88&1          \\
28&G340.829$+$31.460& 15 08 10.6 &$-$21 10 00.3&  -3.9&  229.865& 3.41&  3.46& 1.78&1          \\
29&G011.025$+$53.268& 15 08 25.8 &$ $09 36 18.3&   6.8&  313.500& 1.99&  1.82& 1.59&1, 4       \\
20&G011.025$+$53.268& 15 08 25.8 &$ $09 36 18.4& -10.2&  313.500& 1.99&  1.82& 1.59&7, 8       \\
31&G066.876$+$47.996& 16 05 28.9 &$ $42 10 29.4&   0.3&         &     &  1.97& 1.46&4          \\
32&G003.227$+$22.945& 16 32 24.6 &$-$13 12 01.3& -28.4&  332.000&     &  3.83& 1.49&9          \\
33&G007.656$+$18.016& 16 58 46.7 &$-$12 43 46.9& -62.7&         &     &  4.79& 1.48&2          \\
34&G018.305$+$21.632& 17 08 10.3 &$-$02 20 22.6& -40.4&         &     &  6.19& 2.28&4          \\
35&G008.400$-$18.623& 19 19 09.6 &$-$29 43 18.0&  33.7&         &     &  6.16&-1.97&7          \\
36&G012.358$-$25.239& 19 53 21.8 &$-$28 30 39.0& -16.7&  343.000& 3.62&  4.42&-1.54&7          \\
37&G019.002$-$39.495& 21 02 20.8 &$-$27 05 14.7& -51.8&  277.540& 2.07&  1.13&-1.32&1          \\
38&G058.520$-$26.960& 21 14 29.6 &$ $07 48 33.7&  28.6&  429.000&     &  5.08&-2.30&2          \\
39&G045.734$-$38.770& 21 31 06.5 &$-$07 34 20.4&  26.3&  311.350& 3.03&  3.72&-1.90&1          \\
40&G021.513$-$53.023& 22 03 45.8 &$-$28 03 03.3&-100.8&  271.350& 1.91&  2.47&-1.53&1          \\
41&G019.509$-$56.308& 22 18 00.2 &$-$29 36 13.5& -29.0&  294.950& 1.69&  1.69&-1.41&1          \\
42&G085.581$-$67.859& 23 57 54.1 &$-$08 57 31.2&  50.9&  258.910& 1.86&  2.15&-1.72&1          \\
\hline
\multicolumn{10}{l}{{{\bf Note:} Column 1 are ID of sources; Column 2 are Galactic coordinate notated source names; column 3 and 4 are equatorial coordinates;}}\\
\multicolumn{10}{l}{{column 5 are $V_{\rm LSR}$; column 6 are periods of variables; column 7 and 8 are PLR and WISE distances; column 9 are off plane distances;}}\\
\multicolumn{10}{l}{{column 10 denote references.}}\\
\multicolumn{10}{l}{Reference: (1) this paper; (2) \cite{2012PASJ...64....4D}; (3) \cite{2007PASJ...59..559D};(4) \cite{2001A+A...376..112I} (5) \cite{1990ApJS...74..911B};} \\
\multicolumn{10}{l}{(6) \cite{2010PASJ...62..525D}; (7) \cite{2004PASJ...56..765D} (8) \cite{2010ApJS..188..209K}; (9) \cite{2005PASJ...57L...1M}} \\
\end{tabular}
\end{table*}

\subsection{Kinematics}\label{subsec:disc_star}

The Galactic distribution shown in Fig.~\ref{fig-7} indicates that the majority
of Galactic SiO maser sources are within the Galactic disc. Regarding the
off-plane SiO sources listed in Table \ref{tab-3}, we test in this section 
whether these are halo/stream stars or thick-disc stars based on comparison of
their line of sight velocities and those predicted by disc star models.

For disc stars, it is well known that late-type stars with higher velocity
dispersions tend to have slower Galactocentric rotational speeds. This
phenomenon is called ``asymmetric drift'' \citep{1998MNRAS.298..387D}.
\citet{2012A+A...547A..70P} studied the kinematics of thick disc stars as part
of the RAdial Velocity Experiment (RAVE) and derived a rotational lag of
49$\pm$6 \kms\ with respect to the LSR. \citet{2015ApJ...809..145T} determined the
Solar motion using LAMOST DR1 data, and found a $\sim$3~\kms\ asymmetric motion
of stars with $T_{eff}<$6000~$K$ with respect to stars with $T_{eff}>$6000~$K$.
Before investigating the kinematics of off-plane sources, we first studied the
circular rotational speed and asymmetric drift of disc O-rich Miras. 

\begin{table*}
 \caption{Variables used in kinematic analysis}
 \begin{tabular}{ll}
 \hline
 Variables & Definitions \\
 \hline
$D$ & heliocentric distance \\
$R$ & Galactocentric distance \\
$X$,$Y$,$Z$ & Galactocentric Cartesian coordinates \\
$R_0$ & Galactocentric radius of the Sun \\
$\Theta_0$ & rotation velocity of the $\rm LSR$ around the Galactic centre \\
$U_\odot$, $V_\odot$, $W_\odot$, & velocity of the Sun with respect to the $\rm LSR$ \\
$\Theta$ & mean rotation velocity of stars around the Galactic centre \\
$\overline{U_s}$, $\overline{V_s}$,  $\overline{W_s}$ & peculiar (non-circular) motion of stars in cylindrical Galactocentric coordinates \\
$V_{\rm LSR}$ & line-of-sight velocity of stars with respect to the $\rm LSR$\\
$V_{\rm helio}$ & heliocentric line-of-sight velocity of stars \\
$V_{\rm obs}$ & observed heliocentric line-of-sight velocity of stars\\
$V_{\rm model}$ & heliocentric line-of-sight velocity of stars calculated by model\\
$S$ & dispersion of heliocentric line-of-sight velocity (Eq. \ref{eq-3})\\
\hline
\end{tabular}
\end{table*}

\begin{table*}
\caption{velocity dispersion and mean velocity of $R$-$Z$ binned groups.}
 \label{tab-5}
\begin{tabular}{cccccccc}
\hline
$Z$(kpc) & $R$(kpc) & Number & \textit{S}  & $\Theta+\overline{V_s}$  & $\overline{U_s}$  & $\overline{W_s}$ & ID\\
      & &        & (\kms) & (\kms) & (\kms) & (\kms) & \\
\hline
-0.1~$<Z<$~+0.1&+5.0~$<R<$~+9.0  &  192 &  +44.8 $\pm$  1.1 & +236.7 $\pm$  7.6 &   -9.6 $\pm$  6.2 &   +3.9 $\pm$ 10.7 & \\  
-0.3~$<Z<$~-0.1&+5.0~$<R<$~+9.0  &   82 &  +36.6 $\pm$  0.7 & +225.2 $\pm$  6.1 &   +6.7 $\pm$  6.2 &   -7.8 $\pm$  8.6 & \\
+0.1~$<Z<$~+0.3&+5.0~$<R<$~+9.0  &  100 &  +34.4 $\pm$  0.6 & +215.4 $\pm$  5.6 &   +3.1 $\pm$  6.5 &   +7.6 $\pm$  8.5 & \\
-0.5~$<Z<$~-0.3&+5.0~$<R<$~+9.0  &   47 &  +31.9 $\pm$  0.9 & +220.1 $\pm$  5.3 &   -4.7 $\pm$  7.0 &  -15.3 $\pm$  7.3 & \\
+0.3~$<Z<$~+0.5&+5.0~$<R<$~+9.0  &   67 &  +37.4 $\pm$  1.0 & +217.1 $\pm$  6.6 &  -13.8 $\pm$  5.7 &  +10.1 $\pm$  8.3 & \\
-0.8~$<Z<$~-0.5&+5.0~$<R<$~+9.0  &   33 &  +36.4 $\pm$  0.7 & +217.4 $\pm$  6.7 &  -11.9 $\pm$  6.0 &   -5.3 $\pm$  7.7 & \\
+0.5~$<Z<$~+0.8&+5.0~$<R<$~+9.0  &   56 &  +35.2 $\pm$  1.0 & +213.3 $\pm$  6.5 &  -11.4 $\pm$  6.5 &   +1.1 $\pm$  7.7 & \\
+0.8~$<Z<$~+1.4&+5.0~$<R<$~+9.0  &   34 &  +39.8 $\pm$  0.6 & +208.5 $\pm$  7.7 &   -2.8 $\pm$  6.9 &   +2.0 $\pm$  7.2 & \\
-1.4~$<Z<$~-0.8&+5.0~$<R<$~+9.0  &   23 &  +34.5 $\pm$  1.6 & +225.7 $\pm$  8.3 &   -4.8 $\pm$  6.8 &  +18.9 $\pm$  6.4 &3\\
+1.4~$<Z<$~+2.5&+5.0~$<R<$~+9.0  &   20 &  +33.3 $\pm$  3.8 & +189.9 $\pm$  8.9 &  -12.2 $\pm$  7.9 &  +33.6 $\pm$  5.9 &A\\
\hline                                                                                                                    
-0.1~$<Z<$~+0.1&+9.0~$<R<$~+15.0 &   32 &  +24.5 $\pm$  0.9 & +209.8 $\pm$  6.5 &   -5.1 $\pm$  4.2 &   -4.1 $\pm$ 11.1 & \\
+0.1~$<Z<$~+0.3&+9.0~$<R<$~+15.0 &   24 &  +25.7 $\pm$  1.0 & +213.0 $\pm$  6.2 &   +7.3 $\pm$  5.0 &   +3.1 $\pm$ 12.6 & \\
-0.3~$<Z<$~-0.1&+9.0~$<R<$~+15.0 &   21 &  +24.1 $\pm$  0.8 & +214.5 $\pm$  6.0 &   +2.9 $\pm$  4.8 &  -36.5 $\pm$ 19.1 & \\
+0.3~$<Z<$~+0.5&+9.0~$<R<$~+15.0 &   19 &  +30.1 $\pm$  2.3 & +214.3 $\pm$  8.1 &  +35.1 $\pm$ 11.9 & +112.3 $\pm$ 31.8 &1\\
-0.5~$<Z<$~-0.3&+9.0~$<R<$~+15.0 &   19 &  +35.5 $\pm$  1.1 & +190.8 $\pm$  8.3 &  +17.2 $\pm$  6.5 &  -33.6 $\pm$ 17.8 & \\
+0.5~$<Z<$~+0.8&+9.0~$<R<$~+15.0 &   23 &  +35.2 $\pm$  1.7 & +235.5 $\pm$  7.0 &   -9.4 $\pm$  6.3 &   +9.4 $\pm$ 10.2 & \\
-0.8~$<Z<$~-0.5&+9.0~$<R<$~+15.0 &   12 &  +29.6 $\pm$  5.7 & +199.6 $\pm$  7.3 &  -71.5 $\pm$  9.6 & +175.3 $\pm$ 21.8 &2\\
-0.8~$<Z<$~-0.3&+9.0~$<R<$~+15.0 &   27 &  +36.1 $\pm$  0.9 & +201.1 $\pm$  7.8 &   -2.9 $\pm$  5.5 &  +33.6 $\pm$ 10.6 & \\
+0.3~$<Z<$~+0.8&+9.0~$<R<$~+15.0 &   38 &  +32.4 $\pm$  0.8 & +227.8 $\pm$  7.0 &  +13.4 $\pm$  6.8 &  +43.7 $\pm$ 14.0 &C\\
-2.0~$<Z<$~-0.8&+9.0~$<R<$~+15.0 &   18 &  +39.7 $\pm$  2.3 & +176.0 $\pm$ 13.0 &  +25.2 $\pm$  9.3 &  -54.3 $\pm$ 11.0 &B\\
+0.8~$<Z<$~+2.0&+9.0~$<R<$~+15.0 &   15 &  +31.7 $\pm$  3.1 & +219.8 $\pm$  8.3 &  +16.1 $\pm$  7.3 &  -21.3 $\pm$  8.3 & \\
\hline
\multicolumn{8}{l}{{\bf Note:} column 1 and 2 are binned ranges of $R$ and $Z$; column 3 are numbers of sources in $R$-$Z$ binned regions; }\\
\multicolumn{8}{l}{column 4 to 7 are \textit{S}, $\overline{U_s}$, $\Theta$+$\overline{V_s}$, $\overline{W_s}$. The last column gives IDs of moving groups and high-speed stars.}\\
\end{tabular}
\end{table*}
\begin{table*}
\caption{Source list of the northern (group A)  and southern (group B) off-plane moving groups}
\label{tab-6}
\begin{tabular}{rccccccccccc}
\hline
ID & Source & Other  & R.A.(J2000)&DEC.(J2000)& $V_{\rm LSR}$ &$\Lambda_\odot$ & $B_\odot$ & $V_{GSR}$  & Period & PLR Distance &WISE Distance\\
 &Name  &Name&(h:m:s)& (d:m:s)  &(km s$^{-1}$)  & ($\degr$) & ($\degr$)& (km s$^{-1}$) & (day) & (kpc) & (kpc) \\
\hline
 1&G248.033+76.318 & R Com      &12 04 15.5&  18 47 00.0&    -4.0 &  249.84 &   -1.65   &  -52.3  &        &        &  1.64   \\
 2&G288.878+34.294 & V0450 Hya  &11 59 19.1& -27 09 03.5&    72.5 &  273.40 &  +39.65   &  -99.5  & 301.0  &    4.49&  4.59   \\
 3&G286.549+55.660 & T Vir      &12 14 36.7& -06 02 08.9&     6.6 &  263.98 &  +19.58   & -112.4  & 339.5  &    1.39&  2.04   \\
 4&G325.570+85.690 & T Com      &12 58 38.9&  23 08 21.1&    27.0 &  258.91 &  -11.29   &   17.7  & 406.0  &    1.84&  1.88   \\
 5&G315.565+57.522 & VY Vir     &13 18 30.4& -04 41 05.1&    72.0 &  277.45 &  +10.70   &  -10.7  & 278.5  &    1.58&  2.13   \\
 6&G064.549+76.014 & RT CVn     &13 48 44.7&  33 43 34.3&    26.3 &  262.52 &  -26.00   &   74.3  & 253.6  &    3.65&  4.74   \\
 7&G330.755+45.262 & Z Vir      &14 10 21.4& -13 18 14.8&    66.8 &  293.15 &  +11.66   &   -8.9  & 304.1  &    3.17&  3.02   \\
 8&G334.779+50.121 & IO Vir     &14 11 17.6& -07 44 49.8&   -26.8 &  290.35 &   +6.77   &  -86.9  & 500.0  &    2.46&         \\
 9&G325.327+25.630 &            &14 28 09.3& -33 00 03.4&   -19.5 &  308.26 &  +26.02   & -132.3  & 454.0  &    6.46&         \\
10&G356.642+59.618 & AP Vir     &14 28 30.3&  07 17 37.1&    37.2 &  286.03 &   -8.40   &   30.7  & 306.5  &        &  1.96   \\
11&G339.224+44.663 & KS Lib     &14 32 59.9& -10 56 03.4&    69.2 &  296.61 &   +6.84   &   13.7  & 375.5  &    3.79&  3.08   \\
12&G334.109+36.043 & LY Lib     &14 37 29.1& -20 19 41.2&   -25.4 &  302.61 &  +14.34   & -103.1  & 283.5  &    3.38&  4.11   \\
13&G336.532+38.006 & V Lib      &14 40 22.2& -17 39 26.9&    18.1 &  301.74 &  +11.72   &  -50.9  & 255.6  &    3.05&  2.71   \\
14&G340.829+31.460 & YY Lib     &15 08 10.6& -21 10 00.3&    -3.9 &  309.29 &  +11.46   &  -65.5  & 229.9  &    3.41&  3.46   \\
15&G011.025+53.268 & FV Boo     &15 08 25.8&  09 36 18.3&     6.8 &  293.35 &  -15.42   &   32.0  & 313.5  &    1.99&  1.82   \\
16&G066.876+47.996 & V1012 Her  &16 05 28.9&  42 10 29.4&     0.3 &  281.55 &  -49.51   &  135.7  &        &        &  1.97   \\
17&G004.482+32.104 & BD Oph     &16 05 46.3& -06 42 27.8&    -7.5 &  314.40 &   -7.82   &    7.1  & 340.4  &    2.00&  2.79   \\
18&G056.375+43.529 & V0697 Her  &16 27 51.4&  34 48 10.5&    54.5 &  293.92 &  -46.80   &  187.3  & 486.0  &    1.47&  2.14   \\
19&G003.227+22.945 & V0720 Oph  &16 32 24.6& -13 12 01.3&   -28.4 &  323.25 &   -4.73   &  -17.0  & 332.0  &        &  3.83   \\
\hline               
 1&G128.642-50.107 &WX Psc      &01 06 26.0&  12 35 53.0&     7.5 &   99.17 &  -19.03   &  117.7  & 660.0  &    1.59&         \\
 2&G141.939-58.536 &R Psc       &01 30 38.3&  02 52 53.0&   -57.5 &   99.10 &   -7.64   &   13.3  & 345.2  &    0.96&  1.48   \\
 3&G149.396-46.550 &S Ari       &02 04 37.7&  12 31 36.9&     9.5 &  111.52 &  -11.71   &   86.5  & 291.0  &        &  2.34   \\
 4&G166.966-54.751 &R Cet       &02 26 02.3& -00 10 42.0&    37.0 &  109.29 &   +1.92   &   65.6  & 166.2  &    0.62&  1.38   \\
 5&G146.989-21.317 &TV Per      &02 43 48.5&  36 15 02.0&   -25.4 &  133.26 &  -26.89   &   86.3  & 358.0  &    1.35&  2.30   \\
 6&G161.474-41.919 &RU Ari      &02 44 45.5&  12 19 02.9&    20.4 &  119.81 &   -6.59   &   72.4  & 353.5  &    1.23&         \\
 7&G165.616-40.899 &YZ Ari      &02 57 27.5&  11 18 05.2&    14.2 &  121.94 &   -4.18   &   55.5  & 447.0  &    2.69&  2.63   \\
 8&G195.025-53.735 &SS Eri      &03 11 53.1& -11 52 32.4&    33.2 &  112.84 &  +17.83   &   -0.5  & 316.7  &    3.07&  3.33   \\
 9&G180.069-36.185 &V1083 Tau   &03 43 43.9&  06 55 30.5&    58.0 &  129.65 &   +5.13   &   57.8  & 343.0  &    3.32&  4.45   \\
10&G182.006-35.653 &V1191 Tau   &03 49 27.7&  06 04 40.4&    61.0 &  130.49 &   +6.54   &   54.7  & 338.0  &    2.37&  1.65   \\
11&G174.233-28.113 &UY Tau      &03 54 23.6&  16 01 01.9&    45.5 &  136.30 &   -1.79   &   65.0  & 334.5  &    2.70&  1.96   \\
12&G206.048-43.665 &WZ Eri      &04 02 08.7& -13 44 56.0&     9.4 &  123.05 &  +25.71   &  -60.5  & 401.5  &    1.47&         \\
13&G195.134-24.895 &EP Ori      &04 50 57.3&  03 08 32.3&   -13.2 &  143.16 &  +15.89   &  -65.3  & 358.5  &    2.92&  1.97   \\
14&G195.853-23.965 &V1648 Ori   &04 55 30.3&  03 04 28.1&    47.0 &  144.21 &  +16.41   &   -7.9  & 396.0  &    3.37&  3.45   \\
15&G227.912-25.120 &RT Lep      &05 42 33.2& -23 41 41.0&    65.2 &  142.94 &  +46.20   &  -82.6  & 400.0  &    1.12&  3.43   \\
16&G238.258-26.921 &            &05 47 58.7& -33 05 10.9&   110.8 &  137.14 &  +55.22   &  -56.0  & 199.3  &    4.21&  7.27   \\
\hline
\multicolumn{12}{l}{{{\bf Note:} Column 1 are ID of sources; Column 2 are Galactic coordinate notated source names; column 3 are Bayer designation names of variables; column 4 and 5}}\\
\multicolumn{12}{l}{{are equatorial coordinates; column 6 are $V_{\rm LSR}$; column 7, 8 are Sgr stellar stream coordinates; column 9 are radial velocities with respect to the Galactic centre;}}\\
\multicolumn{12}{l}{{column 10 are periods of variables; column 11and 12 are PLR and WISE distances.}}\\

\end{tabular}
\end{table*}
\begin{table*}
\caption{List of very high speed Sources }
\label{tab-7}
 \begin{tabular}{ccccccccccc}
 \hline
 Source & Other & R.A.(J2000) & DEC.(J2000) & $V_{\rm LSR}$    & Period & PLR Distance &WISE Distance & ID\\
 Name  & Name & (h:m:s) & (d:m:s)& (km s$^{-1}$)&(day) &(kpc)&(kpc)\\
\hline
G002.216-37.218 & RV Mic      & 20 40 29.9 & $-$39 37 42.0 & -96.55 &    327.0     &        &  2.29  &1 \\
G196.675+21.331 & S Gem       & 07 43 02.5 & $+$23 26 58.2 &   94.8 &    292.1      &   1.41&  1.14  &2 \\
G138.089-26.921 & T Tri       & 01 56 47.1 & $+$34 01 10.7 & -121.9 &    324.0      &   1.60&  1.87  &3 \\
\hline
 \end{tabular}
\end{table*}

\begin{figure}
\includegraphics[width=8cm]{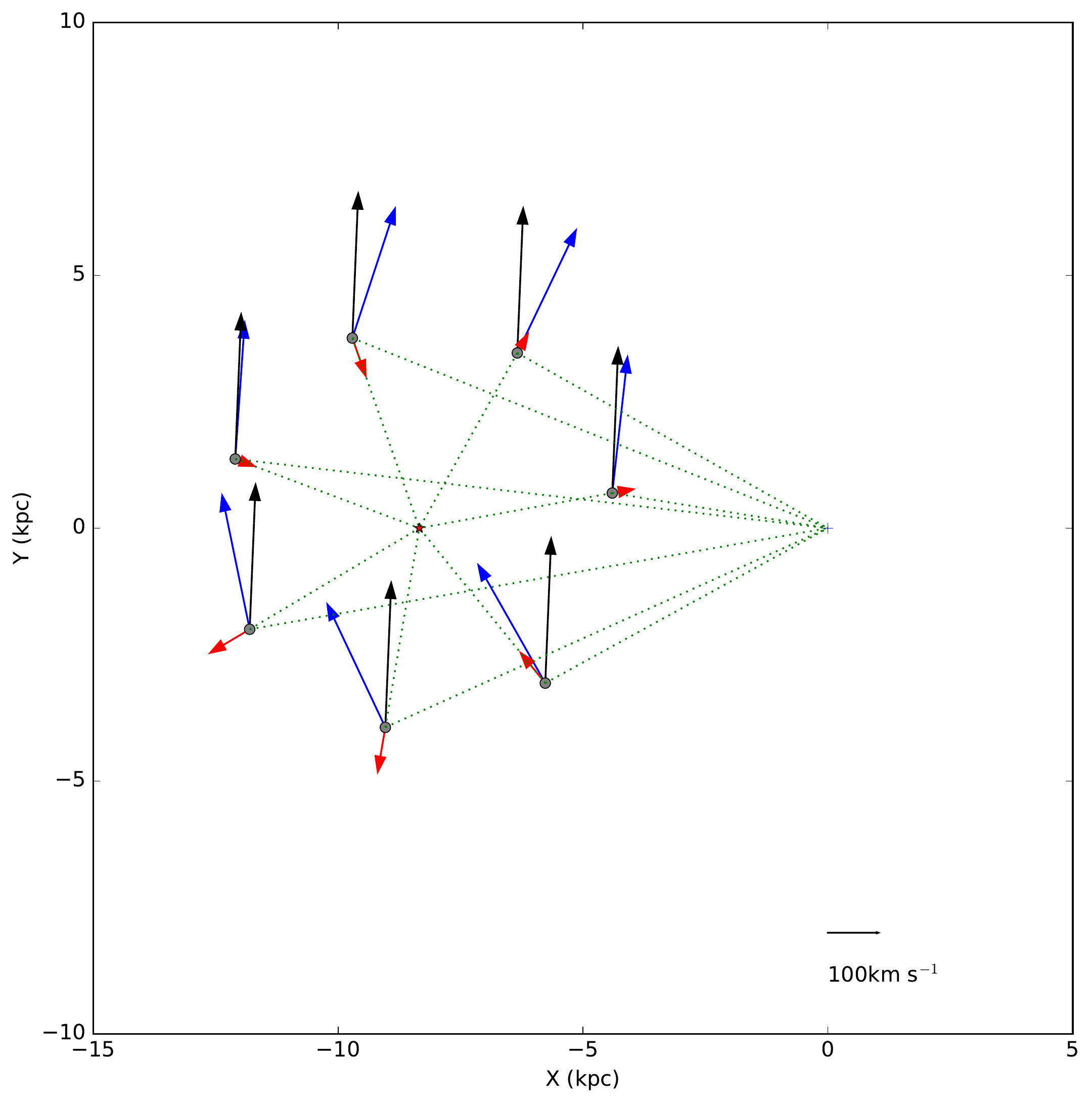}
\caption{Sketch of de-projection of the line-of-sight velocity.
Blue and black arrows are velocity vectors of stars and the Sun
with respect to the Galactic centre. Red arrows
denote the line of sight velocities of stars with respect to the
Sun\label{fig-8}.}
\end{figure}

The de-projection method is a useful method to study Galactic
kinematics using radial velocity data \citep{2012A+A...547A..70P,
2015ApJ...809..145T}. Here we used a similar
de-projection method to model the $V_{\rm LSR}$ of SiO masers. A sketch of the
de-projection is shown in Fig.~\ref{fig-8}, where, blue and black arrows are velocity
vectors of the star ($\overrightarrow{V_{\star}}$) and the Sun
($\overrightarrow{V_{\odot}}$) with respect to the Galactic centre. The line of
sight velocity of the star towards the Sun, \textit{V$_{\rm helio}$}, denoted
by the red arrow, is the projection of
($\overrightarrow{V_{\star}}$-$\overrightarrow{V_{\odot}}$) on 
$\overrightarrow{r_{\star}}$, where $\overrightarrow{r_{\star}}$ is the
direction vector from the Sun towards the star (Equation \ref{eq-1}). The
observed $V_{\rm LSR}$ can be converted to \textit{V$_{\rm helio}$} by adding
back the component of the standard Solar motion in the line-of-sight direction
that had been removed from the observed Doppler shift to calculate $V_{\rm
LSR}$ using Equation \ref{eq-2}, where (\textit{U}$^{Std}_\odot$,
\textit{V}$^{Std}_\odot$, \textit{W}$^{Std}_\odot$) are not the best values
available today, but the (old) standard Solar motion,  a value of 20 km
s$^{-1}$ toward $\alpha$(1900)~=~18$^{h}$, $\delta$(1900)~=~+30$^d$ generally
adopted by observatories \citep{2014ApJ...783..130R}. In this study, for values
of the Solar motion and Galactic parameters, we adopted the A5 model from
\citet{2014ApJ...783..130R}, where (\textit{U}$_\odot$,
\textit{V}$_\odot$,\textit{W}$_\odot$)= (10.7$\pm$1.8, 15.6$\pm$6.8,
8.9$\pm$0.9)~\kms, $\Theta_0$~=~240$\pm$8~\kms\ and $R_0$~=~8.34 kpc. 

For modelling radial velocities, we assumed that a group of stars with
similar Galactic locations (similar $R$ and $Z$) share a mean 3D velocity of
$\overrightarrow{V_{\star}}$~=~($\overline{U_s}$,$\Theta$+$\overline{V_s}$,
$\overline{W_s}$) with respect to the Galactic centre, where ($\overline{U_s}$,
$\overline{V_s}$, $\overline{W_s}$) are peculiar velocities in the
direction of the Galactic centre, Galactic rotation and toward the North
Galactic Pole (NGP), $\Theta$ is the mean circular rotational velocity. 
For the Sun,
$\overrightarrow{V_{\odot}}$~=~($U_{\odot}$,$\Theta_0$+$V_{\odot}$,
$W_{\odot}$).
It is
noted that the, $\overline{V_s}$ and $\Theta$ are two highly correlated
parameters, which can not be separated.  Instead, $\Theta$+$\overline{V_s}$ can
be well constrained, as have been pointed out by several previous studies
\citep{2010MNRAS.402..934M, 2012PASJ...64..136H, 2014ApJ...783..130R}. Thus, in
our study, we treat $\Theta$+$\overline{V_s}$ as one parameter that we need to
fit.

\begin{equation} 
V_{\rm helio}=V_{\rm model}=(\overrightarrow{V_{\star}}-\overrightarrow{V_{\odot}})\cdot \overrightarrow{r_{\star }}
\label{eq-1}
\end{equation}   

\begin{equation} 
V_{\rm helio}=V_{\rm obs}=V_{\rm LSR}-(\textit{U}^{Std}_\odot~\mathrm{cos}~\textit{l}~+ \textit{V}^{Std}_\odot~\mathrm{sin}~\textit{l})\mathrm{cos}~\textit{b}- \textit{W}^{Std}_\odot~\mathrm{sin}~\textit{b}
\label{eq-2}
\end{equation}   

\begin{equation} 
S^2=\frac{\sum (V_{\rm obs} - V_{\rm model})^2}{N}
\label{eq-3}
\end{equation}   

A program\footnote{contact yuanwei.wu@ntsc.ac.cn for the program code.}  was developed for investigating kinematics with
(\textit{l},\textit{b},\textit{D},$V_{\rm LSR}$) of a group of stars as inputs
and (\textit{S}, $\overline{U_s}$, $\Theta$+$\overline{V_s}$, $\overline{W_s}$)
of the group as outputs. \textit{S}
denotes the dispersion of the line-of-sight velocity, which is expressed in
Equation \ref{eq-3}. We used the Markov Chain Monte Carlo (MCMC) method in the
procedure to minimize the dispersion and estimate the uncertainties of
(\textit{S}, $\overline{U_s}$, $\Theta$+$\overline{V_s}$, $\overline{W_s}$).
We first calculate a prior $S$, by using a prior of ($\overline{U_s}$,
$\Theta$+$\overline{V_s}$, $\overline{W_s}$)=(0, 220, 0)~\kms. Then we let
($\overline{U_s}$, $\Theta$+$\overline{V_s}$, $\overline{W_s}$) walk randomly
in an sampling window of (30, 30, 30)~\kms\ to calculate a posterior $S$ until
we find better solution of ($\overline{U_s}$, $\Theta$+$\overline{V_s}$,
$\overline{W_s}$) with smaller $S$.  Once a better solution is found, we use
these values to replace the prior, and resample again to search for better
solutions. Usually, $S$ can converge within 10 interations. When S converges,
($\overline{U_s}$, $\Theta$+$\overline{V_s}$, $\overline{W_s}$) and $S$ are
recorded and the program stops. We repeat the above s 10000 times, and
estimate the mean and standard deviation of ($\overline{U_s}$,
$\Theta$+$\overline{V_s}$, $\overline{W_s}$) and $S$ as the final outputs.

Using this procedure, we obtained the values of $\Theta$+$\overline{V_s}$~=~225.6
\kms, $S$~=~34.1~\kms\ for disc sources with $D<$~1.2~kpc; and
$\Theta$+$\overline{V_s}$~=~185.4~\kms, $S$~=~41.3~\kms\ for off-plane
($Z>$~1.2~kpc) sources.  Adopting $\Theta_0$~=~240 \kms, the velocity lag of
disc and off-plane SiO masers with respect to the $\rm LSR$ are 15 \kms\
and 55 \kms\ respectively.

\subsection{Moving groups and high speed stars}\label{subsec:deprojection} With
the aim of identifying potential stream motions and/or high-speed sources, we
further apply this procedure to a series of ($R$,$Z$)-binned subsamples.  The
ranges of ($R$, $Z$) bins are given in Table~\ref{tab-5}.  Regarding the group
strategy, we initially separated $R>$~9 and $R<$~9 sources into two groups,
since (1) sources are not uniformly distributed in the Galactic plane; there
are fewer sources outside of $R>9$~kpc, and (2) the relatively large
uncertainties in the rotation curve at $R>$~9kpc \citep{2000PASJ...52..275N,
2007A+A...473..143D, 2013RAA....13..849X}.  We only consider sources with
$R>5$~kpc, since (1) the kinematics in the inner disc could not be well modeled
by circular motions due to the influence of the Galactic bar; (2) distance
uncertainties for sources with $R<$~5kpc tend to be high (Fig.~\ref{fig-5} and
Fig.~\ref{fig-7}).

\begin{figure*}
\includegraphics[width=18cm]{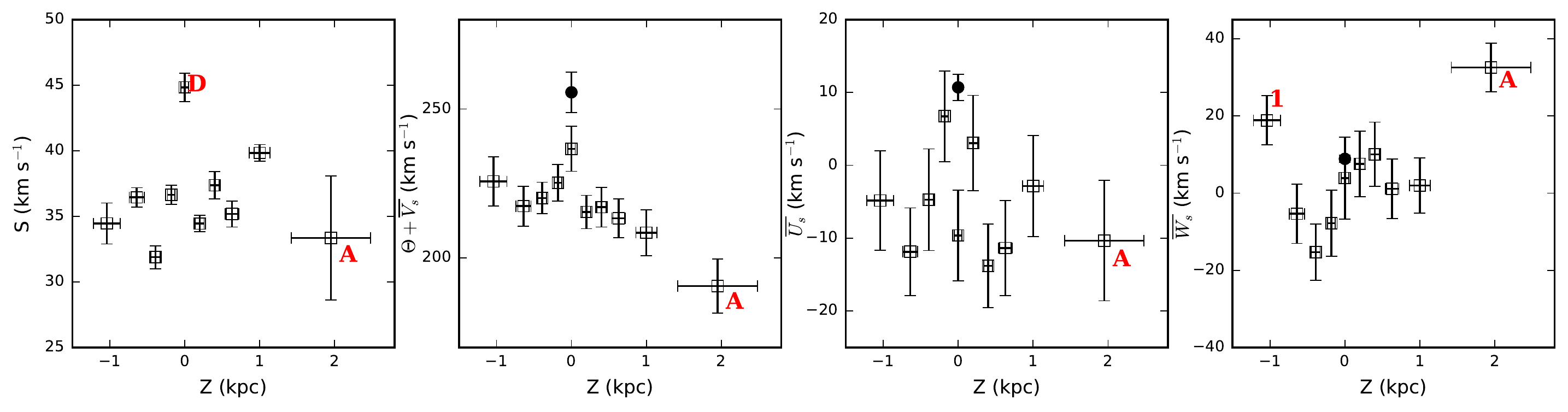} 
\includegraphics[width=18cm]{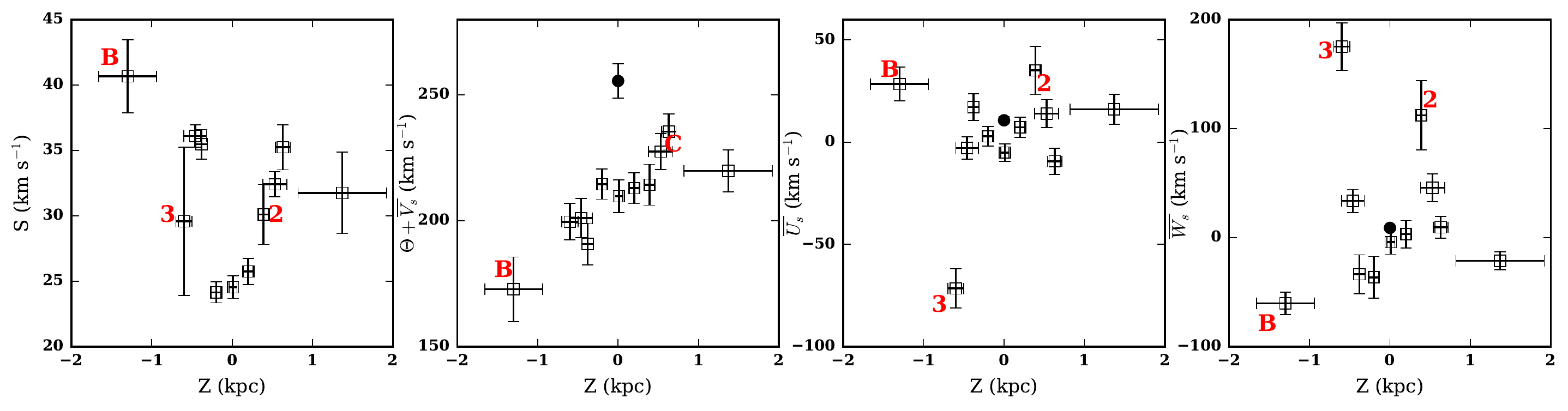} 
\caption{Visualization of best estimated $S$, $\Theta+\overline{V_s}$,
$\overline{U_s}$, $\overline{W_s}$ of each spatially binned
subsamples.{Upper panels are $R$-$Z$ binned groups of 5~$<R<$~9~kpc,
lower panels are $R$-$Z$ binned groups of $R>$~9~kpc. Red colour A, B, C and 1, 2, 3
denote three moving groups and 3 high speed sources. The filled circle denotes
Solar motions where (\textit{U}$_\odot$,
\textit{V}$_\odot$,\textit{W}$_\odot$)~=~(10.7$\pm$1.8, 15.6$\pm$6.8,
8.9$\pm$0.9)~\kms \citep{2014ApJ...783..130R}. Horizonal ``errorbars'' denote
 the min($Z$) and max($Z$) of stars in the 
binned range of $Z$; vertical errorbars denote 1$\sigma$ uncertainties of $S$,
$\Theta+\overline{V_s}$, $\overline{U_s}$, $\overline{W_s}$.  }\label{fig-11}}
\end{figure*}

Outputs for each of the $R$-$Z$ binned groups obtained with the above procedures
are listed in Table \ref{tab-5} and visualized in Fig.~\ref{fig-11}. Top panels
plot how (\textit{S}, $\overline{U_s}$, $\Theta$+$\overline{V_s}$,
$\overline{W_s}$) varies with $Z$ for sources with 5~$<R<$~9~kpc, while bottom
panels plot the same but for sources at $R>$9~kpc. Horizonal ``errorbars'' indicate 
the min($Z$) and max($Z$) of stars in the binned range of $Z$; vertical
errorbars show 1$\sigma$ uncertainties of $S$, $\Theta+\overline{V_s}$,
$\overline{U_s}$, $\overline{W_s}$. Figure \ref{fig-11} is very useful for
identifying potential stream motions and/or high velocity stars. Careful
inspection of these diagrams allows us to identify three bulk motions and three
very high speed stars. 

\begin{figure*}
\includegraphics[width=16cm]{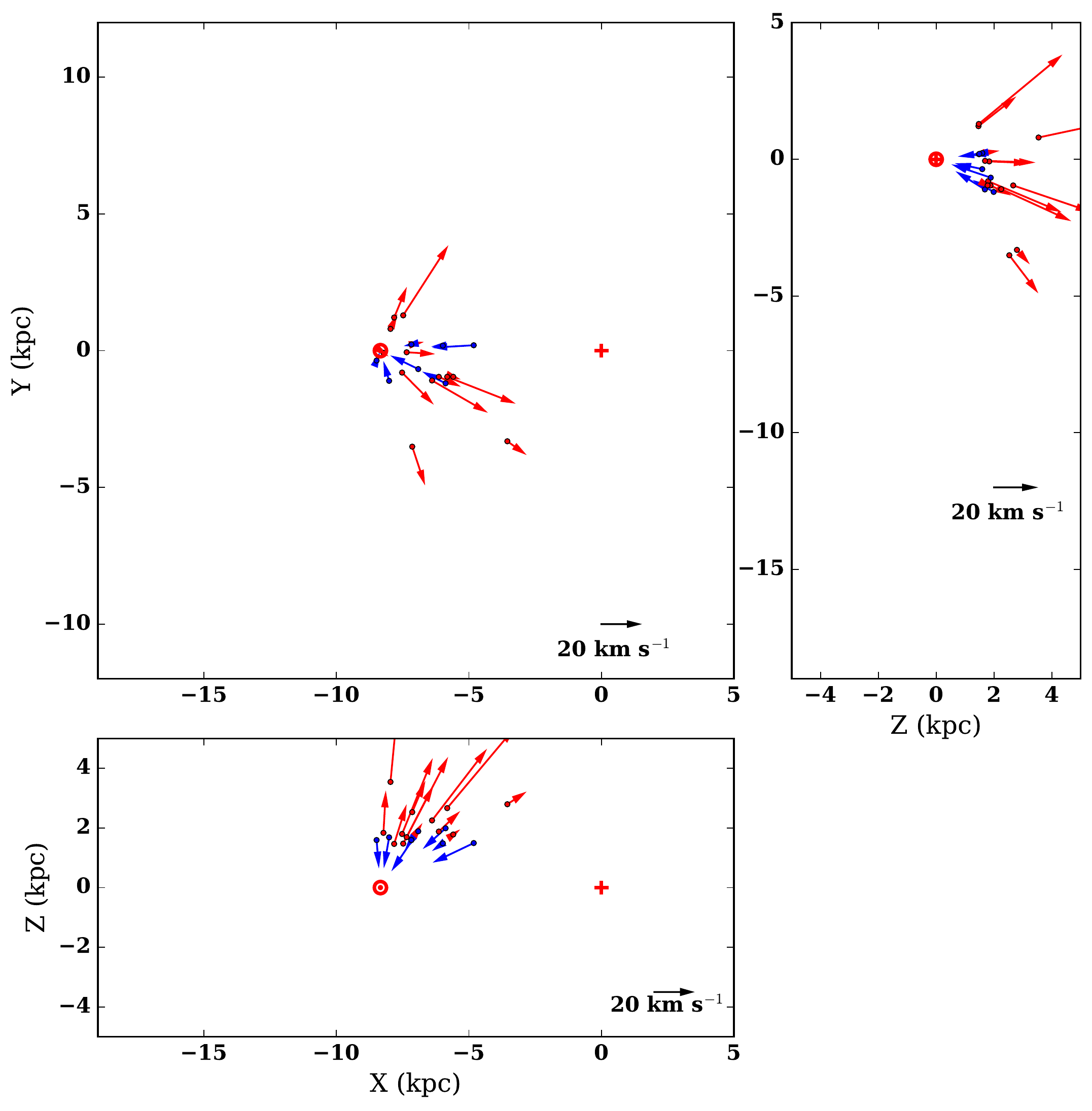} 
\caption{2D Projection of $V_{\rm obs}$-$V_{\rm model}$ in Galactocentric Cartesian
coordinate for group A, with 5~$<R<$~9~kpc, 1.4~$<Z<$~2.5~kpc. $V_{\rm model}$
was calculated by using a rotational speed of 180 \kms. Blue/red arrows denote
negative/positive $V_{\rm obs}$-$V_{\rm model}$.\label{fig-12}}
\end{figure*}

In the top panels of Fig.~\ref{fig-11}, for the 5~$<R<$~9~kpc,
1.4~$<Z<$~2.5~kpc group, one can see large deviations of $\overline{W_s}$ and
$\Theta$+$\overline{V_s}$, and a very large uncertainty in \textit{S}.  Further
inspection of $V_{\rm obs}$-$V_{\rm model}$ (Fig.~\ref{fig-12}) reveals the bulk motion
of this group. For further discussion, we denote this group as
group A. In Fig.~\ref{fig-12}, dots with arrows denote the 2D projection of
$V_{\rm obs}$-$V_{\rm model}$, where a model with a rotational speed of 180 \kms\ was
used. Blue/red colours denote negative/positive values of $V_{\rm obs}$-$V_{\rm model}$. 

\begin{figure*}
\includegraphics[width=16cm]{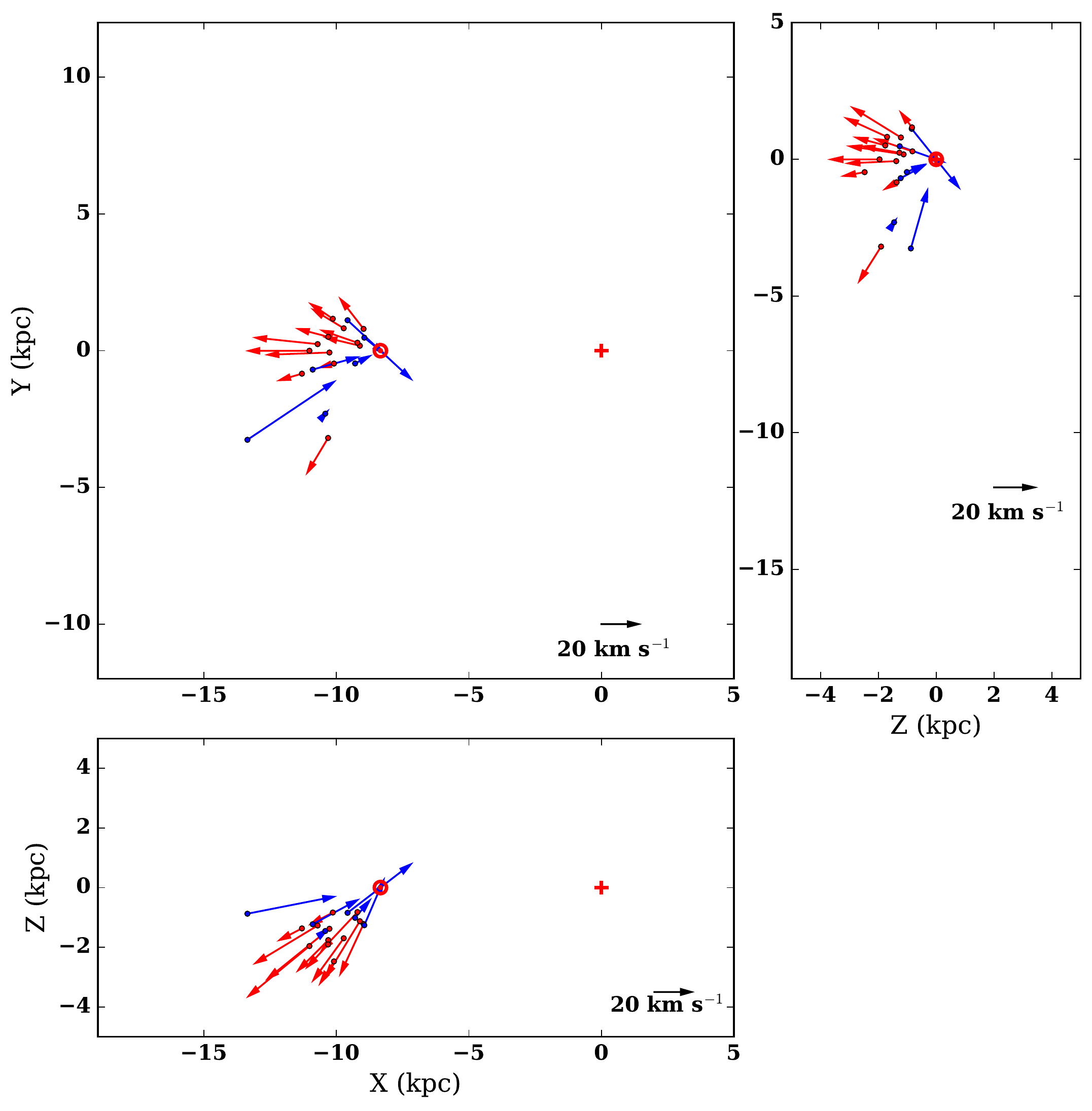} 
{\caption{2D Projection of $V_{\rm obs}$-$V_{\rm model}$ in Galactocentric Cartesian
coordinate for group B, with $R>$~9~kpc,
-2.0~$<Z<$~$-$0.8~kpc. $V_{\rm model}$ was calculated by using
$\Theta+\overline{V_s}$~=~180~\kms.\label{fig-13}}}
\end{figure*}

In the lower panel of Fig.~\ref{fig-11}, the -2.0~$<Z<$~$-$0.8~kpc group has
the highest \textit{S} and very low $\Theta$+$\overline{V_s}$, and
$\overline{W_s}$.  Further inspection of the value of $V_{\rm obs}$-$V_{\rm
model}$ confirms the systematic peculiar motion of this group as shown in
Fig.~\ref{fig-13}.  For further discussion, we denote this group as group B.

\begin{figure*}
\includegraphics[width=16cm]{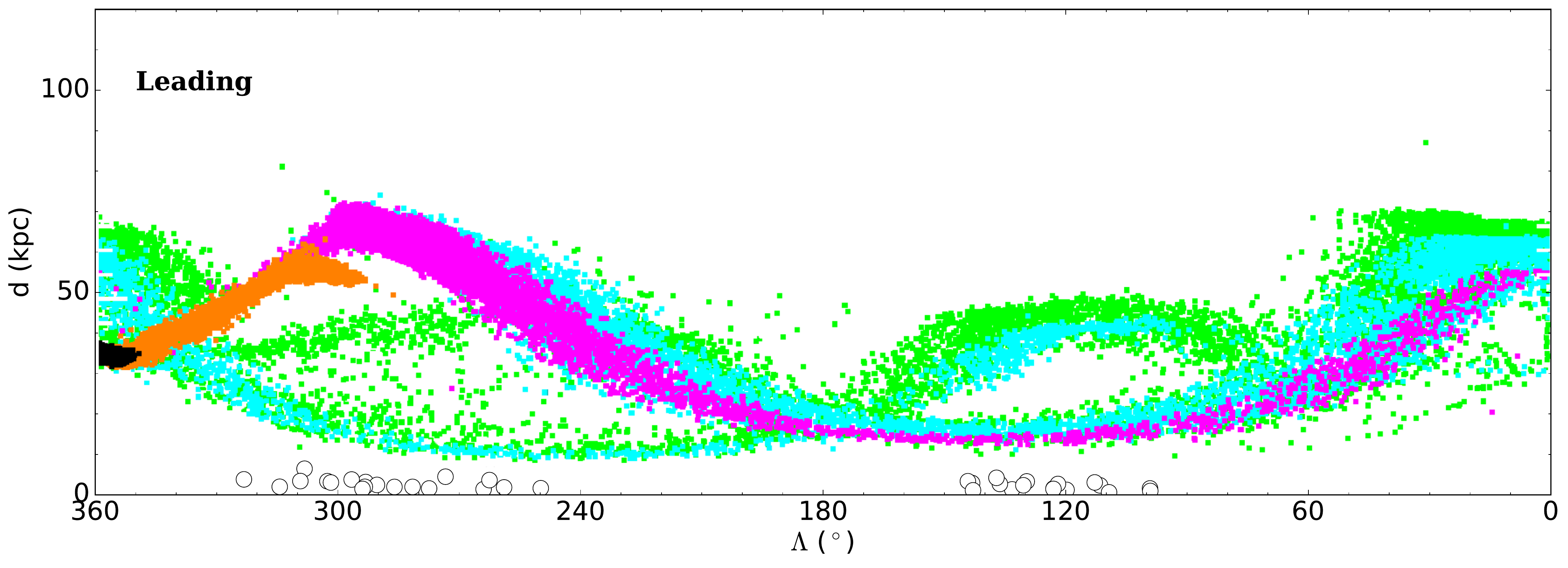} 
{\caption{Position-distance
diagrams of the Leading arm of the Sgr stellar stream. Color dots are HM10
Leading arm model (Figure 7 of \citet{2010ApJ...718.1128L}). Open circles are groups A and B offplane sources listed in Table \ref{tab-6}. 
\label{fig-14}}}
\end{figure*}

In Table \ref{tab-6}, we list the detailed source information of these two
groups.  Both groups are made up of off-plane sources. Radial velocities
of stars within the northern group (group A) can be modelled by stars with a rotational
speed $\Theta$+$\overline{V_s}$~=~189$\pm$9~\kms, with peculiar motion
$\overline{U_s}$~=~12$\pm$8~\kms, $\overline{W_s}$~=~33$\pm$6~\kms\ away from
the Galactic plane.  Radial velocities of stars within the southern group (group B) can
be modeled by stars with rotational speed
$\Theta$+$\overline{V_s}$~=~176$\pm$13~\kms, with peculiar motion
$\overline{U_s}$~=~25$\pm$9~\kms, and $\overline{W_s}$~=~$-$54$\pm$11~\kms\
away from the Galactic plane. Regarding their Sgr stellar stream coordinates
(columns 7 and 8 of Table \ref{tab-6}), they seem to be aligned with the orbital
plane of the Sgr stellar stream. For a comparison with the Sgr stellar stream, in
Fig.~\ref{fig-14}, we overlay these moving group sources on a
longitude-distance diagram of the L1/L2 wrap of LM10 model. These off-plane sources
are within 5 kpc of the Sun. 

In the LM10 model, the Sgr stream is not expected to be located within 13~kpc of the Sun, while in the model of
\citet{2004PASA...21..197M}, the intersection between L2 wrap of the Sgr
stellar stream and the Galactic plane is very close to the Sun. It should be
mentioned that in the LM10 model, all Leading arm SDSS Constrains are beyond
18~kpc \citep[Table 1 of][]{2010ApJ...714..229L}. This can be due to method
bias, as the majority of halo stream features are identified by stellar
overdensity, which can be overlooked at nearest distances due to severe disc
contaminations. Although observational constraints of stream features at near
distances are rare, \citet{2002ApJ...576L.125K} found eight giant stars with
kinematics, abundances and locations roughly consistent with leading tidal arm
of the Sagittarius arm. 6 out of these 8 giant stars are within 5~kpc of the
Sun. Even for LM10 model, they did not exclude the possibility of the existence
of Sgr stream stars in nearby ($D<$10~kpc) region \citep[see discussions in
Section 8.2 of][]{2010ApJ...714..229L}. Taking into account of the Galactic
location and kinematics of groups A and B sources, they may be thick disc
stars with their kinematics affected by the halo stellar stream or very old Sgr
stream debris.  Future measurements of proper motions and determinations of the
full 3D motions and Galactic orbits of these sources will be crucial for
testing these scenarios.


In the lower panel of Fig.~\ref{fig-11}, there exist two ``high" peculiar
motion groups, groups 1 and 2 in Fig.~\ref{fig-11}, one is the 0.3~$<Z<$~0.5~kpc
group and the other is the -0.8~$<Z<$~$-$0.5~kpc group.  Further investigation
of the values of $V_{\rm obs}$-$V_{\rm model}$ of these two groups reveals most
of sources within these groups are actually normal disc stars, the large
peculiar motion seen in Fig.~\ref{fig-11} are due to two very high speed
sources, which were listed in Table \ref{tab-7}.  With same method, another high
speed source was identified in the -1.4~$<Z<$~$-$0.8~kpc, 5~$<R<$~9~kpc group,
which was also listed in Table \ref{tab-7}. In Figure \ref{fig-11}, groups
including these high speed sources are highlighted with red labels 1, 2 and 3.

In the upper panels of Fig.~\ref{fig-11}, another outlier is the
-0.1~$<Z<$~0.1~kpc group (group D), with the largest value of $S$ and a relatively high rotation
speed of 237~$\pm$~7~\kms\ compared to other groups. The large \textit{S} of this
group is due to inclusion of a large number of objects from extended regions.
For comparison, in the lower panel, for the 9~$<R<$~13~kpc, -0.1~$<Z<$~0.1~kpc
group, the value of \textit{S} is only $\sim$25.0 \kms. Un-modeled peculiar motions due
to non-axisymmetric perturbations from spiral arms and/or the bar could explain
the large value of \textit{S} in group D.  

In Fig.~\ref{fig-11}, we noticed dependences of the rotational speed and the
velocity dispersion on the Galactic plane distance. In the 1st lower left panel
of Fig.~\ref{fig-11}, it can be seen that the velocity dispersion \textit{S}
increases with $Z$.  In the second column of Fig.~\ref{fig-11}: in the group of
$Z>$~0, 5~$<R<$~9~kpc, and group of $Z<$~0, $R>$~9~kpc, there is a clear trend
that rotational speed decreases with $Z$. These phenomena are consistent
with stellar dynamics theory: old thick disc stars which are dynamically
evolved sources tend to have slower rotational speeds and higher velocity
dispersions. 

\begin{figure*}
\includegraphics[width=16cm]{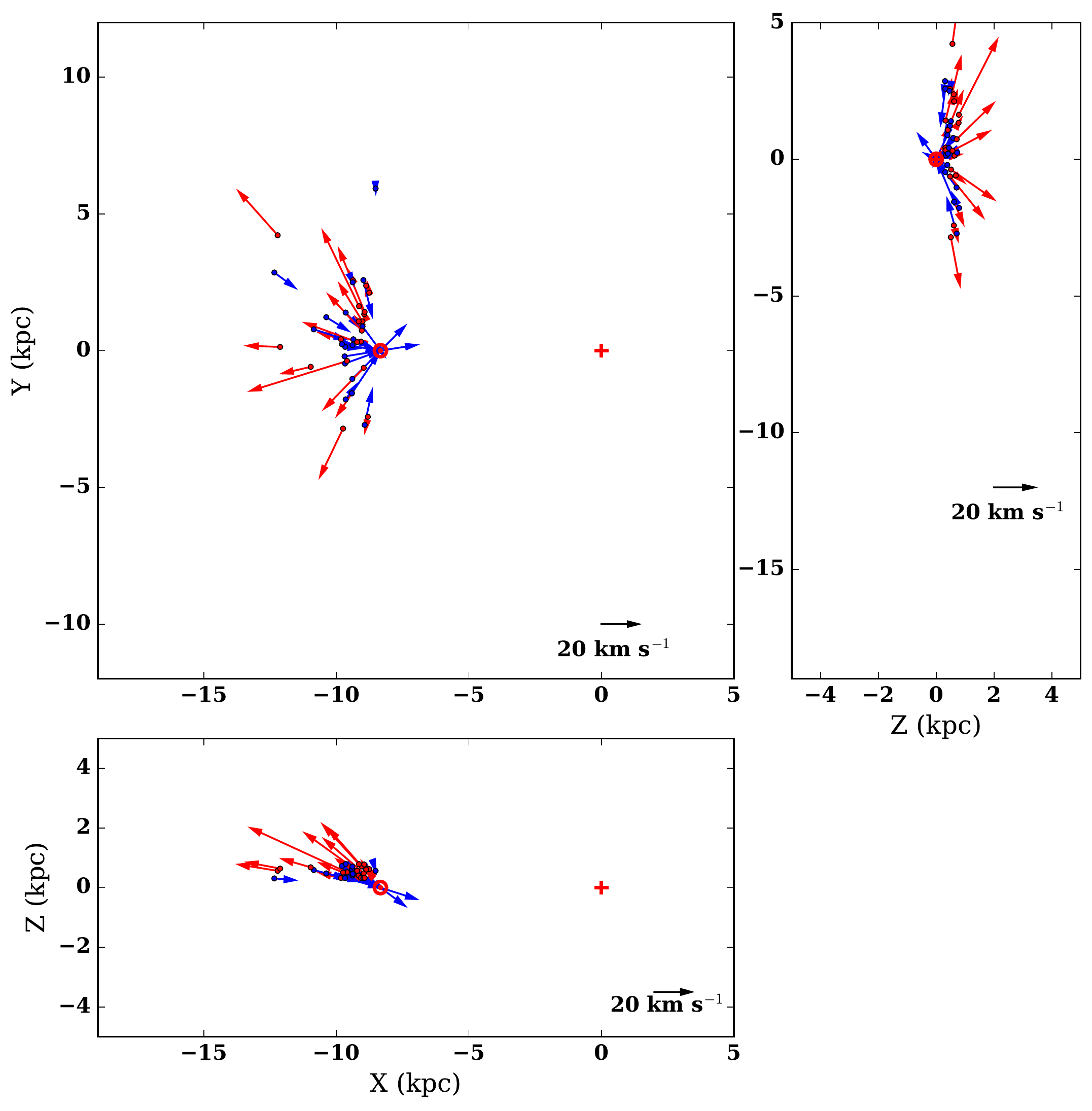} 
{\caption{2D Projection of $V_{\rm obs}$-$V_{\rm model}$ in Galactocentric Cartesian
coordinate for group C, with $R>$~9~kpc,
0.3~$<Z<$~0.8~kpc. $V_{\rm model}$ was calculated by adopting
$\Theta+\overline{V_s}$~=~220~\kms.\label{fig-16}}}
\end{figure*}

In the 0.3~$<Z<$~0.8~kpc, $R>9$~kpc region, which we denote as group C, there
can be seen an increasing of $\Theta$+$\overline{V_s}$. Such an enhancement in
rotation speed is hard to explain by the stellar dynamics mentioned above. Fig.~\ref{fig-16}
illustrates $V_{\rm obs}$-$V_{\rm model}$ for the sources in this group. In the
Figure, one can see outward and upward bulk motions for many sources.
Considering the locations of these sources, it is possible that they trace peculiar motions or
flows associated with the Perseus spiral arm.  \citet{2017arXiv170502197K}
found that high-mass star forming regions (HMSFRs) in the Perseus arm traced by
6.7~GHz methanol masers between $85^\circ<l<124^\circ$, and
$173^\circ<l<196^\circ$ are moving outward and rotate about the Galaxy at a
higher velocity with respect to the gas tracers (CO and CS). As can be seen in
Figure 7 of \citet{2017arXiv170502197K}, the kinematics of these HMSFRs are
somewhat simiar to our group C SiO masers, which supports our viewpoint that there
exist large scale peculiar motions in the Perseus arm.

In order to evaluate whether the systematic motions identified in groups A, B
and C were statistically significant, we made simulations to test the null
hypothesis that the systematic peculiar motions are not true but within
statistical errors. In the simulation, we generate "trial" stars with identical
Galactic coordinates.  Distance errors were simulated by varying the nominal
distance values up to 20\% as a 1~$\sigma$ error.  For kinematics, we assume
they follow the kinematic model of disc stars, with rotational speed, $\Theta$,
varying as function of Galactic plane distances ranging from 225 \kms\ to 180
\kms, plus normal random motions ($\overline{U_s}$, $\overline{V_s}$,
$\overline{W_s}$) with 1$\sigma$ errors ranging from 25 \kms\ to 40 \kms. Then
(\textit{S}, $\overline{U_s}$, $\Theta$+$\overline{V_s}$, $\overline{W_s}$)
were estimated by the same procedure as applied to the observational data. 

We conducted 1000 trials, and then calculated the probability distributions of
(\textit{S}, $\overline{U_s}$, $\Theta$+$\overline{V_s}$, $\overline{W_s}$),
which are shown in Figure \ref{fig-A4}. In Fig.~\ref{fig-A4}, black vertical lines
denote 1$\sigma$ (68\%) and 2$\sigma$ (95\%) boundaries of the simulations, red
dashed lines denote the 1$\sigma$ range of (\textit{S}, $\overline{U_s}$,
$\Theta$+$\overline{V_s}$, $\overline{W_s}$) estimated with real data.
\textit{Top}, \textit{middle} and \textit{bottom} panels are results of the groups A, B
and C respectively. For all three groups, the value of $\overline{W_s}$ estimated by
real data were within 1$\sigma$ to 2$\sigma$ range of simulations.  Thus,
statistically, we were $\sim$70\% confident that the systematic peculiar
motions vertical to the Galactic plane of all three groups are real.
Especially, for the group C, the disagreement between simulations and real data is
significant for all the four (\textit{S}, $\overline{U_s}$,
$\Theta$+$\overline{V_s}$, $\overline{W_s}$) parameters. Future measurements of
proper motions and determinations of full 3D motions and Galactic orbits of
these sources will be crucial for better understanding these bulk motions.

\section{Summary} \label{sec:summary}

We conducted an SiO maser survey towards 221 O-rich AGBs projected on the Sgr
stellar stream region. In total, we detected maser emission in 44 AGBs, of
which 35 were new detections. We found that the WISE colour-colour diagram method
provides an effective way to select O-rich Miras for SiO maser survey.

We compiled a Galactic SiO maser catalogue that includes $\sim$2300 sources. We
then cross matched this maser catalogue with the GCVS and AAVSO variable catalogues
and the 2MASS and WISE point souces catalogues and calculated the PLR distance and
WISE luminosity distances for $\sim$1000 sources. By studying the Galactic
distribution of sources, we determined a typical height scale (0.40~kpc) of disc SiO
masers, and identified 42 off-plane SiO masers. 50\% of these off-plane SiO
masers were discovered in this work.

We identified three very high speed stars and found systematic peculiar motions
in three groups (70\% confidence) by comparing the radial velocities with
models of disc stars. The most statistically significant group is within the
thick disc, with 0.3~$<Z<$~0.8~kpc, $R>9$~kpc, which might be tracing an
outward (13$\pm$7 \kms) and upward (44$\pm$14 \kms) flow in the Perseus arm.
The other two moving groups are mainly made up of off-plane sources. The
northern group can be modeled by disc stars with rotational speed $\sim$190
\kms, and peculiar motion $\overline{W_s}$~=~34$\pm$6~\kms\ away from the
Galactic plane; the southern group can be modeled by disc stars with rotational
speed $\sim$180 \kms, and peculiar motion of $\overline{U_s}$~=~25$\pm$9~\kms,
$\overline{W_s}$~=~$-$54$\pm$11~\kms\ away from the Galactic plane. As these
two off-plane groups are aligned with the Sgr orbital plane, we suspect that
they could be thick disc stars whose kinematics might be affected by the halo
stellar stream or very old stream debris. Future measurements of proper motions and determinations of full 3D
motions and Galactic orbits of these sources will be crucial to confirm/deny
our conclusions.

For disc SiO maser sources, model fitting on radial velocities allows us to
reveal dependences of the rotational speed and the velocity dispersion on the distances
from the Galactic plane, i.e., stars with higher Galactic plane distances tend
to have large velocity dispersion and rotate slowly within the Galactic disc. These
trends are consistent with the theory of stellar dynamics. With
$\Theta_0$=~240~\kms, we derived a velocity lag of 15 \kms\ and 55 \kms\ for
disc and off-plane SiO maser sources respectively.

\section*{Acknowledgements}
\addcontentsline{toc}{section}{Acknowledgements}
We would like to thank Prof. Majewski from the University of Virginia and Dr.
Mark Claussen from NRAO for useful discussions on the Sgr stellar stream.  We
would like to thank Prof. Honma from the NAOJ who carefully read the manuscript
and share valuable comments and suggestions with us. We would like to thank
NRO45m staffs for their helps and supports during observations. The Nobeyama
45-m radio telescope is operated by Nobeyama Radio Observatory, a branch of
National Astronomical Observatory of Japan.



\appendix
\section{Distance estimate}\label{sect:Appendix}
Distances is the key parameter to study the Galactic distribution. For Miras
with known period, their distances can be calculated based on the period luminosity
relation (PLR). For AGB stars without period or Ks magnitude but
with the WISE colour available, we derived WISE luminosity distances by
using an empirical formula calibrated with O-rich Miras, in which the distance is function of the $W4$
magnitude and a dereddened colour [$W1$$^\prime$-$W4$]. Formulas used to derive the PLR
distances and WISE luminosity distances are given in the section \ref{subsec:A1} and \ref{subsec:A2} respectively.
.

\subsection{PLR Distance}\label{subsec:A1}

First we collect K$_s$ magnitudes from 2MASS catalogue
\citep{2003tmc..book.....C} and periods from the General Catalog of Variable
Stars (GCVS, \citet{2003AstL...29..468S}) and/or International Variable Star
Index Catalog (AAVSO, \citet{2006SASS...25...47W}). Then the absolute K
magnitudes of O-rich Miras were derived by using the infrared (K)
period-luminosity relation formula \citep[from][]{2008MNRAS.386..313W} of 
\begin{equation} 
M_K = -3.51[\log~P-2.38]-7.25.
\label{eq-A1}
\end{equation}   

Comarison between the observed and absolute magnitudes gives the initial
estimation of distance without interstellar extinction taking into account.

\begin{equation} 
D = 10^{(m_{Ks}-M_K+5)/5}
\label{eq-A2}
\end{equation}   

With this distance, the interstellar extinction (\textit{A$_V$}) is estimated
using the three-dimensional Galactic extinction model
\citep{2003A+A...409..205D}. The extinction \textit{A$_V$} is solved by using
the FORTRAN script Av3$\_$FEB03.f given by \citet*{2003A+A...409..205D}. The
$A_V$ is converted to $A_{Ks}$ by

\begin{equation} 
A_{Ks} = R(Ks) \times E(B-V) = A_V \frac{R(Ks)}{R(V)}
\label{eq-A3}
\end{equation}
where $R(V)$~=~3.10 and $R(Ks)$~=~0.306 are the V and Ks band extinction
coefficients from \citet{1999PASP..111...63F, 2013MNRAS.430.2188Y}.  The
$A_{Ks}$ is then used to estimate the dereddened magnitude,
\begin{equation} 
m^\prime_{Ks}= m_{Ks}-A_{Ks},
\label{eq-A4}
\end{equation}
and recalculate the distance. Two iterations usually
suffice.

\subsection{WISE Luminosity Distance}\label{subsec:A2}
\begin{figure*}
\centering
\includegraphics[width=15cm]{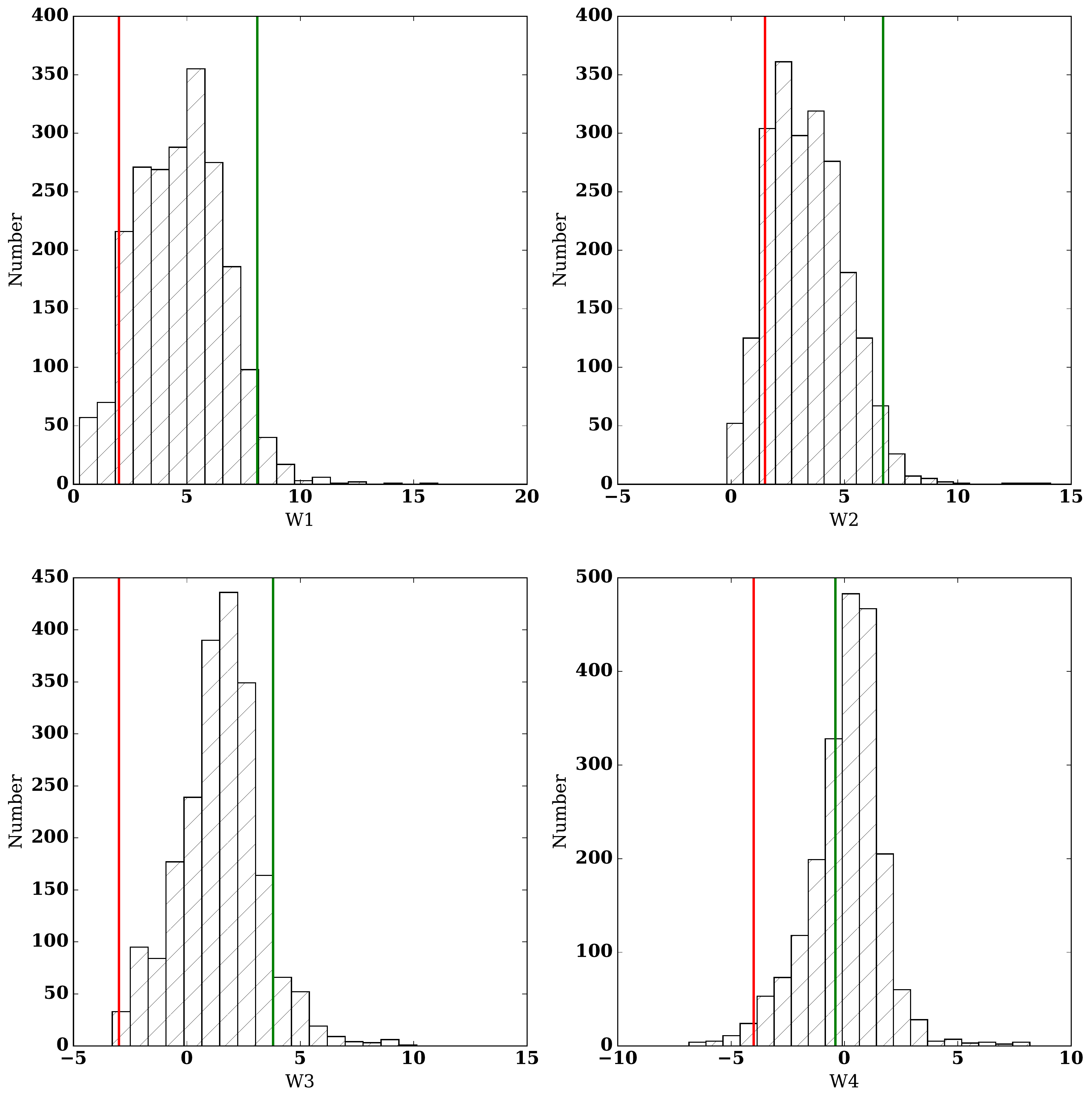}
\caption{Histogram of WISE magnitudes of $W1$ , $W2$, $W3$ and $W4$,
respectively.  Green vertical lines denote the saturation limits for the WISE 4 bands
that are 8.1, 6.7, 3.8, and -0.4 mag, respectively.  Red vertical lines denote the
bright source photometry limits for the WISE 4 bands that are 2, 1.5, $-$3, and
$-$4 mag, respectively.
 \label{fig-A1}} 
\end{figure*}

According to the WISE Explanatory Supplement documentation of All-Sky
Data Release that bright sources will saturate WISE detector \footnote{WISE
Explanatory Supplement documentation for All-Sky Data Release
http://wise2.ipac.caltech.edu/docs/release/allsky/expsup/}. The saturation
limits are 8.1, 6.7, 3.8, and -0.4 mag in 
the $W1$-$W4$ bands, respectively (green lines in Fig.~\ref{fig-A1}). For saturated objects, WISE fits the PSF to the
unsaturated pixels on the images to recover the saturated pixels and yield
photometry for these objects (we call this PSF-fit photometry). The bright
source photometry limits for the WISE 4 bands are 2, 1.5, $-$3, and $-$4 mag
(red lines in Fig.~\ref{fig-A1}). When they are brighter than this limit,
all the pixels are saturated and no unsaturated pixels are available to fit to
the PSF. As can be seen in Fig.~\ref{fig-A1}, $\sim$ 100 bright
nearby AGBs are totally saturated. In WISE 1, 2 and 3 band, photometry are
PSF-fit photometry for most of our targets. $W4$ band is the best band in which
the saturation is not so serious. In addition, for sources within $\sim$2~kpc,
although they are not totally saturated, their PSF-fit photometry tend to be
underestimated, which result in a overestimate of distance. 

Regarding the interstellar extinctions, we consider the method in Section \ref{subsec:A1} that 
makes use of the 3D extinction model to correct for $W1$ band interstellar extinction.

\begin{equation} 
A_{W1} = R(W1) \times E(B-V) = A_V \frac{R(W1)}{R(A)},
\label{eq-A5}
\end{equation}
where $R(V)$~=~3.10 and $R(W1)$~=~0.18 are the $V$ band and WISE $W1$ band extinction
coefficients from \citet{1999PASP..111...63F} and \citet{2013MNRAS.430.2188Y}.
Then we correct $W1^\prime$ with the extinction A$_{W1}$,
\begin{equation} 
W1^\prime= W1-A_{W1}
\label{eq-A6}
\end{equation}

The $W4$ band luminosity,
\begin{equation}
L_{W4} = L_{\star} + L_{dust} = \alpha \times L_{\star}
\label{eq-A7}
\end{equation}
where, L$_{dust}$ is
the W4 band luminosity from the dusty envelope, and L$_{\star}$ is the inherent
 W4 band luminosity of the star, $\alpha$ is a reddening factor.  Here we tried to
correct a de-reddening $W4$ band magnitude to let 

\begin{equation}
W4C = \beta \times (W1^\prime-W4)+ \gamma \times W4 \propto \log (L_\star \times D^{-2})
\label{eq-A8}
\end{equation}
where $\beta$ and $\gamma$ are constants.

If we further assume that L$_\star$ is a constant, then the formula \ref{eq-A8}
can be written as

\begin{equation} 
\log~D = \textit{a}\times W4 + \textit{b} \times (W1^\prime-W4) + \textit{c}.
\label{eq-A9}
\end{equation}

\begin{figure*}
\centering
\includegraphics[width=15cm]{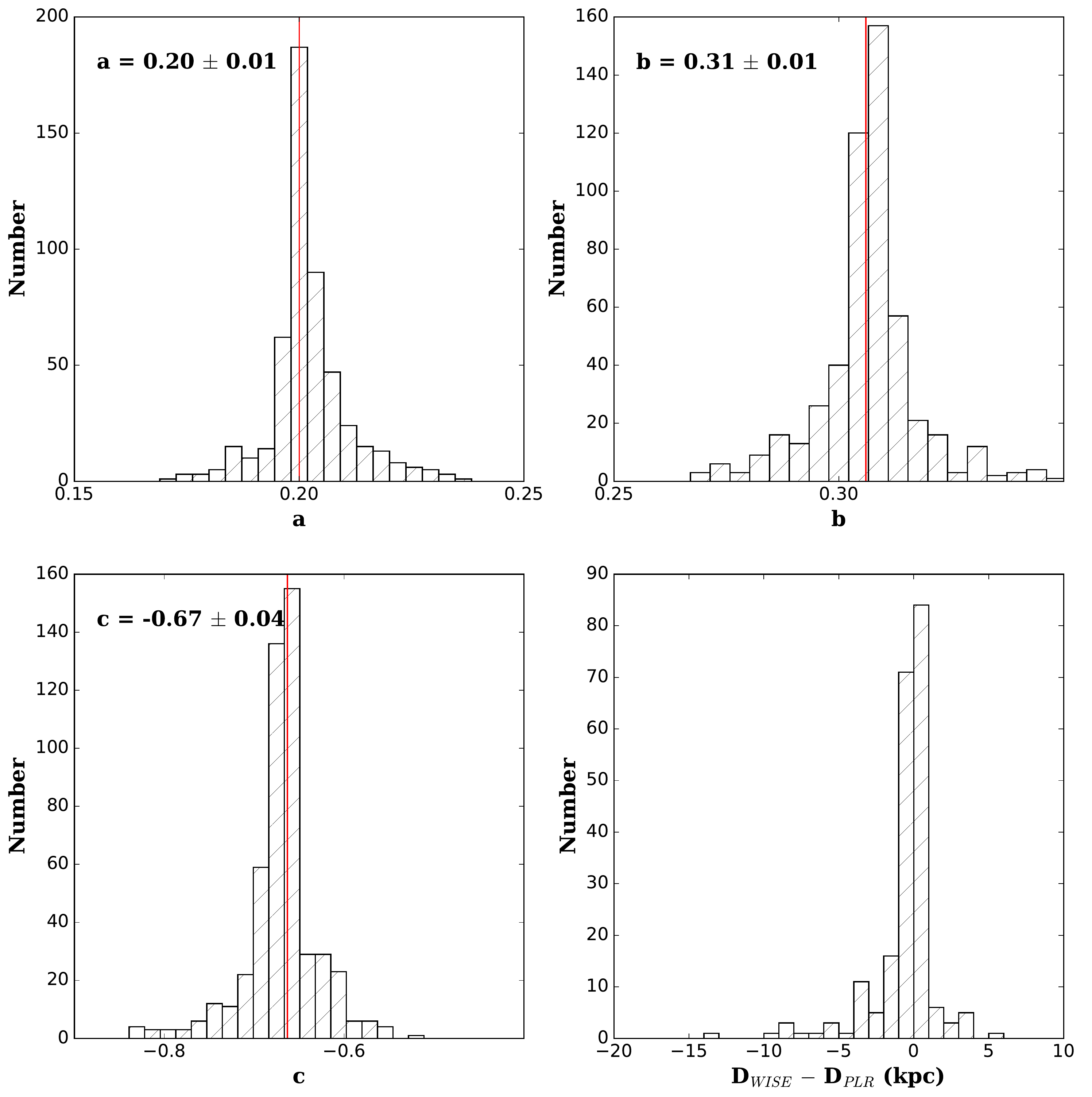}
\caption{\textit{Top left panel} is the histogram of $a$ in
Equation \ref{eq-A9} estimated with the MCMC method.  The best estimated values
and uncertainties are label in the top left corner. \textit{Top right panel}
and \textit {bottom left panel} is same as the \textit{top left panel} but for
the constant $b$ and $c$ in Equation \ref{eq-A9}. Red vertical lines denote the
best estimate values. \textit{bottom right panel} is the histogram for
differences between $D_{WISE}$ and $D_{PLR}$.
\label{fig-A2}}
\end{figure*}

We use the Markov chain Monte Carlo (MCMC) method to fit \textit{a}, \textit{b}
and \textit{c} by minimizing the differences between the model distance,
$D_{WISE}$ and the PLR distance, $D_{PLR}$. In Fig.~\ref{fig-A2}, we present
histogram of the MCMC model fitting results, with the best fitted model,
a~=~0.200, b~=~0.306, c~=~-0.663, indicated by red lines. In the bottom right
panel of Fig.~\ref{fig-A2}, shown is histogram of $D_{WISE}$-$D_{PLR}$. For the
best fitted model, the mean, median and standard deviation of
$D_{WISE}$-$D_{PLR}$ is $-$0.26, 0.03 and 1.52~kpc. It can be seen that
$D_{WISE}$-$D_{PLR}$ is within 2 kpc for most of sources. Most of the outliers are
sources towards the Bulge and Galactic centre region, where the extinction are
usually very large ($A_{Ks}$~$>$~2). Their deviation can also be
seen in lower left panel of Fig.~\ref{fig-5}, i.e., sources with $D_{WISE}$
$\sim$ 8 to 15 kpc, but with $D_{PLR}$ from 10 kpc to more than 25 kpc. The
discrepancy may also be attributed to underestimates of the extinction
estimated by three-dimensional Galactic extinction model
\citep{2003A+A...409..205D} .

\newpage
\begin{table*}
\caption{List of observed Sources}
\label{tab-8}
\begin{tabular}{ccccrrrcc}
\hline
Source & R.A.(J2000) & DEC.(J2000) & Other  & Star   & Spectral & $V^\ast$      & Period & Maser      \\
Name   &     (h:m:s) &     (d:m:s) & Name   & Type   & Type     & (km s$^{-1}$) & (days) & Detection \\
\hline
G000.760$+$51.066  & 15 00 12.98 & $+$03 15 41.7 &           &   * &M5         & ...      &      & NN  \\
G001.856$+$72.531  & 13 53 55.19 & $+$17 16 50.1 & XZ Boo    &LPV* &M5         & ...      &      & NN  \\
G003.378$+$47.338  & 15 16 02.72 & $+$02 10 04.6 & Z Ser     & sr* &M5         & -25.00   &      & NN  \\
G003.602$+$43.085  & 15 29 38.03 & $-$00 23 45.8 & AM Ser    & sr* &M2         & ...      &      & NN  \\
G003.848$+$44.341  & 15 26 10.68 & $+$00 31 56.5 & V380 Ser  & sr* &Me         & ...      &      & NN  \\
G004.482$+$32.104  & 16 05 46.29 & $-$06 42 27.8 & BD Oph    & Mi* &M6e        & -432.72  &      & YY  \\
G006.077$+$30.607  & 16 13 53.44 & $-$06 32 16.4 & V2577 Oph & Mi* &...        & ...      &      & NN  \\
G008.104$+$45.840  & 15 28 43.67 & $+$03 49 43.6 & MW Ser    & Mi* &M8         & ...      &      & YY  \\
G010.937$+$35.609  & 16 07 08.16 & $-$00 18 53.6 & AI Ser    & Mi* &...        & ...      &      & NN  \\
G011.025$+$53.268  & 15 08 25.77 & $+$09 36 18.4 & FV Boo    & Mi? &M9III      & ...      &      & YY  \\
G011.131$+$37.023  & 16 02 49.18 & $+$00 36 40.6 & DW Ser    & Mi* &M1e        & ...      &      & NN  \\
G011.159$-$41.196  & 21 04 36.85 & $-$33 16 47.3 & X Mic     & Mi* &M...       & 19.0     &      & YY  \\
G012.097$+$32.492  & 16 19 37.28 & $-$01 15 47.2 & CM Ser    & Mi* &...        & -36.93   &220.6 & NN  \\
G012.230$-$32.998  & 20 27 29.17 & $-$30 48 37.3 & V5556 Sgr & Mi* &M8         & ...      &      & NN  \\
G012.444$+$38.278  & 16 00 58.08 & $+$02 10 27.9 & BC Ser    & Mi* &M5e        & 49       &      & NN  \\
G012.901$-$63.916  & 22 53 30.71 & $-$32 55 39.9 & SS PsA    & Mi* &...        & ...      &      & NN  \\
G014.428$+$52.730  & 15 14 41.21 & $+$11 03 31.6 &           &   * &...        & ...      &      & NN  \\
G015.405$-$35.139  & 20 40 02.99 & $-$28 47 31.2 & R Mic     & Mi* &M4e        & 10.00    &      & YN  \\
G016.257$+$64.629  & 14 33 28.30 & $+$17 36 46.9 & CO Boo    & Mi* &...        & ...      &280.0 & NN  \\
G016.831$-$45.308  & 21 26 44.10 & $-$29 51 04.7 & S Mic     & Mi* &M3-5.5e    & 49       &      & NN  \\
G017.647$-$37.951  & 20 54 26.42 & $-$27 44 15.9 & RX Mic    & Mi* &...        & -51.01   &      & NN  \\
G017.863$+$72.316  & 14 04 52.15 & $+$21 21 19.1 &           &   * &...        & ...      &      & NN  \\
G019.002$-$39.495  & 21 02 20.78 & $-$27 05 14.9 & RR Cap    & Mi* &M6e:       & -63      &      & YY  \\
G019.408$+$42.783  & 15 56 29.69 & $+$09 01 50.2 & RU Ser    & Mi* &...        & 11.0     &282.2 & NN  \\
G019.509$-$56.308  & 22 18 00.24 & $-$29 36 13.8 & R PsA     & Mi* &M6+e       & -25.0    &      & NN  \\
G020.708$-$31.502  & 20 30 06.77 & $-$23 30 41.5 & AY Cap    & sr* &...        & 46.55    &      & NN  \\
G021.290$-$47.410  & 21 38 41.87 & $-$27 12 34.1 & RV PsA    & Mi* &...        & -24.36   &      & NN  \\
G021.343$-$32.293  & 20 34 07.64 & $-$23 14 58.6 & AK Cap    &LPV* &M4III      & ...      &      & NN  \\
G021.513$-$53.023  & 22 03 45.83 & $-$28 03 04.2 & S PsA     & Mi* &M3e        & -92.0    &      & YN  \\
G022.158$+$40.858  & 16 07 17.66 & $+$09 55 52.5 & U Ser     & Mi* &M4-6e      & -31.00   &      & YY  \\
G022.461$+$33.819  & 16 32 55.55 & $+$06 51 29.7 & SS Her    & Mi* &M0-5+e     & -46.0    &      & NN  \\
G022.943$-$31.448  & 20 32 34.16 & $-$21 41 26.5 & RU Cap    & Mi* &M9         & -3       &      & YY  \\
G023.139$-$38.703  & 21 02 42.82 & $-$23 46 54.8 & CE Cap    & sr* &...        & ...      &      & NN  \\
G023.291$-$34.188  & 20 44 09.74 & $-$22 18 12.9 & CC Cap    & sr* &M6.5       & ...      &      & NN  \\
G023.376$-$39.816  & 21 07 36.63 & $-$23 55 13.4 & V Cap     & Mi* &M5.5-6e    & -36.0    &      & YY  \\
G023.664$+$40.100  & 16 12 09.44 & $+$10 36 26.1 & DN Her    & Mi* &M6.5       & -46.0    &      & NN  \\
G025.952$-$42.117  & 21 19 37.45 & $-$22 42 26.9 & CH Cap    & sr* &...        & ...      &      & NN  \\
G026.343$-$35.418  & 20 52 39.24 & $-$20 20 00.3 & BX Cap    &LPV* &...        & ...      &      & NN  \\
G026.645$-$39.261  & 21 08 33.06 & $-$21 20 51.5 & X Cap     & Mi* &M2e...     & -37.55   &      & NN  \\
G028.805$-$31.578  & 20 40 32.08 & $-$17 03 28.2 & TX Cap    & Mi* &M4         & 9.0      &      & NN  \\
G032.106$-$32.402  & 20 48 08.58 & $-$14 47 01.0 & U Cap     & Mi* &M5.5e      & ...      &203.8 & NN  \\
G032.852$-$33.566  & 20 53 35.02 & $-$14 39 50.4 & XX Cap    &LPV* &M5         & -15.68   &      & NN  \\
G033.078$-$37.937  & 21 10 37.52 & $-$16 10 24.8 & Z Cap     & Mi* &M2e:       & -64      &      & NN  \\
G033.245$-$56.048  & 22 23 12.94 & $-$22 03 25.5 & RT Aqr    & Mi* &M5         & -34.00   &      & NN  \\
G033.519$-$53.484  & 22 12 50.93 & $-$21 09 51.9 & AQ Aqr    & Mi* &...        & -90.68   &      & NN  \\
G035.634$-$40.082  & 21 22 00.82 & $-$15 09 33.1 & T Cap     & Mi* &M6e:       & 42.00    &      & NN  \\
G036.242$-$55.590  & 22 23 30.38 & $-$20 18 49.2 & KU Aqr    &LPV* &M3III      & ...      &      & NN  \\
G037.206$-$55.581  & 22 24 13.46 & $-$19 47 42.7 & AV Aqr    & Mi* &M          & -73.0    &      & NN  \\
G037.250$-$44.253  & 21 39 53.43 & $-$15 40 35.4 & CK Cap    & LP? &...        & ...      &      & NN  \\
G038.070$+$66.469  & 14 37 11.58 & $+$26 44 11.7 & R Boo     & Mi* &M4-8e      & -58.00   &      & YY  \\
G038.666$-$42.354  & 21 34 22.91 & $-$13 58 29.3 & Y Cap     & Mi* &M8.5       & ...      &      & NN  \\
G039.040$-$42.707  & 21 36 10.46 & $-$13 52 04.9 & UU Cap    & sr* &M3/4III    & -4.08    &      & NN  \\
G041.021$+$62.253  & 14 56 41.05 & $+$27 30 25.2 & NP Boo    & Mi* &...        & ...      &      & NN  \\
G041.037$-$51.361  & 22 11 13.36 & $-$16 07 47.2 & YY Aqr    & sr* &...        & ...      &      & NN  \\
G041.058$-$50.181  & 22 06 44.98 & $-$15 38 40.3 & BM Aqr    & sr* &M3(III)    & -21.00   &      & NN  \\
G041.181$-$31.735  & 20 59 00.66 & $-$07 32 29.4 & VV Aqr    & Mi* &...        & -183.18  &140.7 & NN  \\
G041.307$-$63.037  & 22 57 06.46 & $-$20 20 35.7 & S Aqr     & Mi* &M6e        & -58      &      & NN  \\
G041.346$-$64.747  & 23 04 00.56 & $-$20 54 24.0 & MN Aqr    & Mi* &M7:        & ...      &      & NN  \\
G045.006$-$52.540  & 22 19 54.43 & $-$14 24 07.0 & SS Aqr    & Mi* &M2         & 2        &      & NN  \\
G045.032$-$36.798  & 21 23 03.67 & $-$07 06 29.4 & RZ Aqr    & Mi* &M9         & ...      &      & NN  \\
\hline
\multicolumn{9}{l}{Column 1 are Galactic coordinate notated source names; column 2 and 3 are equatorial coordinates; column 4 are Bayer}\\
\multicolumn{9}{l}{designation names of variables; column 5 and 6 are stellar and Spectral types; column 7 are radial velocities,}\\
\multicolumn{9}{l}{column 8 are periods; column 9 denote detections for v=1 and 2 SiO J=1-0 maser lines.}
\end{tabular}
\end{table*}

\begin{table*}
\contcaption{List of observed Sources}
\begin{tabular}{ccccrrrcc}
\hline
Source & R.A.(J2000) & DEC.(J2000) & Other  & Star   & Spectral & $V^\ast$      & Period & Maser      \\
Name   &     (h:m:s) &     (d:m:s) & Name   & Type   & Type     & (km s$^{-1}$) & (days) & Detection \\
\hline
G045.734$-$38.770  & 21 31 06.50 & $-$07 34 20.4 & HY Aqr    & Mi* &M8         & -20      &      & YN  \\
G051.533$-$62.801  & 23 04 17.19 & $-$16 00 35.9 & EQ Aqr    & sr* &M3/4       & ...      &      & NN  \\
G051.533$-$62.801  & 23 04 17.19 & $-$16 00 35.9 & EQ Aqr    & sr* &M3/4       & ...      &      & NN  \\
G051.594$-$50.205  & 22 19 35.29 & $-$09 40 32.6 & ZZ Aqr    & sr* &...        & ...      &      & NN  \\
G055.342$-$64.356  & 23 13 24.09 & $-$15 19 16.0 & UX Aqr    & Mi* &M4e        & ...      &      & NN  \\
G058.824$+$57.868  & 15 17 14.71 & $+$36 21 33.4 & RT Boo    & Mi* &M6.5e      & 35.00    &      & NN  \\
G060.246$-$57.280  & 22 54 55.48 & $-$09 22 27.7 & TT Aqr    & sr* &M3III      & ...      &      & NN  \\
G064.549$+$76.014  & 13 48 44.69 & $+$33 43 34.2 & RT CVn    & Mi* &M5e:       & -12      &      & YN  \\
G069.300$-$72.233  & 23 52 14.54 & $-$15 51 17.2 & Z Aqr     & sr* &M1/2Ib/II  & 68.90    &      & NN  \\
G070.786$-$60.449  & 23 18 58.18 & $-$07 18 50.9 & DM Aqr    & Mi* &...        & ...      &      & NN  \\
G073.394$-$77.357  & 00 10 57.96 & $-$18 34 23.4 & AC Cet    &LPV* &M3III      & -12.90   &      & NN  \\
G077.779$-$73.062  & 00 02 07.39 & $-$14 40 33.1 & W Cet     & S*  &S5.5-7/1.5 & 13.0     &      & NN  \\
G083.632$-$79.140  & 00 22 30.89 & $-$18 32 45.2 &           &LPV* &M          & ...      &201.3 & NN  \\
G085.581$-$67.859  & 23 57 54.07 & $-$08 57 31.2 & V Cet     & Mi* &M3/5(III)e & 51       &      & YN  \\
G093.803$-$63.776  & 00 01 38.63 & $-$03 45 23.3 & DU Psc    & sr* &...        & -14.09   &      & NN  \\
G094.565$-$80.456  & 00 32 21.61 & $-$18 39 08.5 & ET Cet    & sr* &M6         & ...      &      & NN  \\
G096.925$+$58.455  & 14 22 52.92 & $+$53 48 37.3 & S Boo     & Mi* &M5-6e      & -17.00   &      & NN  \\
G120.894$-$64.174  & 00 47 53.14 & $-$01 18 58.7 & SX Cet    & sr* &...        & ...      &      & NN  \\
G131.720$-$64.091  & 01 06 45.20 & $-$01 28 51.8 & Z Cet     & Mi* &M5-6e      & 3        &      & YN  \\
G133.797$-$53.388  & 01 17 34.56 & $+$08 55 52.0 & S Psc     & Mi* &M5+-7e     & 14.6     &      & YN  \\
G134.760$-$49.278  & 01 22 58.48 & $+$12 52 04.0 & U Psc     & Mi* &M4e        & -35.0    &      & NN  \\
G135.687$-$52.241  & 01 22 59.11 & $+$09 50 50.2 &           &   * &...        & ...      &      & NN  \\
G141.402$+$31.072  & 08 03 59.72 & $+$73 24 30.6 & SW Cam    & Mi* &M5e        & ...      &      & NN  \\
G141.940$-$58.536  & 01 30 38.35 & $+$02 52 52.5 & R Psc     & Mi* &M4-8e      & -45      &      & YN  \\
G144.537$-$70.053  & 01 20 37.11 & $-$08 24 52.6 & CU Cet    & sr* &M2         & ...      &      & NN  \\
G146.496$-$59.340  & 01 38 30.12 & $+$01 21 40.1 & SW Cet    &LPV* &M5         & ...      &      & NN  \\
G146.657$-$43.306  & 02 02 28.33 & $+$16 16 11.2 & RY Ari    &LPV* &M6         & ...      &      & NN  \\
G147.548$-$60.663  & 01 38 32.50 & $-$00 03 43.6 &           &LPV* &...        & ...      &168.8 & NN  \\
G149.396$-$46.550  & 02 04 37.67 & $+$12 31 36.9 & S Ari     & Mi* &M4-5e      & -27.0    &      & NY  \\
G150.794$-$47.560  & 02 06 27.29 & $+$11 12 46.1 &           &   * &M0         & ...      &      & NN  \\
G152.650$-$53.296  & 02 00 42.61 & $+$05 31 53.7 & TT Psc    & sr* &M4         & ...      &      & NN  \\
G153.782$-$52.608  & 02 04 24.28 & $+$05 50 17.1 &           &   * &...        & ...      &      & NN  \\
G156.088$-$52.393  & 02 09 46.83 & $+$05 21 41.7 &           &   * &M2         & ...      &      & NN  \\
G156.914$-$37.397  & 02 42 46.87 & $+$18 01 13.4 &           &   * &...        & ...      &      & NN  \\
G158.558$-$71.111  & 01 35 47.92 & $-$11 22 30.2 & FY Cet    & sr* &M3/4III    & ...      &      & NN  \\
G158.741$-$40.144  & 02 41 44.10 & $+$14 56 12.3 &           &   * &...        & ...      &      & NN  \\
G159.484$-$38.173  & 02 48 12.39 & $+$16 16 28.3 & BD Ari    & sr* &M7         & ...      &      & NN  \\
G159.820$-$47.007  & 02 29 17.65 & $+$08 44 08.4 &           &   * &...        & ...      &      & NN  \\
G160.874$-$49.439  & 02 26 18.94 & $+$06 18 52.5 &           &   * &M4         & ...      &      & NN  \\
G162.261$-$42.711  & 02 44 51.63 & $+$11 20 05.7 &           &   * &...        & ...      &      & NN  \\
G162.904$-$49.384  & 02 30 49.91 & $+$05 36 58.3 &           &   * &...        & ...      &      & NN  \\
G163.265$-$46.216  & 02 39 00.42 & $+$08 03 41.2 &           &   * &M4.5       & ...      &      & NN  \\
G165.494$-$43.684  & 02 50 14.10 & $+$09 09 16.2 &           &   * &...        & ...      &      & NN  \\
G165.616$-$40.899  & 02 57 27.52 & $+$11 18 05.3 & YZ Ari    & Mi* &M8         & 25       &433.4 & YY  \\
G166.965$-$54.751  & 02 26 02.31 & $-$00 10 42.0 & R Cet     & Mi* &M4-5e      & 42.00    &      & YN  \\
G168.980$+$37.738  & 08 40 49.50 & $+$50 08 11.9 & X UMa     & Mi* &M4e        & -83      &      & YY  \\
G173.516$-$38.103  & 03 23 35.71 & $+$09 23 55.0 &           &   * &M6.5       & ...      &      & NN  \\
G175.479$+$46.580  & 09 30 56.58 & $+$44 41 01.7 &           &   * &M6         & 25.80    &      & NN  \\
G175.507$+$50.992  & 09 55 19.92 & $+$44 00 29.5 & YZ UMa    &LPV* &M5V:       & ...      &      & NN  \\
G175.651$-$77.193  & 01 34 25.98 & $-$18 58 28.2 & AP Cet    & sr* &M7         & ...      &      & NN  \\
G176.154$-$30.578  & 03 51 44.23 & $+$13 06 28.8 &           &   * &M6.5       & ...      &      & NN  \\
G177.272$-$37.906  & 03 32 32.90 & $+$07 25 32.2 &           &   * &M5.5       & ...      &      & NN  \\
G177.576$-$33.940  & 03 44 59.43 & $+$09 56 36.0 & CH Tau    & sr* &M1         & ...      &      & NN  \\
G179.335$+$42.704  & 09 08 47.86 & $+$42 09 21.6 & DH Lyn    & sr* &M7         & ...      &      & NN  \\
G179.379$+$30.743  & 08 05 03.70 & $+$40 59 08.1 &           & IR  &...        & ...      &      & YY  \\
G179.553$-$33.716  & 03 50 06.63 & $+$08 52 09.0 &           &   * &M7.5       & ...      &      & NN  \\
G180.069$-$36.185  & 03 43 43.89 & $+$06 55 30.5 & V1083 Tau & Mi* &M9         & 82       &      & YY  \\
G180.829$+$32.784  & 08 16 46.88 & $+$40 07 53.3 & W Lyn     & Mi* &M6         & ...      &      & YY  \\
G181.209$-$73.392  & 01 50 33.86 & $-$17 39 00.9 & DH Cet    & sr* &M5         & ...      &      & NN  \\
G181.889$-$44.366  & 03 22 31.61 & $+$00 31 48.0 &           &   * &M5.5       & ...      &      & NN  \\
G182.006$-$35.653  & 03 49 27.68 & $+$06 04 40.4 & V1191 Tau & Mi* &M8.5       & ...      &      & YY  \\
G182.059$-$45.211  & 03 20 15.56 & $-$00 06 29.0 &           &   * &M0         & ...      &      & NN  \\
G183.100$+$30.845  & 08 08 50.31 & $+$37 52 20.6 &           &   * &...        & ...      &      & NN  \\
G183.385$+$34.661  & 08 28 08.04 & $+$38 20 23.0 & RX Lyn    & sr* &M          & ...      &      & NN  \\
\hline
\end{tabular}
\end{table*}

\begin{table*}
\contcaption{List of observed Sources}
\begin{tabular}{ccccrrrcc}
\hline
Source & R.A.(J2000) & DEC.(J2000) & Other  & Star   & Spectral & $V^\ast$      & Period & Maser      \\
Name   &     (h:m:s) &     (d:m:s) & Name   & Type   & Type     & (km s$^{-1}$) & (days) & Detection \\
\hline
G183.614$+$31.966  & 08 14 50.64 & $+$37 40 11.7 & RT Lyn    & Mi* &M6e        & ...      &      & NY  \\
G184.565$-$83.387  & 01 16 52.83 & $-$23 50 40.1 & RT Cet    &LPV* &M2         & ...      &      & NN  \\
G184.714$+$33.826  & 08 24 55.36 & $+$37 06 53.1 &           &AGB* &M7         & ...      &      & NN  \\
G185.796$+$37.989  & 08 46 12.58 & $+$36 54 42.8 &           &   * &...        & ...      &      & NN  \\
G185.814$+$62.448  & 10 48 34.26 & $+$36 17 35.2 &           &   * &M8         & ...      &      & NN  \\
G186.761$+$33.619  & 08 25 31.37 & $+$35 24 13.9 & X Lyn     & Mi* &M5         & 7.0      &      & NN  \\
G187.196$+$36.638  & 08 40 25.60 & $+$35 36 23.7 &           &AGB* &M          & ...      &      & NN  \\
G188.099$+$48.741  & 09 40 15.16 & $+$36 06 19.0 & Z LMi     &LPV* &M          & ...      &      & NN  \\
G188.152$+$51.648  & 09 54 38.71 & $+$36 05 22.8 & U LMi     & sr* &M6         & -32      &      & NN  \\
G188.344$+$50.247  & 09 47 42.79 & $+$35 58 15.1 &           &   * &...        & ...      &      & NN  \\
G188.664$-$39.190  & 03 51 15.85 & $-$00 15 53.8 &           & V*  &M7         & ...      &      & NN  \\
G189.897$-$30.567  & 04 21 45.82 & $+$03 53 50.1 &           &   * &M7.5       & ...      &      & NN  \\
G190.865$+$51.122  & 09 52 09.00 & $+$34 23 29.3 &           &   * &...        & ...      &      & NN  \\
G190.894$+$48.491  & 09 39 25.61 & $+$34 14 53.0 & VZ LMi    & Mi* &M          & ...      &292.2 & NN  \\
G195.025$-$53.735  & 03 11 53.14 & $-$11 52 32.4 & SS Eri    & Mi* &M5...      & 48       &      & YY  \\
G196.576$+$45.020  & 09 25 17.07 & $+$29 58 47.6 & TW Leo    &LPV* &Me         & ...      &216.4 & NN  \\
G197.990$+$33.262  & 08 34 28.05 & $+$26 13 47.2 &           &   * &...        & ...      &      & NN  \\
G198.081$+$74.222  & 11 40 43.82 & $+$30 03 17.0 & AY UMa    &LPV* &M3         & ...      &      & NN  \\
G198.593$-$69.596  & 02 16 00.08 & $-$20 31 10.5 & RY Cet    & Mi* &M6+e:      & 16.0     &      & YY  \\
G203.330$+$30.788  & 08 30 22.54 & $+$21 09 27.4 &           &   * &M...       & ...      &      & NN  \\
G203.660$+$52.559  & 10 02 41.02 & $+$26 41 36.0 & SV Leo    & Mi* &M7         & ...      &      & NN  \\
G204.101$+$32.634  & 08 38 46.63 & $+$21 09 32.7 & UV Cnc    &LPV* &M0         & ...      &      & NN  \\
G204.907$+$32.225  & 08 38 06.76 & $+$20 22 50.2 & DK Cnc    & sr* &M3         & ...      &      & NN  \\
G206.317$+$31.191  & 08 35 46.31 & $+$18 53 44.7 & U Cnc     & Mi* &M2e        & 72       &      & NN  \\
G206.346$+$47.275  & 09 41 34.11 & $+$23 50 33.9 &           &   * &...        & ...      &      & NN  \\
G211.919$+$50.661  & 10 00 01.99 & $+$21 15 43.9 & V Leo     & Mi* &M5-6e      & -23.00   &      & YY  \\
G212.108$+$46.532  & 09 43 25.67 & $+$19 51 40.0 & RS Leo    & Mi* &M5e        & ...      &      & NN  \\
G212.164$+$57.683  & 10 29 21.56 & $+$23 03 44.9 & UY Leo    &LPV* &M7III:     & ...      &      & NN  \\
G212.195$+$63.282  & 10 53 09.43 & $+$24 21 31.1 & RU Leo    &LPV* &M3         & ...      &      & NN  \\
G214.232$+$43.058  & 09 31 51.13 & $+$17 15 05.5 &           &LPV* &...        & ...      &177.7 & NN  \\
G217.372$+$50.948  & 10 05 58.80 & $+$18 06 04.9 &           & V*  &M8         & ...      &      & NN  \\
G218.786$+$51.178  & 10 08 14.78 & $+$17 21 30.6 & DD Leo    & sr* &M8         & ...      &      & NN  \\
G220.968$+$59.976  & 10 44 40.35 & $+$19 25 23.8 & EW Leo    & sr* &M5         & ...      &      & NN  \\
G223.002$+$48.943  & 10 04 15.94 & $+$13 58 57.7 & RY Leo    & sr* &M2         & 22.00    &      & NN  \\
G234.525$+$52.514  & 10 31 27.87 & $+$09 21 07.0 &           &   * &M7         & ...      &      & NN  \\
G235.246$+$67.258  & 11 23 40.03 & $+$16 51 07.0 & TZ Leo    & Mi* &M8         & 18.0     &      & YY  \\
G236.412$+$81.845  & 12 18 46.68 & $+$23 38 43.2 & AB Com    & Mi* &M          & ...      &195.6 & NN  \\
G238.855$+$56.995  & 10 52 11.04 & $+$09 48 55.2 &           &   * &...        & ...      &      & NN  \\
G239.757$+$46.896  & 10 20 50.19 & $+$03 21 09.0 & SZ Sex    & Mi* &...        & ...      &147.9 & NN  \\
G243.022$+$46.682  & 10 25 43.90 & $+$01 28 08.7 & SY Sex    & Mi* &...        & ...      &208.7 & NN  \\
G248.071$-$84.665  & 01 11 36.38 & $-$30 06 29.4 & U Scl     & Mi* &M5e        & -8       &      & YY  \\
G250.569$+$57.710  & 11 10 50.78 & $+$05 27 34.7 & S Leo     & Mi* &M6:e       & 106      &      & NN  \\
G254.436$+$65.121  & 11 36 54.77 & $+$09 31 45.9 & ZZ Leo    & sr* &M0         & ...      &      & NN  \\
G261.694$+$46.256  & 11 01 55.14 & $-$07 39 41.8 & RT Crt    & Mi* &M8         & 41.00    &      & YY  \\
G262.094$+$61.213  & 11 37 48.11 & $+$04 19 24.8 & IW Vir    &LPV* &M5         & ...      &      & NN  \\
G264.438$+$71.690  & 12 05 14.81 & $+$12 21 37.9 & SU Vir    & Mi* &M2-5.5e    & 19       &      & NN  \\
G264.623$+$48.097  & 11 12 45.30 & $-$07 17 54.5 & U Crt     & Mi* &M0e        & ...      &      & NN  \\
G268.755$+$80.079  & 12 27 57.87 & $+$18 48 08.5 & TV Com    &LPV* &M2         & ...      &      & NN  \\
G276.361$+$63.050  & 12 04 36.16 & $+$02 37 10.6 & TZ Vir    & sr* &M5         & ...      &      & NN  \\
G283.439$+$80.931  & 12 38 43.06 & $+$18 32 41.7 & DO Com    & sr* &...        & ...      &      & NN  \\
G283.995$+$74.423  & 12 30 58.11 & $+$12 18 30.6 & CV Vir    & Mi* &...        & ...      &148.4 & NN  \\
G289.779$+$69.653  & 12 33 08.38 & $+$07 15 00.3 & CI Vir    &LPV* &M6         & ...      &      & NN  \\
G289.879$+$76.356  & 12 38 51.52 & $+$13 48 13.9 & KM Com    &LPV* &M2         & 23.7     &      & NN  \\
G294.615$+$58.166  & 12 33 52.99 & $-$04 25 19.6 & Y Vir     & Mi* &M5.5e:     & 9        &      & NN  \\
G309.119$+$49.507  & 13 07 55.40 & $-$13 09 58.9 & RV Vir    & Mi* &A5         & 33       &      & NN  \\
G311.686$+$77.932  & 12 58 59.74 & $+$15 11 21.7 & RX Com    & Mi* &...        & ...      &210.5 & NN  \\
G312.284$+$53.849  & 13 13 41.63 & $-$08 37 05.7 & HH Vir    & sr* &...        & ...      &      & NN  \\
G315.566$+$57.522  & 13 18 30.52 & $-$04 41 03.2 & VY Vir    & Mi* &M3pev      & ...      &      & YY  \\
G320.482$+$58.454  & 13 27 48.13 & $-$03 10 22.9 & V Vir     & Mi* &M5-6e      & 33       &      & NN  \\
G325.570$+$85.690  & 12 58 38.90 & $+$23 08 21.0 & T Com     & Mi* &M2         & 15.0     &      & YY  \\
G326.771$+$45.772  & 13 58 59.11 & $-$13 56 59.1 &           &   * &M6         & ...      &      & NN  \\
G330.757$+$45.262  & 14 10 22.10 & $-$13 18 11.8 & Z Vir     & Mi* &M5e        & 68       &      & YN  \\
\hline
\end{tabular}
\end{table*}

\begin{table*}
\contcaption{List of observed Sources}
\begin{tabular}{ccccrrrcc}
\hline
Source & R.A.(J2000) & DEC.(J2000) & Other  & Star   & Spectral & $V^\ast$      & Period & Maser      \\
Name   &     (h:m:s) &     (d:m:s) & Name   & Type   & Type     & (km s$^{-1}$) & (days) & Detection \\
\hline
G331.573+49.548  & 14 04 53.44 & $-$09 11 41.2 & RR Vir    & Mi* &...        & -43.0    &215.7 & NN  \\
G331.781+33.351  & 14 35 49.00 & $-$23 36 29.4 & LX Lib    & Mi* &...        & -10.60   &      & NN  \\
G332.578+54.382  & 13 58 37.59 & $-$04 34 32.8 & SY Vir    & Mi* &M6         & ...      &      & NN  \\
G333.388+33.799  & 14 39 59.87 & $-$22 34 26.2 & EP Lib    & Mi* &...        & ...      &      & NN  \\
G334.109+36.043  & 14 37 29.14 & $-$20 19 41.2 & LY Lib    & Mi* &...        & ...      &      & YY  \\
G335.504+35.524  & 14 42 46.26 & $-$20 12 36.1 & SX Lib    & Mi* &M6e...     & ...      &      & YY  \\
G335.646+44.462  & 14 24 29.54 & $-$12 25 07.3 &           & V*  &M1III      & ...      &      & NN  \\
G335.670+35.438  & 14 43 27.21 & $-$20 12 53.3 & GS Lib    & Mi* &M6         & ...      &      & NN  \\
G336.386+35.873  & 14 44 36.98 & $-$19 32 28.2 & TW Lib    & Mi* &...        & 81.94    &      & NN  \\
G336.532+38.006  & 14 40 22.18 & $-$17 39 26.9 & V Lib     & Mi* &M5e        & 15       &      & YN  \\
G337.373+32.451  & 14 55 21.62 & $-$22 00 19.6 & EG Lib    & Mi* &M5         & 5        &      & YY  \\
G337.755+51.213  & 14 15 44.47 & $-$05 52 06.4 & CF Vir    & Mi* &M5e        & 44.03    &      & NN  \\
G339.224+44.663  & 14 32 59.87 & $-$10 56 03.2 & KS Lib    & Mi* &Me         & ...      &      & YY  \\
G339.938+43.686  & 14 36 54.68 & $-$11 28 40.8 &           &   * &M6.5       & ...      &      & NN  \\
G340.371+66.478  & 13 48 08.49 & $+$07 49 13.2 & HX Boo    & sr* &M5         & ...      &      & NN  \\
G340.829+31.460  & 15 08 10.66 & $-$21 10 00.3 & YY Lib    & Mi* &Me         & ...      &      & YN  \\
G342.142+33.667  & 15 06 26.19 & $-$18 43 56.2 & RT Lib    & Mi* &M2.5-5.5e  & 41.0     &      & NN  \\
G342.197+32.029  & 15 10 44.36 & $-$20 01 08.4 & T Lib     & Mi* &M4         & -48      &      & NN  \\
G344.240+30.203  & 15 21 23.96 & $-$20 23 18.5 & S Lib     & Mi* &M1.5-4e    & 294.0    &      & NN  \\
G345.104+35.879  & 15 08 54.49 & $-$15 29 51.0 & TT Lib    & Mi* &M3e        & -47.27   &      & NN  \\
G346.039+45.775  & 14 46 18.42 & $-$07 15 49.9 & AQ Vir    & Mi* &M5e        & -4.0     &      & NN  \\
G346.087+47.560  & 14 42 00.64 & $-$05 49 57.2 & XY Vir    & Mi* &...        & 29.89    &154.8 & NN  \\
G348.715+50.518  & 14 39 59.19 & $-$02 26 48.6 &           &   * &...        & ...      &      & NN  \\
G349.797+58.296  & 14 21 51.90 & $+$03 54 27.8 & AO Vir    & Mi* &M4         & ...      &      & NN  \\
G350.237+30.144  & 15 38 05.25 & $-$17 01 54.2 & EK Lib    & sr* &M7e        & ...      &      & NN  \\
G350.511+84.894  & 13 07 53.22 & $+$23 37 28.6 &           &   * &...        & ...      &      & NN  \\
G350.719+30.733  & 15 37 39.75 & $-$16 19 02.8 & IRC-20290 & IR  &M7         & ...      &      & NN  \\
G350.868+30.821  & 15 37 48.00 & $-$16 09 57.1 & W Lib     & Mi* &...        & 22.0     &      & NN  \\
G351.579+32.192  & 15 35 41.95 & $-$14 45 16.3 &           &   * &...        & ...      &      & NN  \\
G352.221+57.421  & 14 28 03.54 & $+$04 06 33.3 &           &   * &...        & ...      &      & NN  \\
G353.548+50.630  & 14 48 52.46 & $-$00 19 37.0 &           &   * &M7         & ...      &      & NN  \\
G353.826+42.588  & 15 11 41.26 & $-$06 00 41.2 & Y Lib     & Mi* &M5.5e      & -7.00    &      & YN  \\
G356.016+48.746  & 14 58 42.58 & $-$00 33 16.5 &           &   * &M6         & ...      &      & NN  \\
G356.642+59.618  & 14 28 30.27 & $+$07 17 37.1 & AP Vir    & Mi? &M3         & ...      &      & YY  \\
G358.057+46.657  & 15 08 32.60 & $-$01 00 45.2 &           &   * &M6         & ...      &      & NN  \\
G358.765+67.787  & 14 06 29.61 & $+$13 29 05.5 & Z Boo     & Mi* &M6e        & 40       &      & NN  \\
\hline
\end{tabular}
\end{table*}

\begin{table*}
\caption{Observational results of SiO masers.}
\label{tab-9}
\begin{tabular}{@{\extracolsep{5pt}}crcccrcccc@{}}
\hline
       & \multicolumn{4}{c}{SiO v=1 J=1-0 maser line} & \multicolumn{4}{c}{SiO v=2  J=1-0 maser line} &      \\
\cline{2-5}   \cline{6-9}
Source &\textit{V}$_{\rm LSR}$ & \textit{T}$^\ast_A$(peak) & Int. Flux& rms & \textit{V}$_{\rm LSR}$ & \textit{T}$^\ast_A$(peak) & Int. Flux & rms & Ref.\\
Name & {\footnotesize (K km s$^{-1}$)} & {\footnotesize (K)}   & {\footnotesize (K km s$^{-1}$)} & {\footnotesize (K)}  &  {\footnotesize (K km s$^{-1}$)}    & {\footnotesize (K)}   & {\footnotesize (K km s$^{-1}$)} & {\footnotesize (K)} & {}\\
\hline
G004.482$+$32.104 &   $-$8.8   &  0.18  &  0.79  &  0.03  &  $-$10.8  &  0.10  &  0.14  &  0.02 & 2 \\
G008.104$+$45.840 &     41.0   &  0.37  &  1.39  &  0.05  &     42.8  &  0.16  &  0.72  &  0.04 & 3 \\
G011.025$+$53.268 &      4.5   &  1.61  & 11.76  &  0.08  &  $-$13.7  &  0.54  &  2.36  &  0.07 & 4 \\
G011.159$-$41.196 &     17.8   &  0.72  &  1.64  &  0.09  &     18.2  &  0.28  &  0.94  &  0.05 & 1 \\
G015.405$-$35.139 &     19.9   &  0.20  &  0.87  &  0.04  &      ---  &  ---   &  ---   &  0.04 & 1 \\
G019.002$-$39.495 &  $-$51.8   &  0.94  &  5.27  &  0.07  &  $-$51.3  &  0.40  &  1.52  &  0.07 & 1 \\
G019.509$-$56.308 &  $-$29.0   &  1.26  &  3.76  &  0.07  &  $-$29.7  &  0.36  &  1.10  &  0.06 & 1 \\
G021.513$-$53.023 & $-$100.8   &  0.22  &  0.53  &  0.04  &      ---  &  ---   &  ---   &  0.04 & 1 \\
G022.158$+$40.858 &  $-$15.0   &  0.69  &  1.34  &  0.11  &  $-$15.6  &  1.01  &  1.41  &  0.10 & 1 \\
G022.943$-$31.448 &      9.8   &  1.33  &  4.30  &  0.12  &      7.0  &  1.38  &  2.65  &  0.11 & 4 \\
G023.376$-$39.816 &  $-$20.9   &  1.41  &  2.33  &  0.06  &  $-$16.5  &  0.25  &  0.88  &  0.05 & 1 \\
G038.070$+$66.469 &  $-$42.4   &  0.53  &  2.69  &  0.06  &  $-$44.0  &  0.22  &  0.87  &  0.04 & 3 \\
G045.734$-$38.770 &     26.3   &  0.23  &  0.47  &  0.03  &      ---  &  ---   &  ---   &  0.03 & 1 \\
G064.549$+$76.014 &     26.3   &  0.11  &  0.27  &  0.03  &     26.3  &  0.15  &  0.54  &  0.02 & 1 \\
G085.581$-$67.859 &     50.9   &  0.22  &  1.09  &  0.04  &      ---  &  ---   &  ---   &  0.04 & 1 \\
G131.720$-$64.091 &      4.0   &  0.29  &  0.94  &  0.04  &      ---  &  ---   &  ---   &  0.04 & 1 \\
G133.797$-$53.388 &      4.3   &  1.10  &  5.68  &  0.10  &      4.3  &  1.48  &  4.23  &  0.11 & 1 \\
G141.940$-$58.536 &  $-$57.0   &  0.82  &  2.01  &  0.09  &  $-$56.8  &  0.58  &  1.38  &  0.09 & 5 \\
G149.396$-$46.550 &      ---   &  ---   &  ---   &  0.05  &      9.5  &  0.16  &  0.75  &  0.05 & 1 \\
G165.616$-$40.899 &     13.0   &  0.88  &  1.59  &  0.07  &     12.5  &  0.52  &  0.71  &  0.07 & 6 \\
G166.965$-$54.751 &     35.2   &  0.38  &  0.90  &  0.05  &      ---  &  ---   &  ---   &  0.05 & 7 \\
G168.980$+$37.738 &  $-$82.9   &  0.38  &  0.65  &  0.06  &  $-$83.5  &  0.20  &  0.26  &  0.04 & 1 \\
G179.379$+$30.743 &  $-$10.6   &  0.81  &  2.59  &  0.08  &   $-$9.2  &  0.72  &  2.01  &  0.07 & 2 \\
G180.069$-$36.185 &     58.0   &  0.09  &  0.25  &  0.02  &     57.4  &  0.11  &  0.29  &  0.02 & 1 \\
G180.829$+$32.784 &  $-$24.8   &  0.41  &  0.40  &  0.08  &  $-$25.1  &  0.54  &  0.62  &  0.07 & 1 \\
G182.006$-$35.653 &     61.0   &  0.11  &  0.47  &  0.02  &     61.3  &  0.19  &  0.85  &  0.03 & 1 \\
G183.614$+$31.966 &      ---   &  ---   &  ---   &  0.03  &     27.5  &  0.13  &  0.32  &  0.02 & 1 \\
G195.025$-$53.735 &     33.2   &  0.27  &  0.42  &  0.04  &     35.3  &  0.12  &  0.37  &  0.03 & 1 \\
G198.593$-$69.596 &   $-$2.8   &  0.55  &  1.41  &  0.06  &   $-$3.4  &  0.19  &  0.62  &  0.04 & 1 \\
G211.919$+$50.661 &  $-$25.5   &  1.23  &  4.30  &  0.09  &  $-$24.2  &  0.57  &  2.03  &  0.07 & 3 \\
G235.246$+$67.258 &     12.9   &  0.30  &  0.59  &  0.04  &     12.7  &  0.35  &  0.39  &  0.04 & 1 \\
G248.071$-$84.665 &  $-$10.3   &  0.32  &  1.03  &  0.06  &  $-$10.4  &  0.35  &  0.89  &  0.05 & 1 \\
G261.694$+$46.256 &     32.9   &  0.38  &  1.10  &  0.05  &     32.8  &  0.28  &  0.55  &  0.05 & 1 \\
G315.566$+$57.522 &     72.0   &  0.11  &  0.72  &  0.03  &     70.7  &  0.23  &  0.83  &  0.04 & 1 \\
G325.570$+$85.690 &     26.1   &  2.01  &  5.22  &  0.09  &     25.2  &  2.66  &  8.13  &  0.09 & 3 \\
G330.757$+$45.262 &     66.8   &  0.21  &  0.39  &  0.03  &      ---  &  ---   &  ---   &  0.03 & 1 \\
G334.109$+$36.043 &  $-$25.4   &  0.23  &  1.14  &  0.05  &  $-$25.3  &  0.27  &  0.24  &  0.05 & 1 \\
G335.504$+$35.524 &  $-$31.2   &  0.83  &  3.08  &  0.08  &  $-$31.0  &  0.56  &  2.02  &  0.07 & 1 \\
G336.532$+$38.006 &     18.1   &  0.22  &  0.28  &  0.03  &      ---  &  ---   &  ---   &  0.03 & 1 \\
G337.373$+$32.451 &   $-$7.8   &  1.42  &  4.64  &  0.13  &   $-$9.6  &  0.47  &  0.99  &  0.08 & 2 \\
G339.224$+$44.663 &     70.7   &  0.18  &  0.93  &  0.04  &     69.9  &  0.14  &  0.64  &  0.03 & 2 \\
G340.829$+$31.460 &   $-$3.9   &  0.15  &  0.48  &  0.03  &      ---  &  ---   &  ---   &  0.03 & 1 \\
G353.826$+$42.588 &     14.9   &  0.14  &  0.48  &  0.03  &      ---  &  ---   &  ---   &  0.03 & 2,3 \\
G356.642$+$59.618 &     37.2   &  0.34  &  0.41  &  0.04  &     37.0  &  0.44  &  0.60  &  0.04 & 1 \\
\hline

\multicolumn{10}{l}{{{\bf Note:} column 1 are source name; columns 2, 3, 4, 5 and columns 6, 7, 8, 9 are $V_{\rm LSR}$, peak antennas temperatures (in unit of K),}}  \\
\multicolumn{10}{l}{{integrated flux density (in unit of K~km~s$^{-1}$), and 1$\sigma$ rms of spectra for v=1 and v=2 SiO maser lines. Column 9 denote references.}}  \\
\multicolumn{10}{l}{{Reference: (1) this paper; (2) \cite{2001A+A...376..112I}; (3) \cite{2010ApJS..188..209K}; (4) \cite{2010ApJS..188..209K}; (5) \cite{2012AJ....144..129C}; (6) \cite{2012PASJ...64....4D};}} \\
\multicolumn{10}{l}{{(7) \cite{2014MNRAS.441.3226I}}}\\
\end{tabular}
\end{table*}

\begin{figure*}
\includegraphics[width=5.0cm]{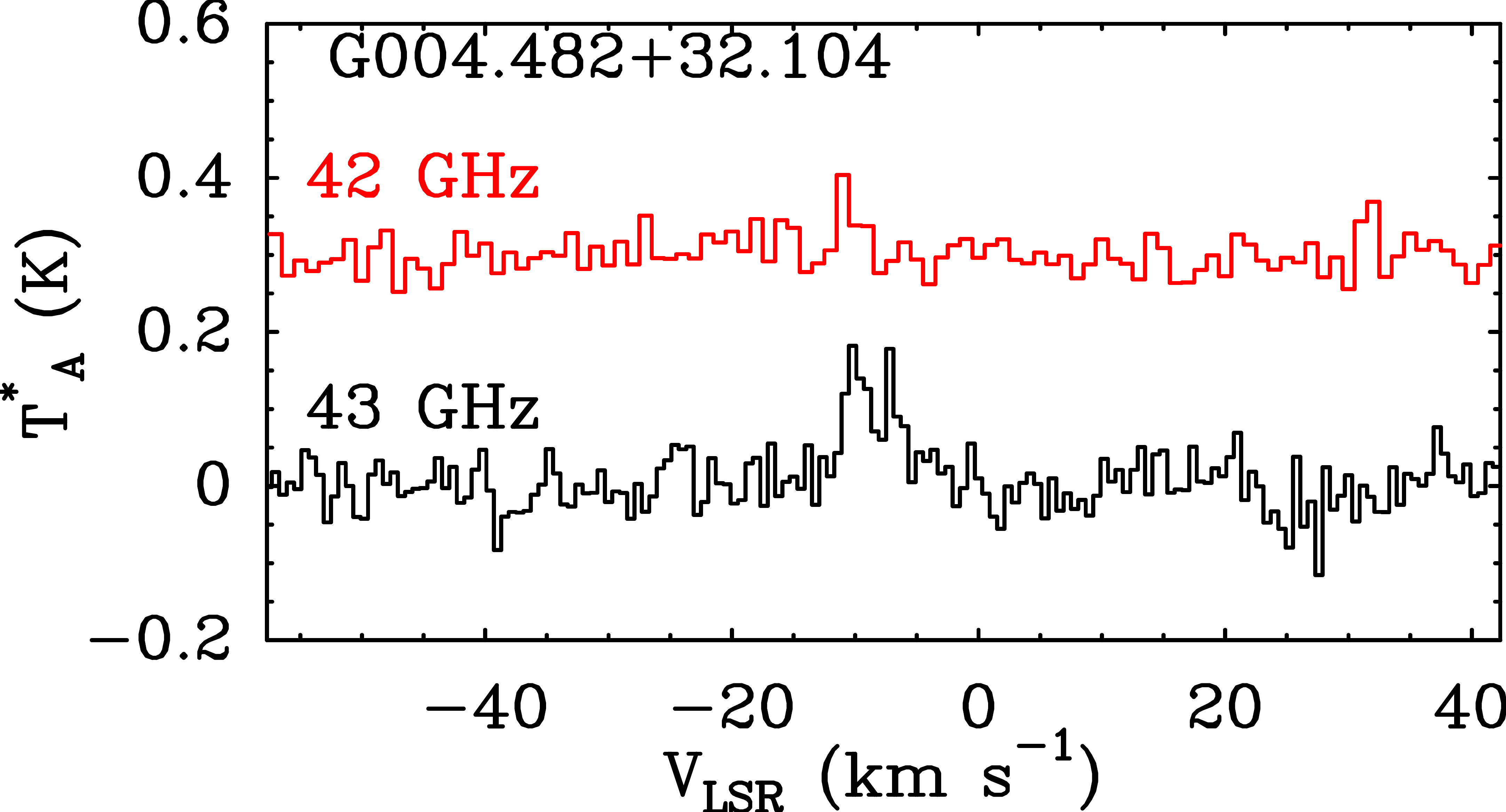}
\includegraphics[width=5.0cm]{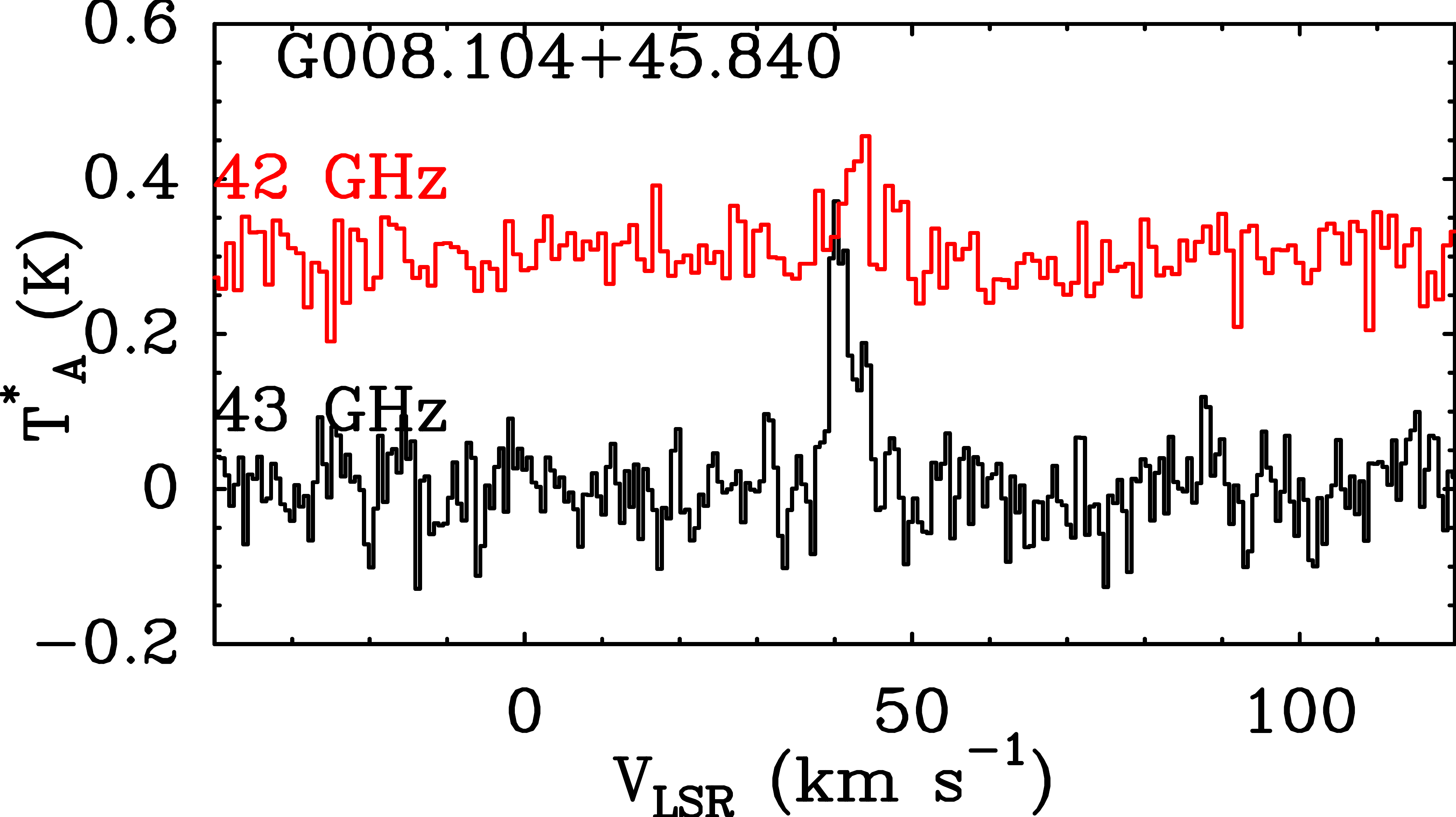} 
\includegraphics[width=5.0cm]{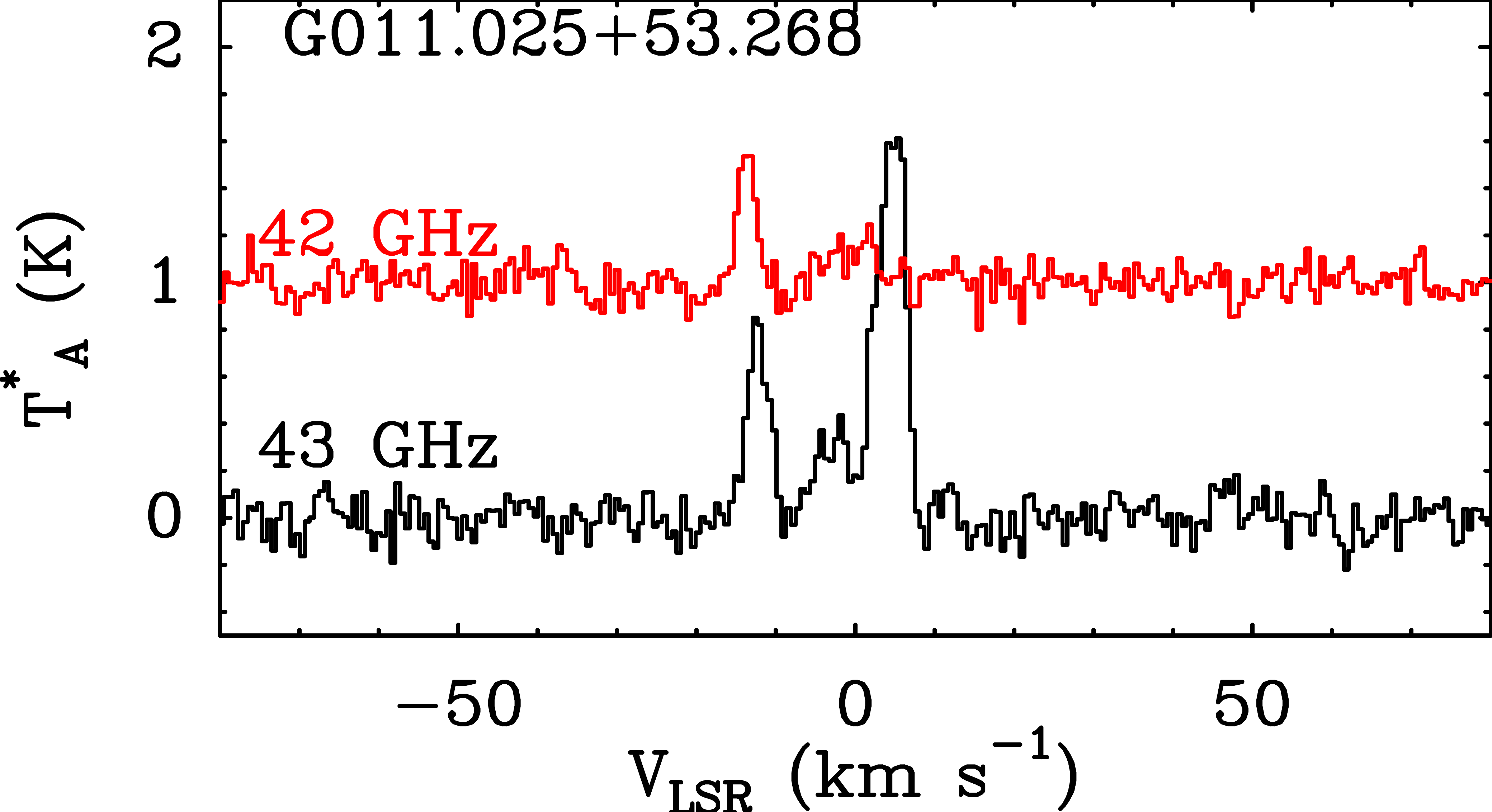} \\
\includegraphics[width=5.0cm]{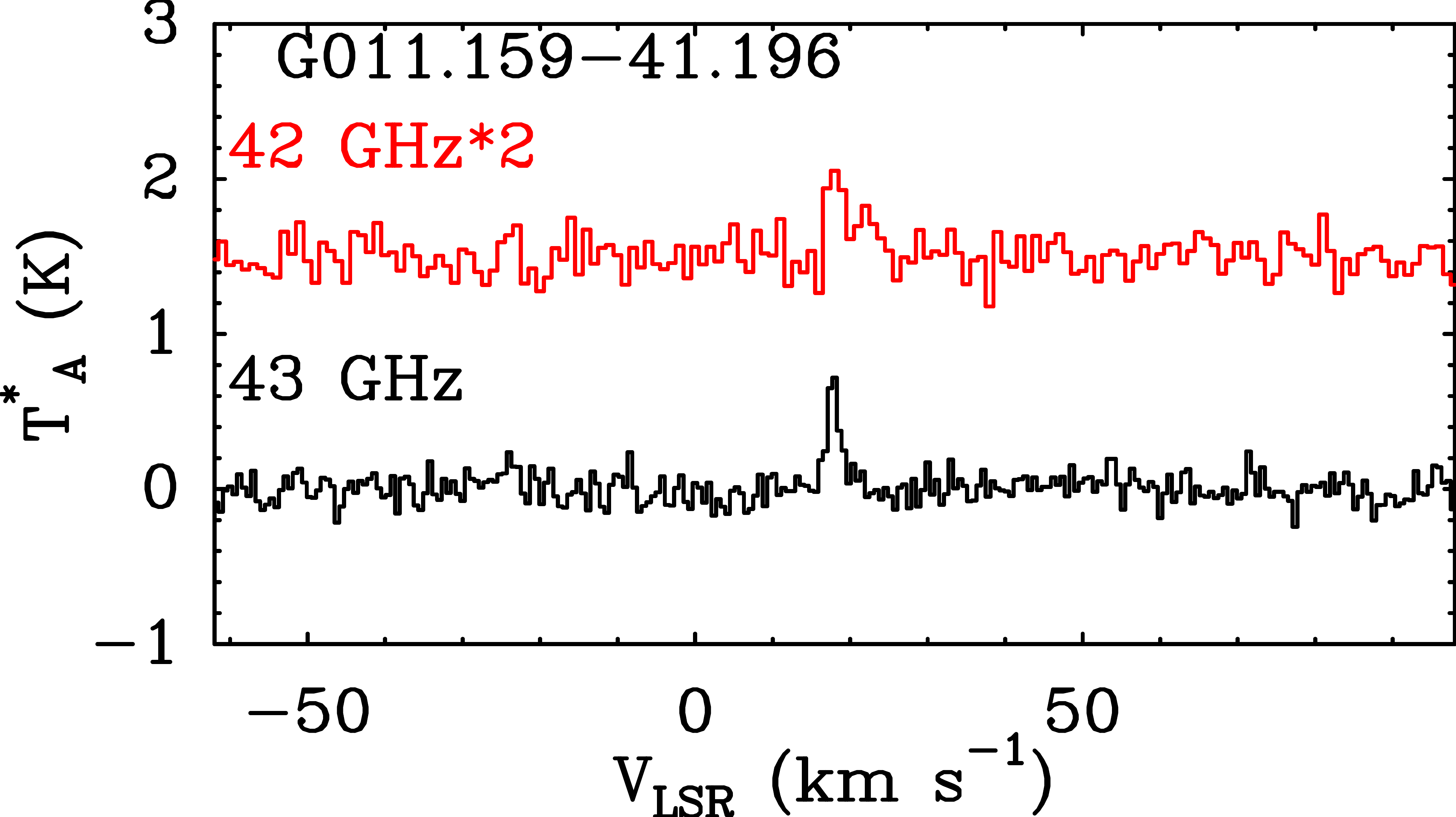} 
\includegraphics[width=5.0cm]{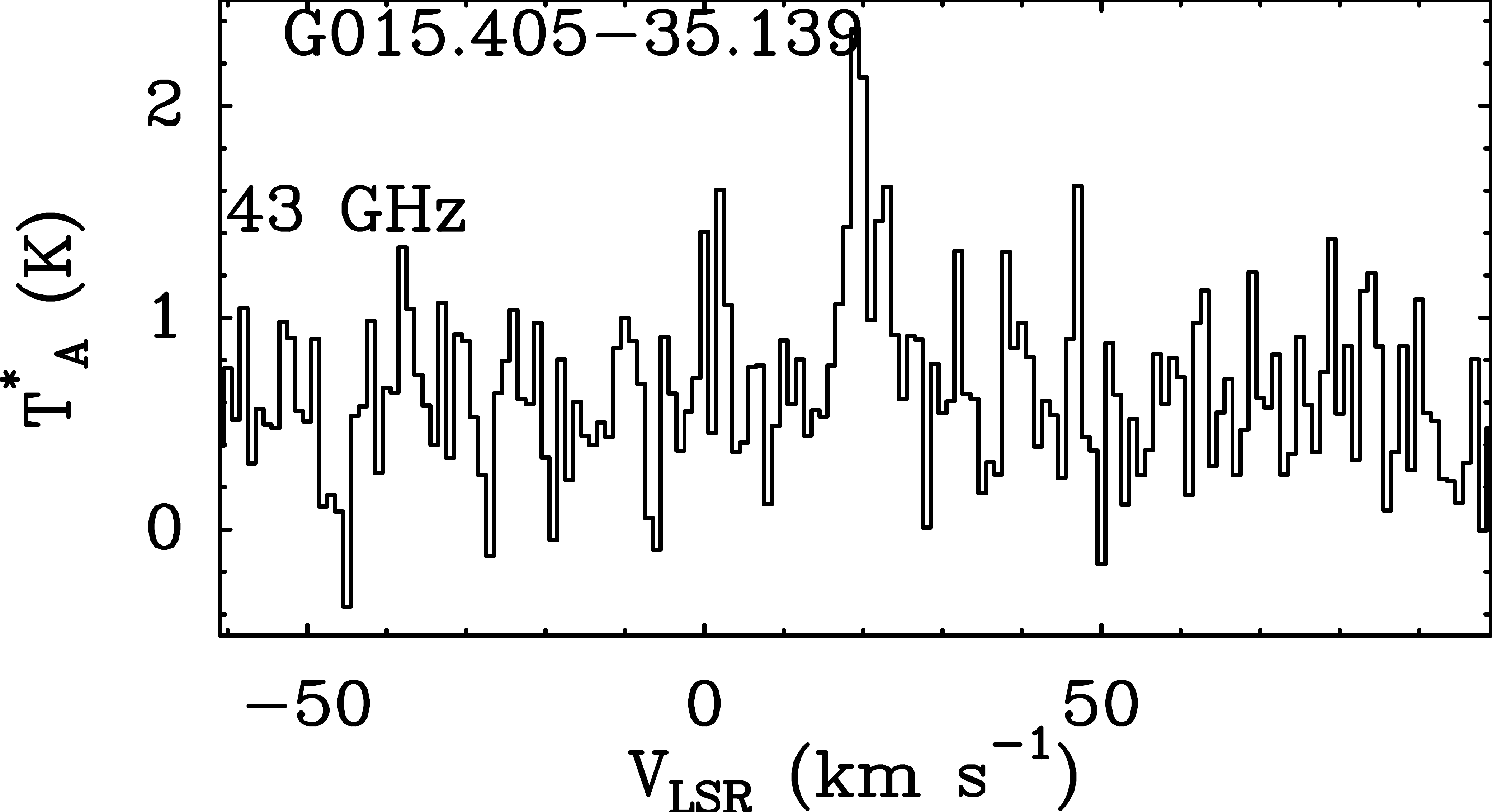}
\includegraphics[width=5.0cm]{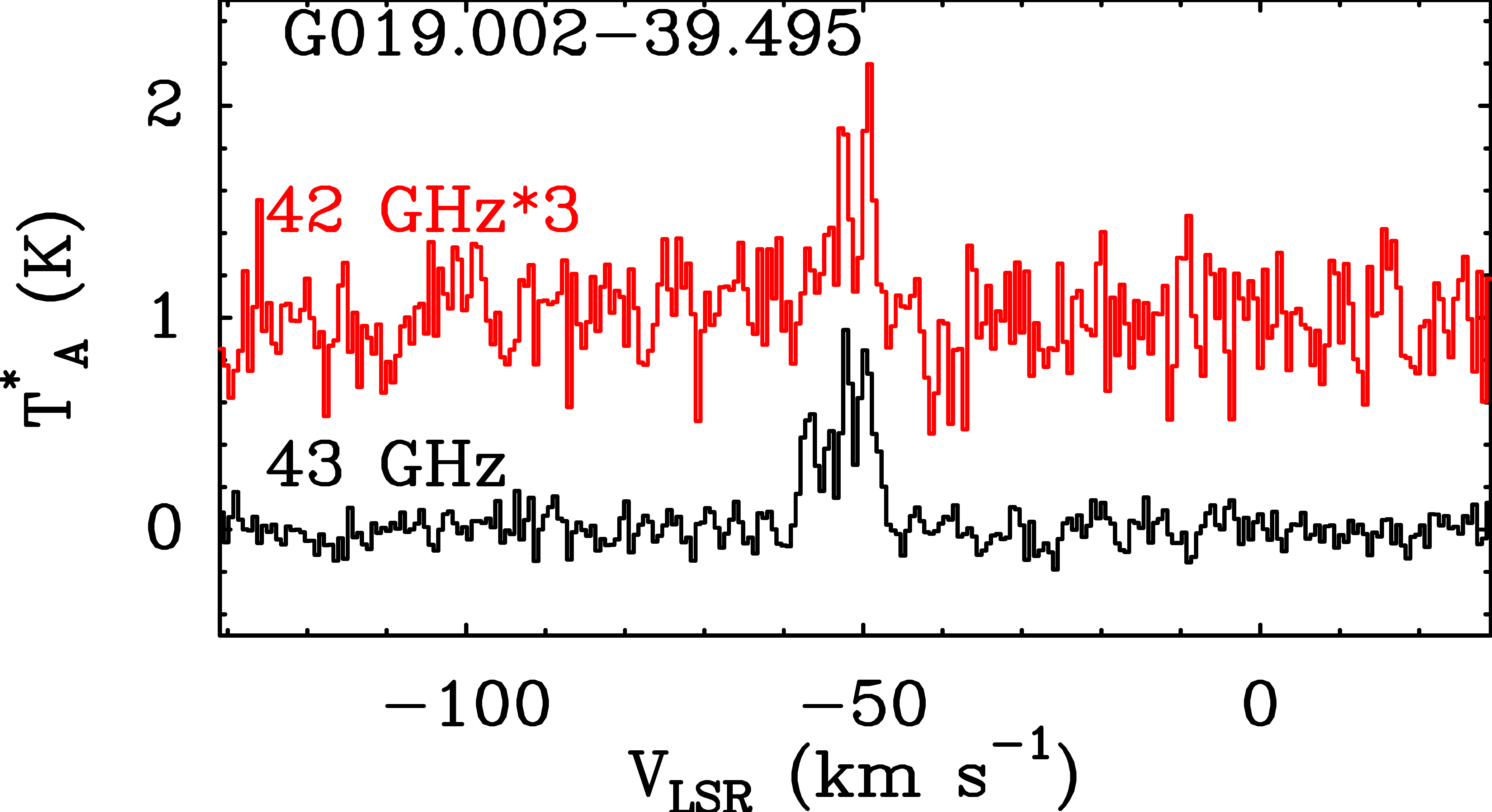}  \\
\includegraphics[width=5.0cm]{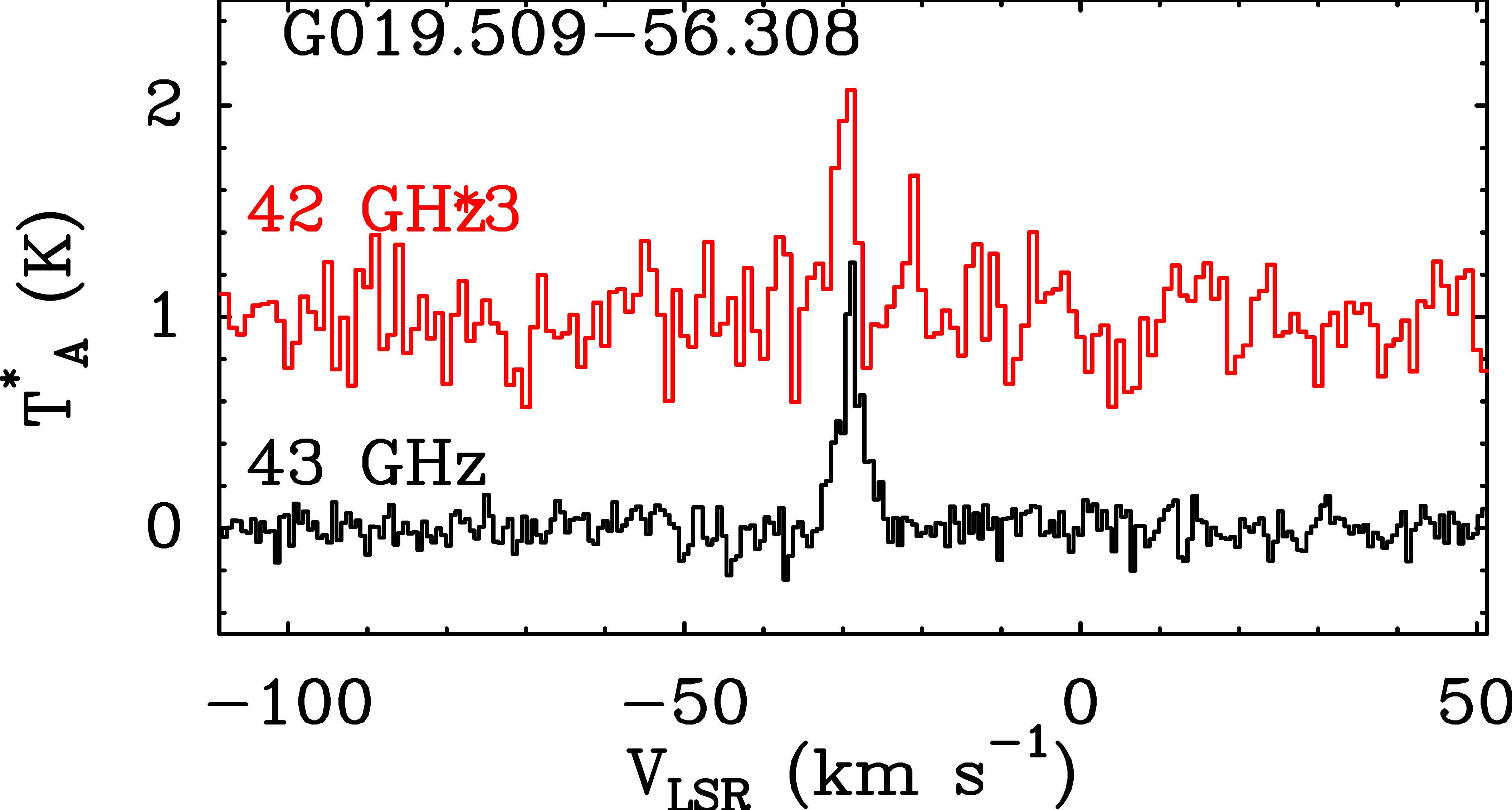} 
\includegraphics[width=5.0cm]{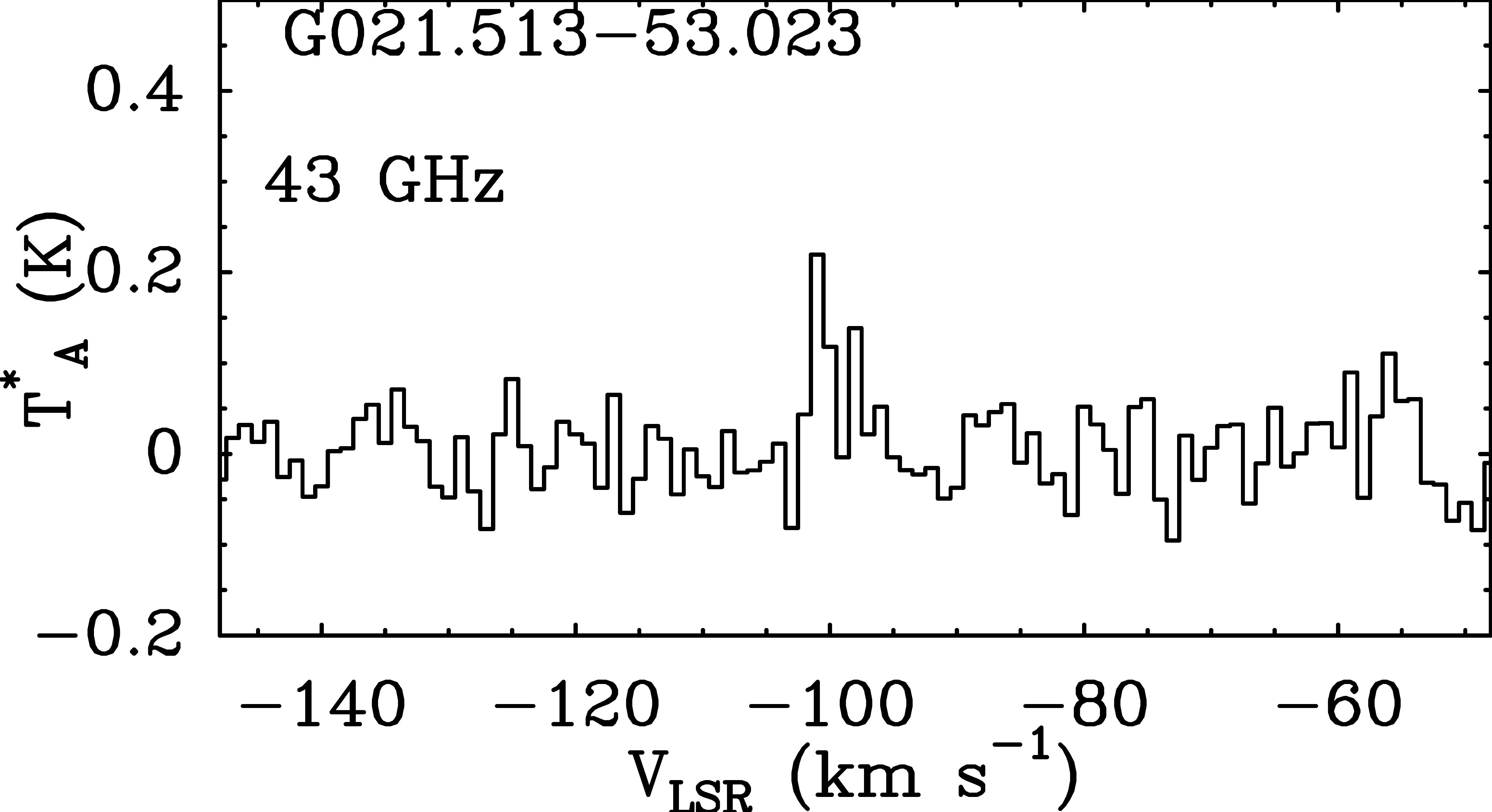} 
\includegraphics[width=5.0cm]{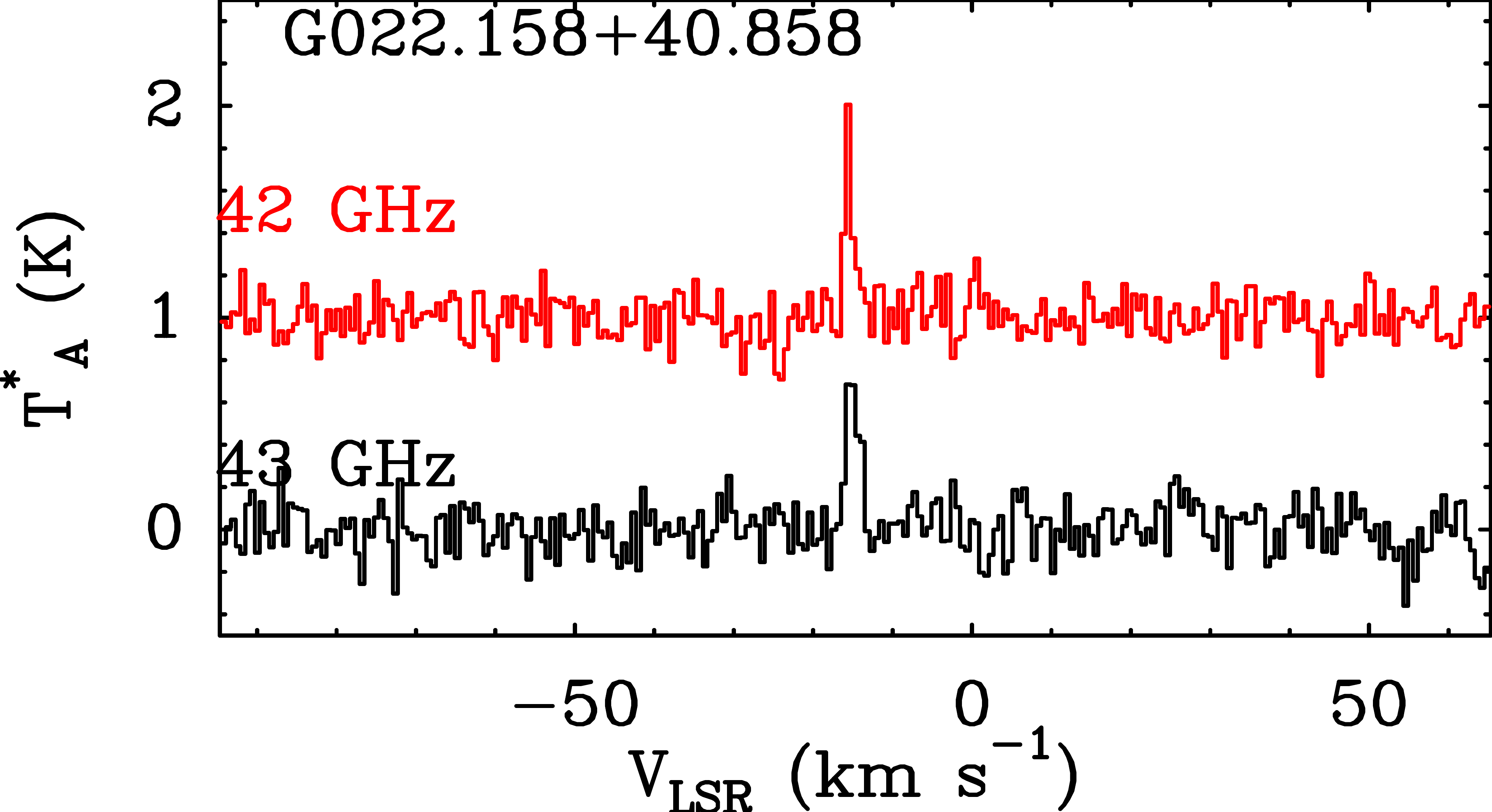} \\
\includegraphics[width=5.0cm]{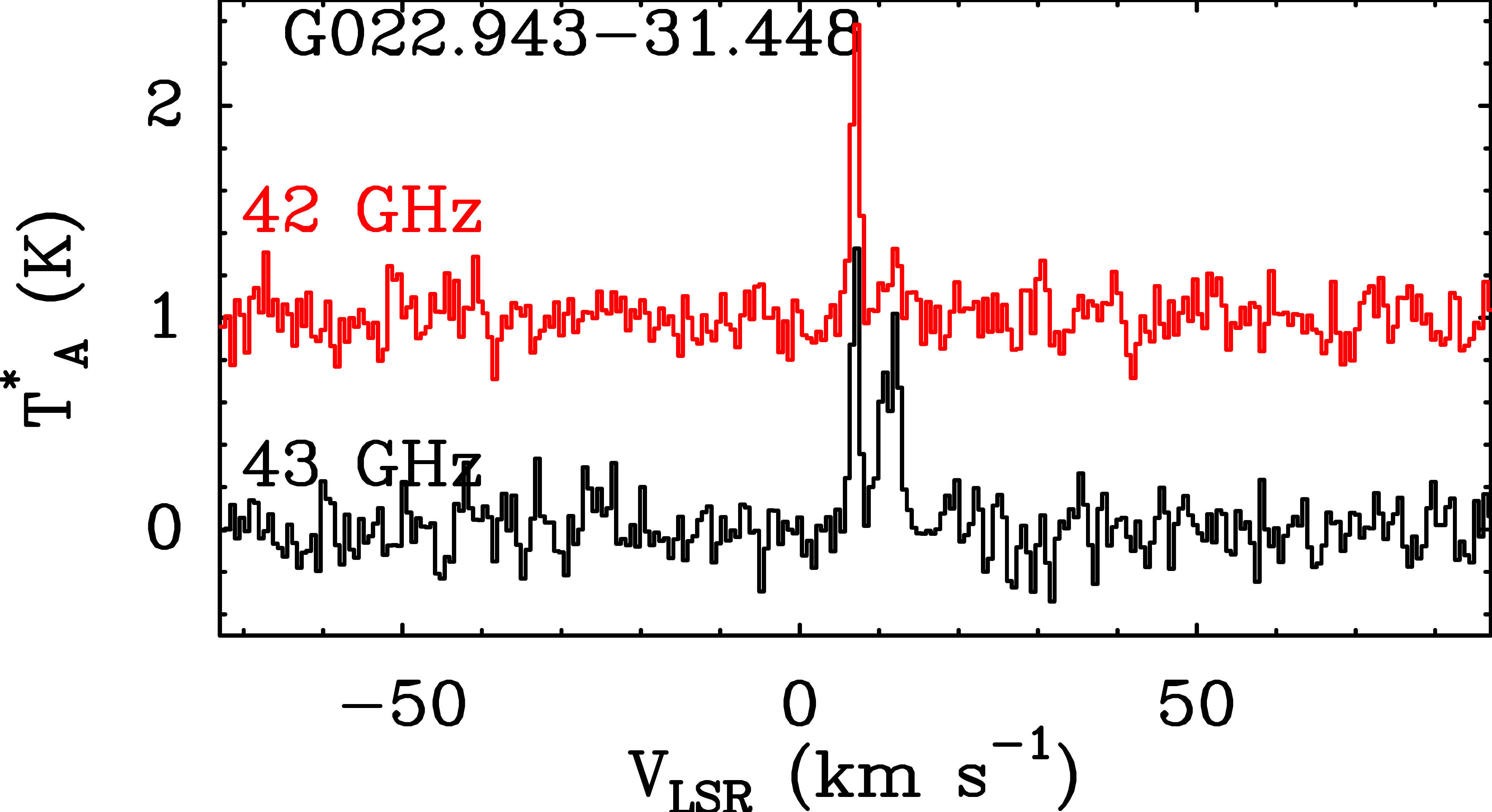} 
\includegraphics[width=5.0cm]{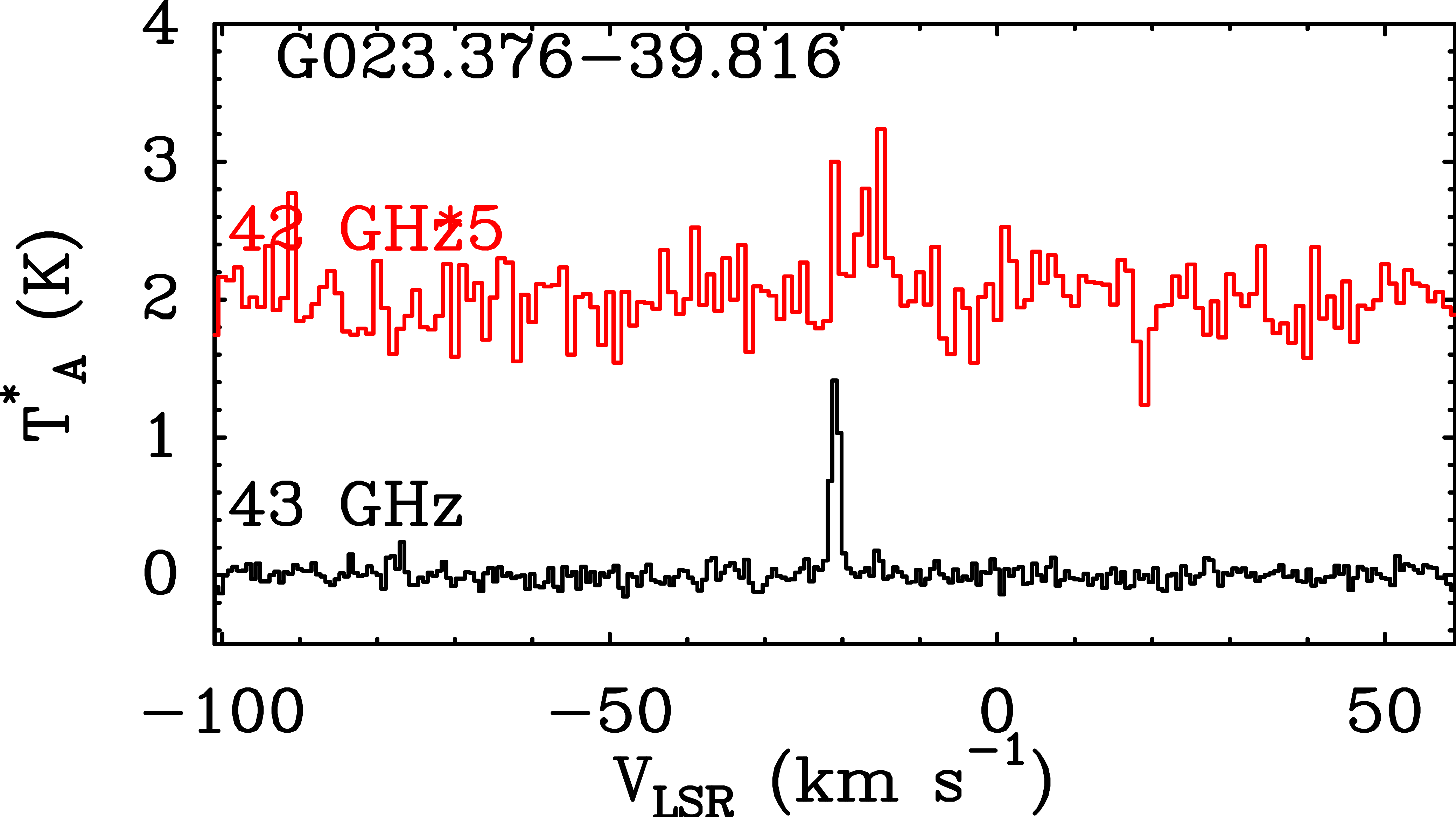}
\includegraphics[width=5.0cm]{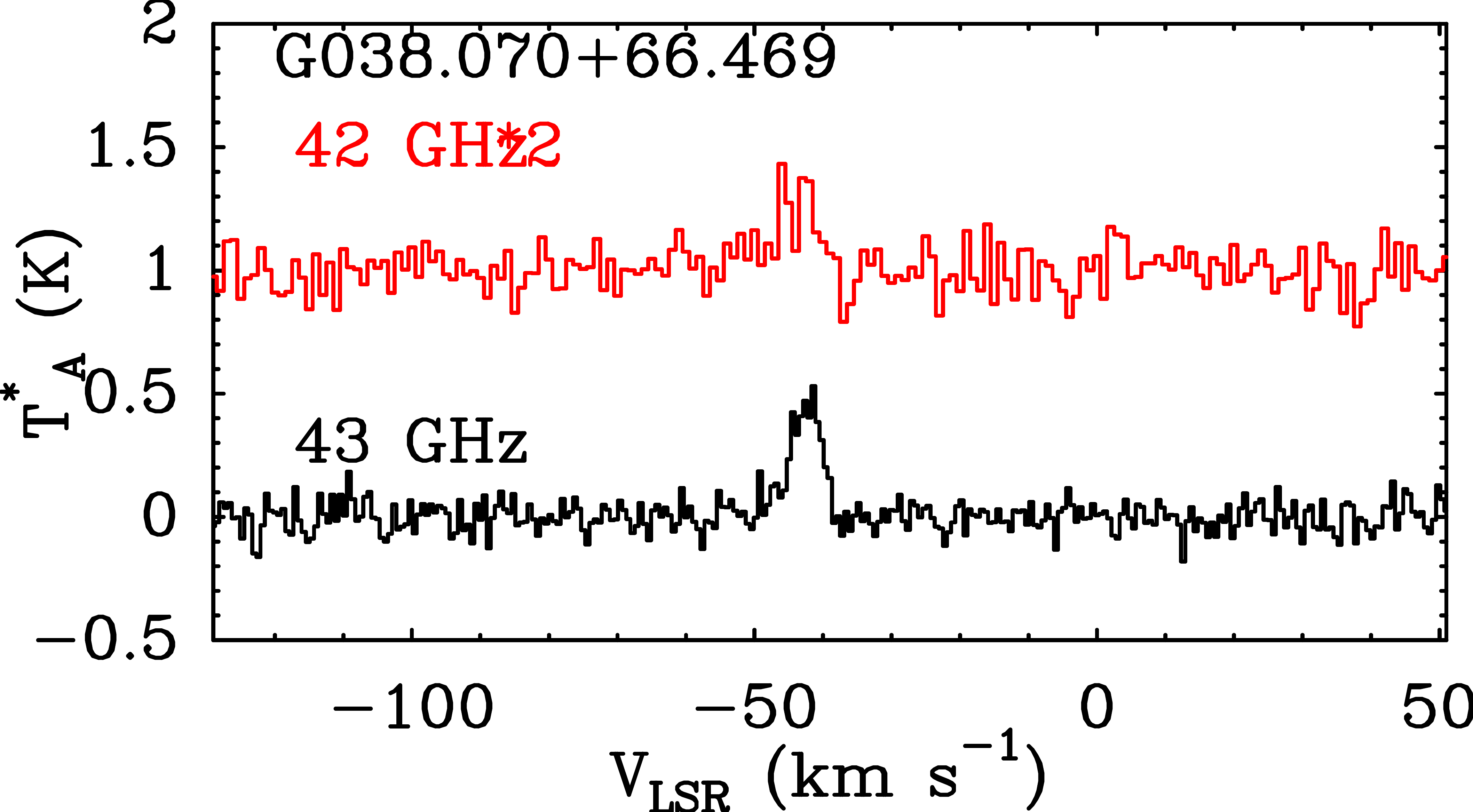} \\
\includegraphics[width=5.0cm]{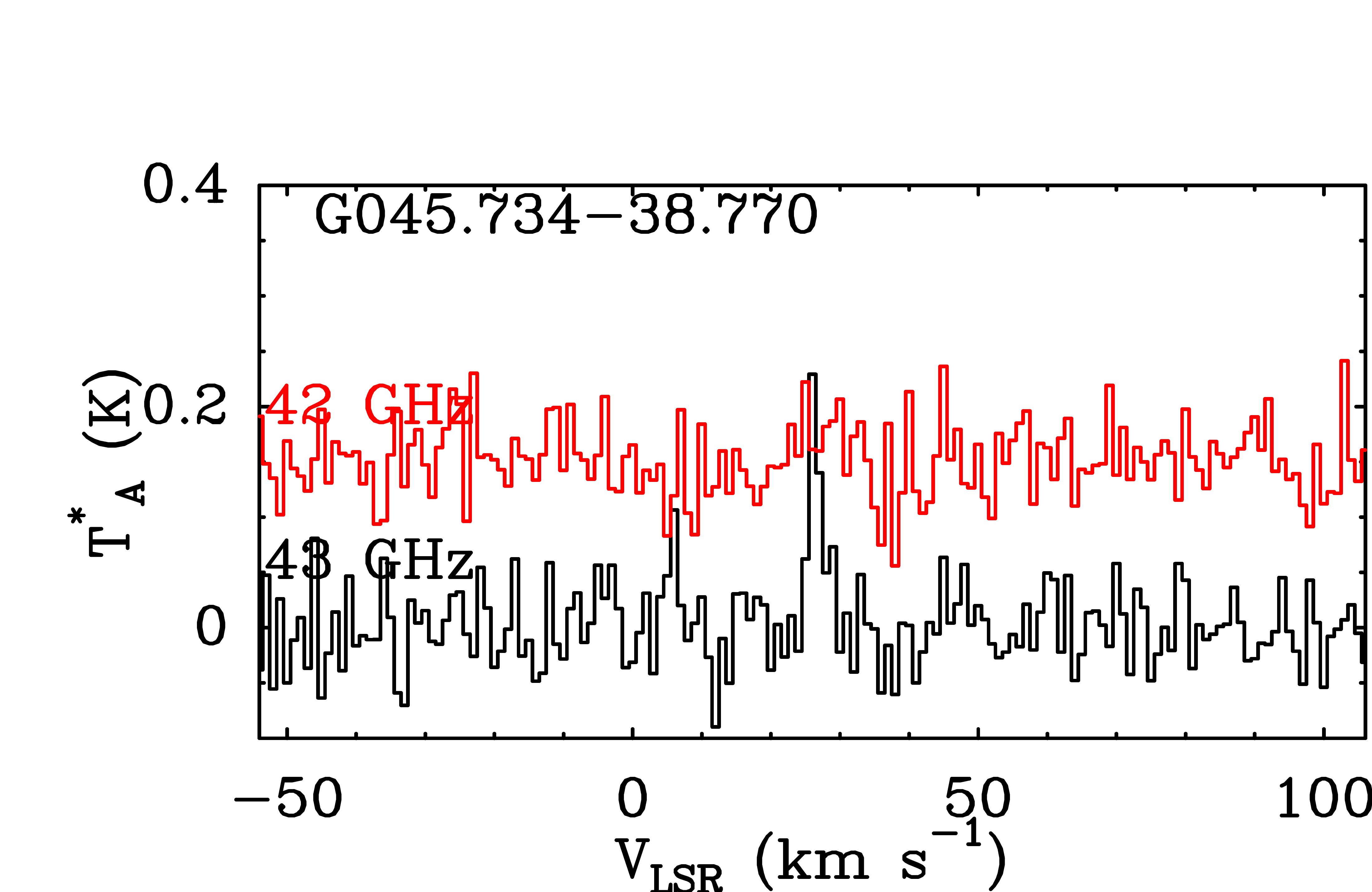}
\includegraphics[width=5.0cm]{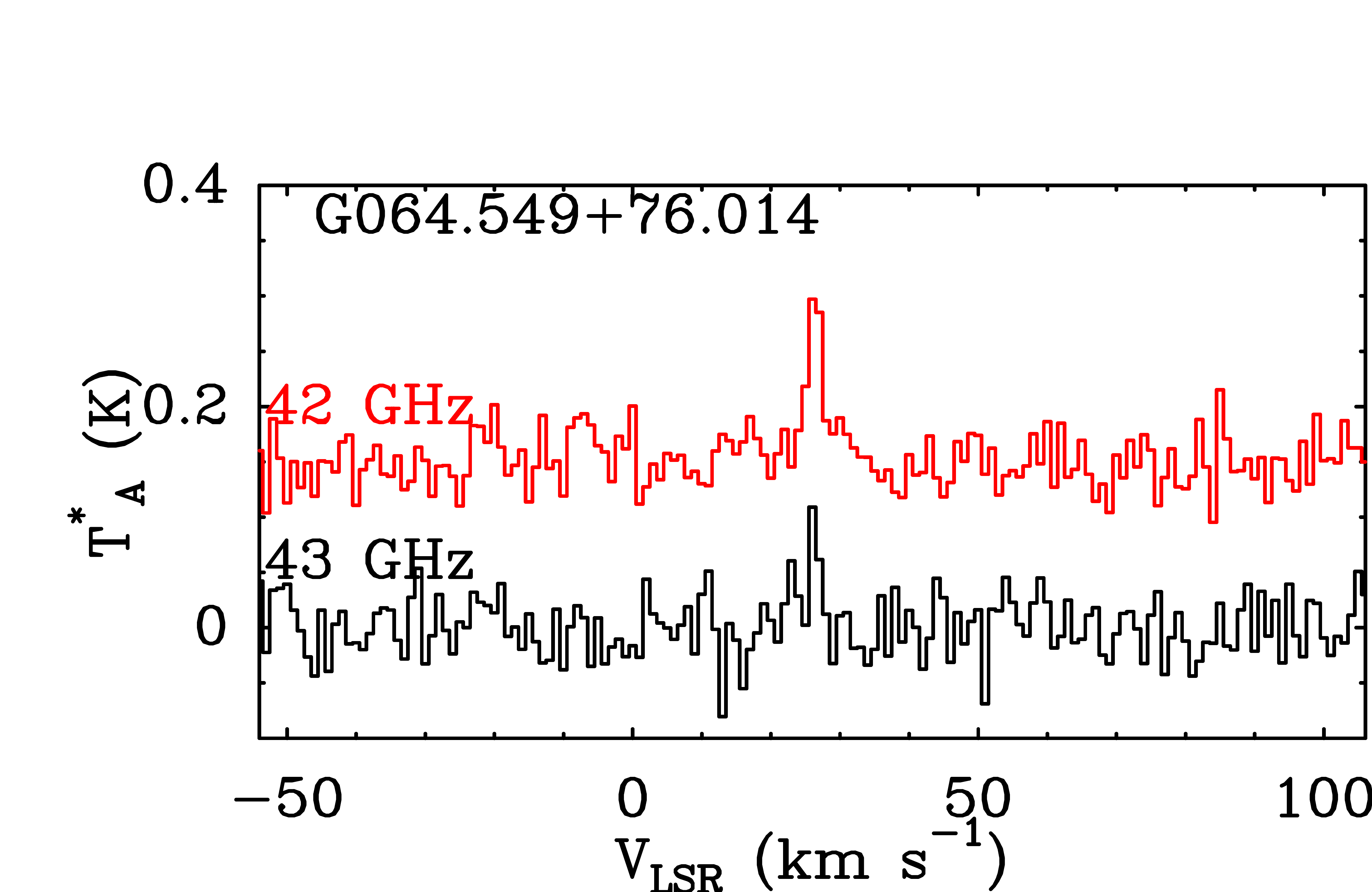}
\includegraphics[width=5.0cm]{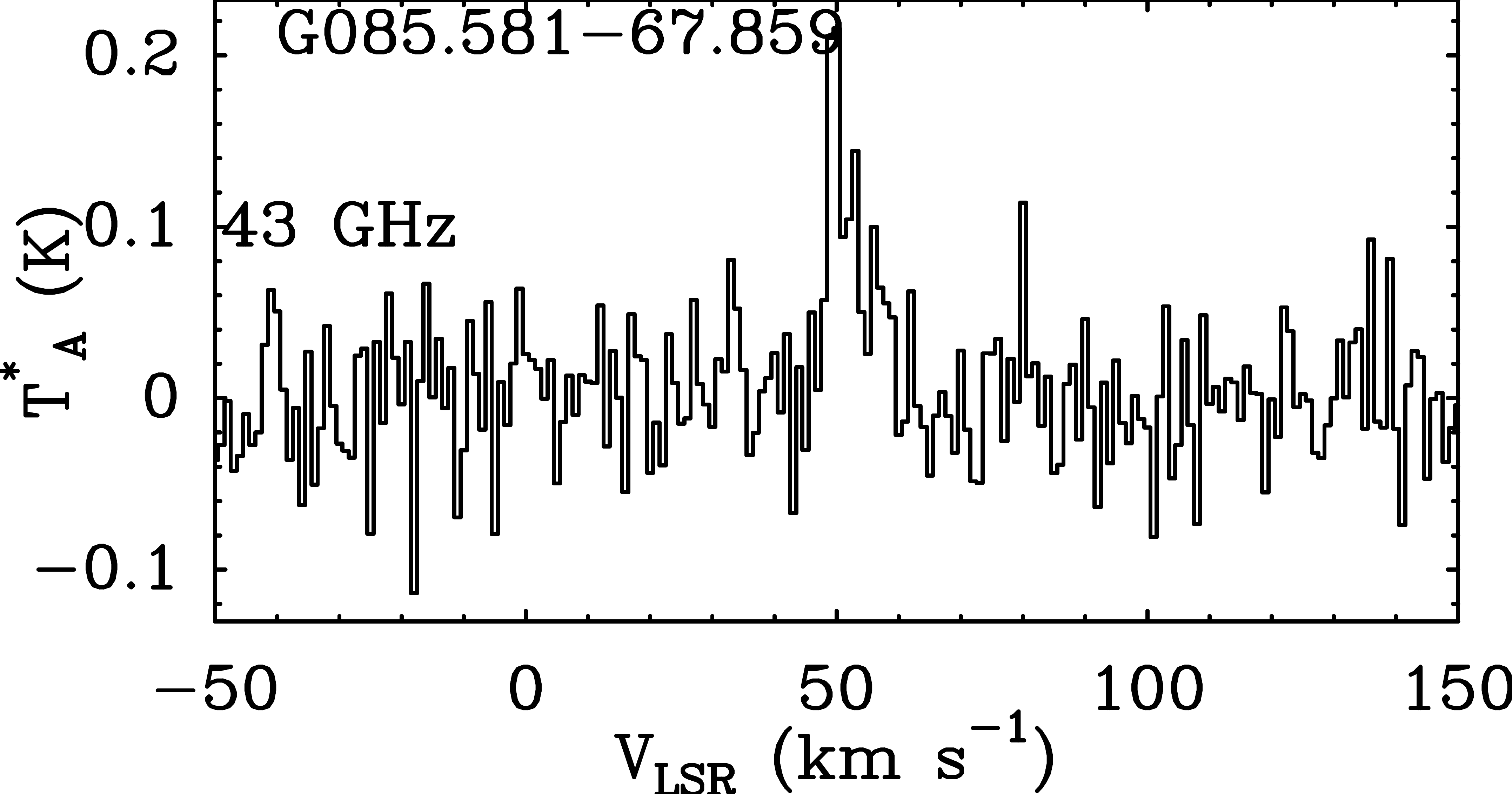} \\
\includegraphics[width=5.0cm]{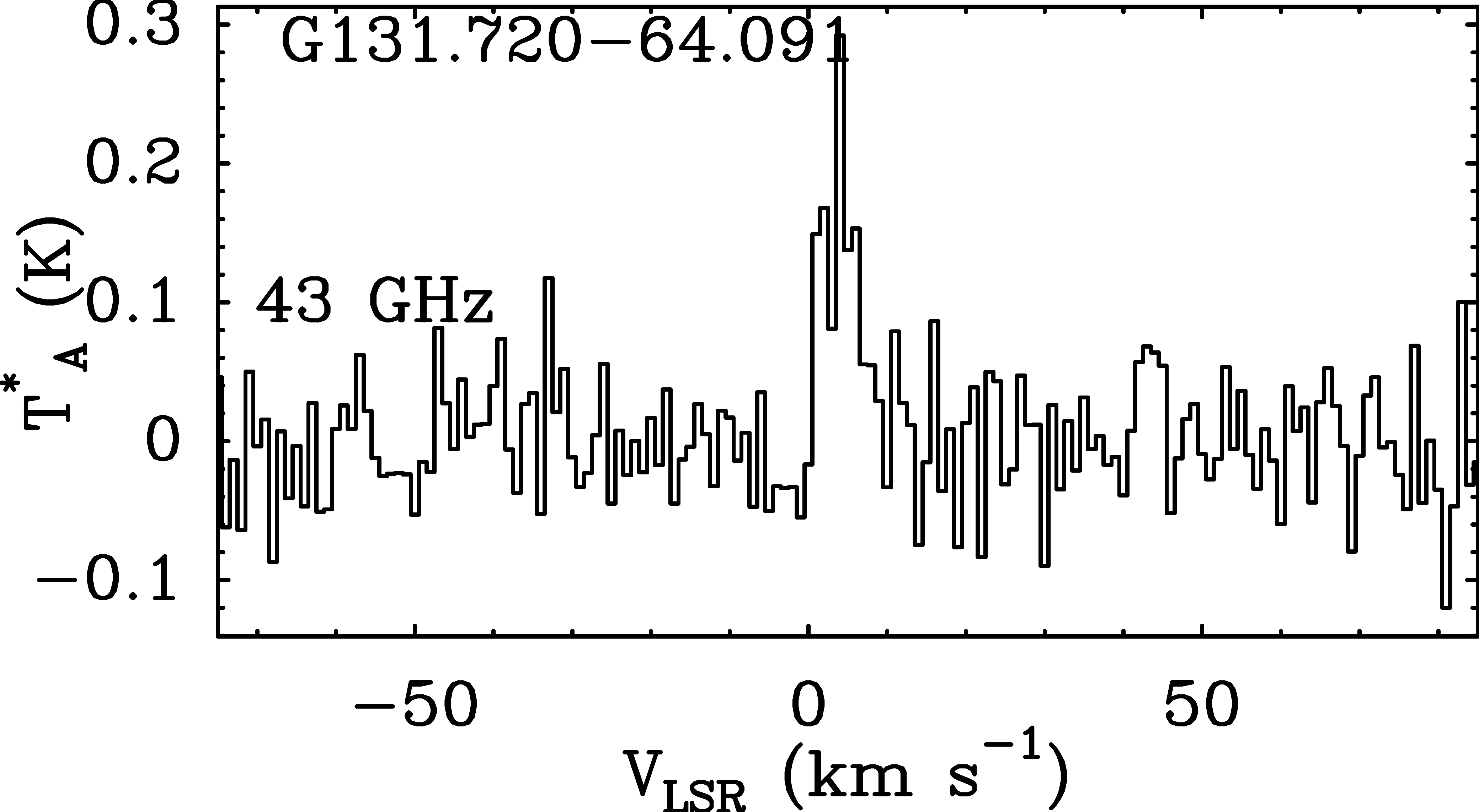}
\includegraphics[width=5.0cm]{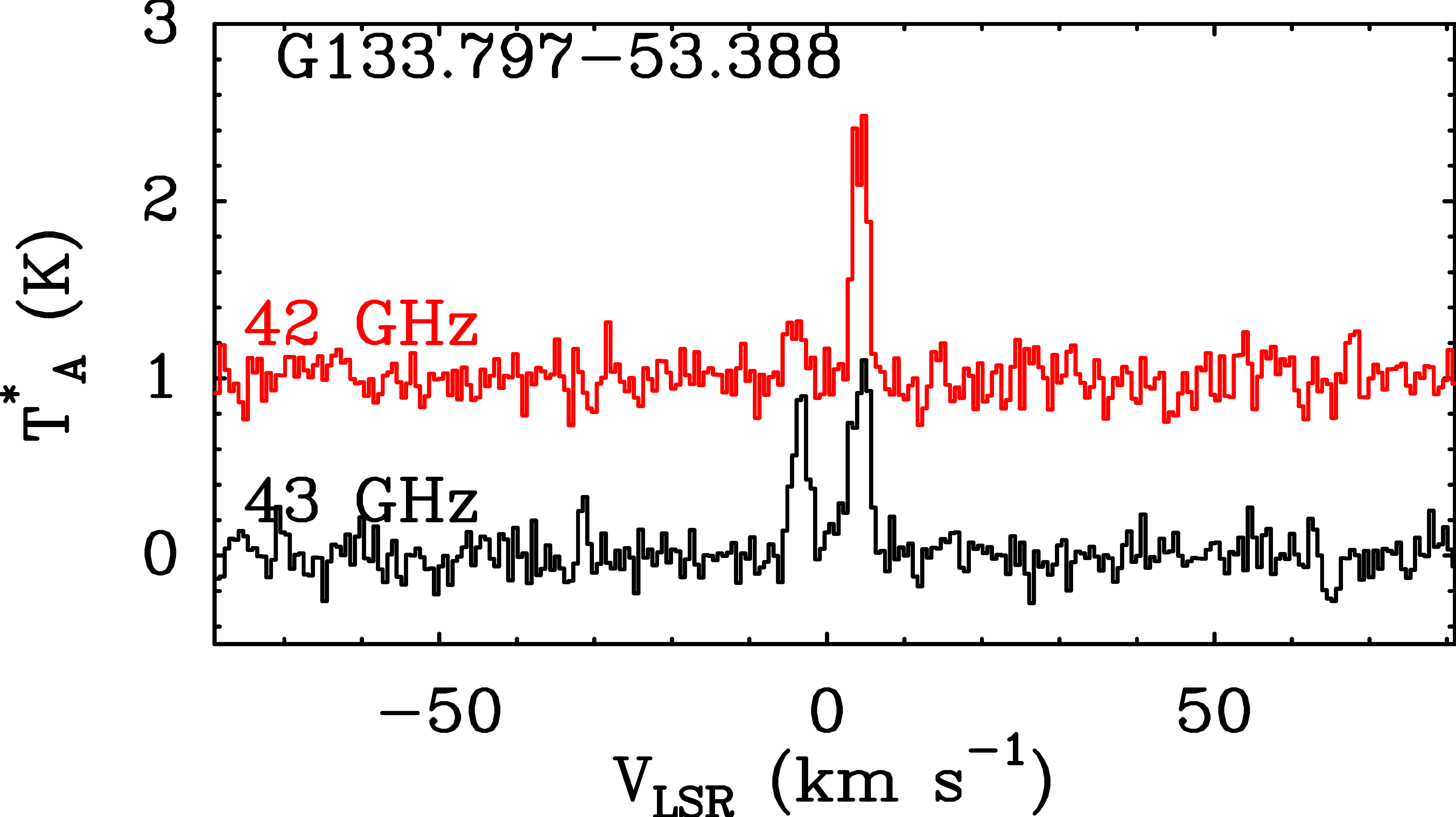} 
\includegraphics[width=5.0cm]{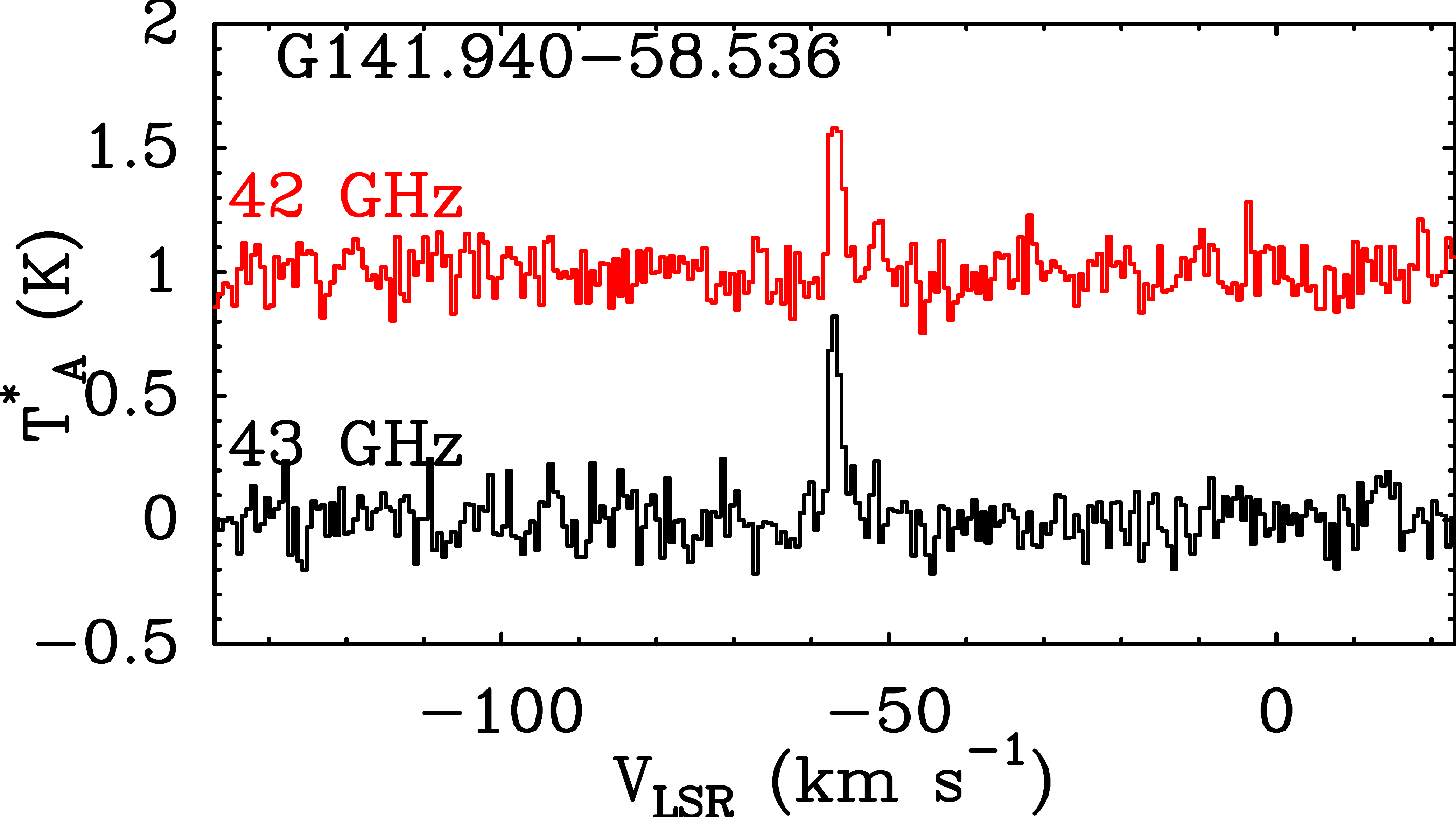} \\
\includegraphics[width=5.0cm]{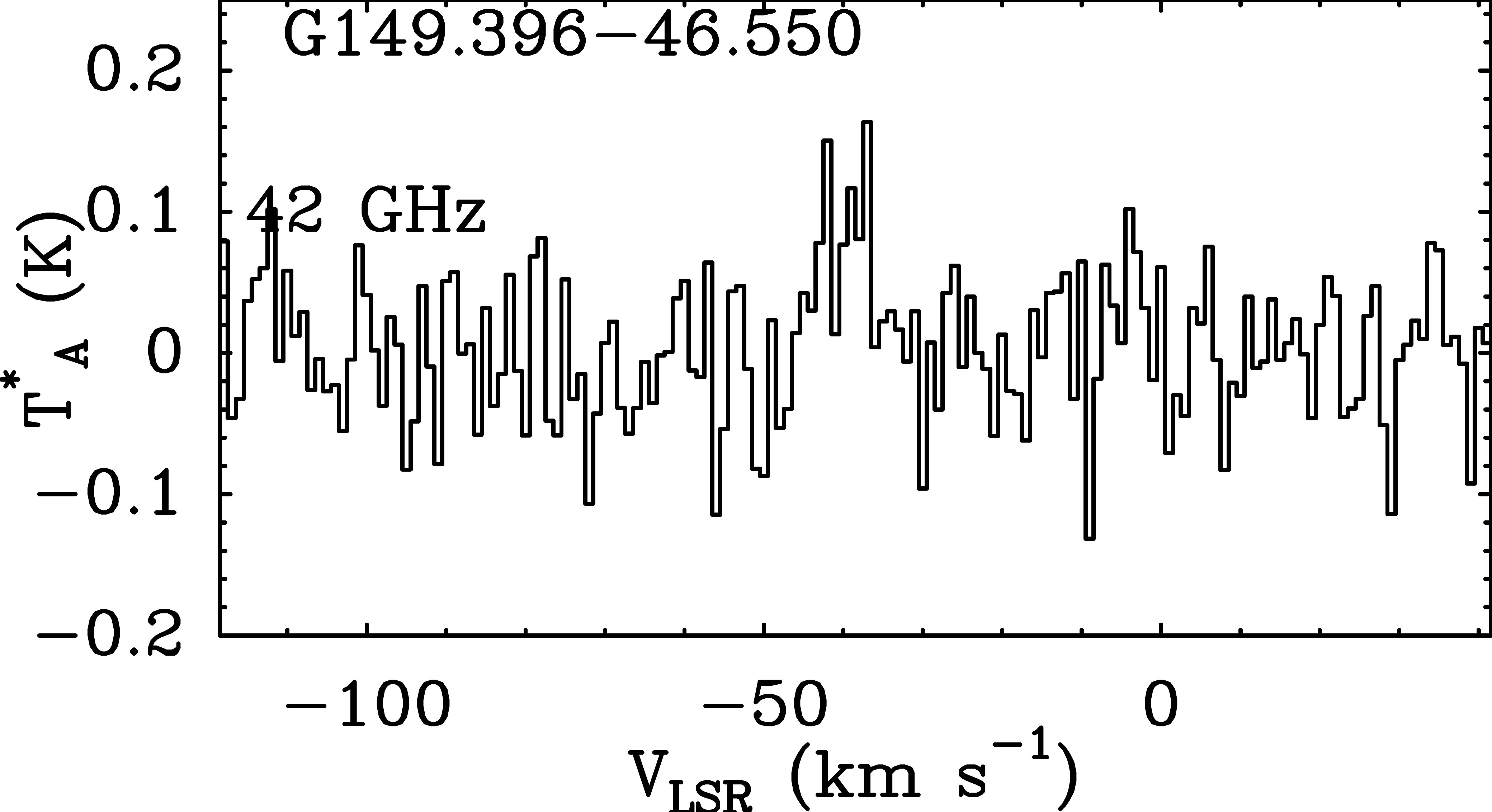}
\includegraphics[width=5.0cm]{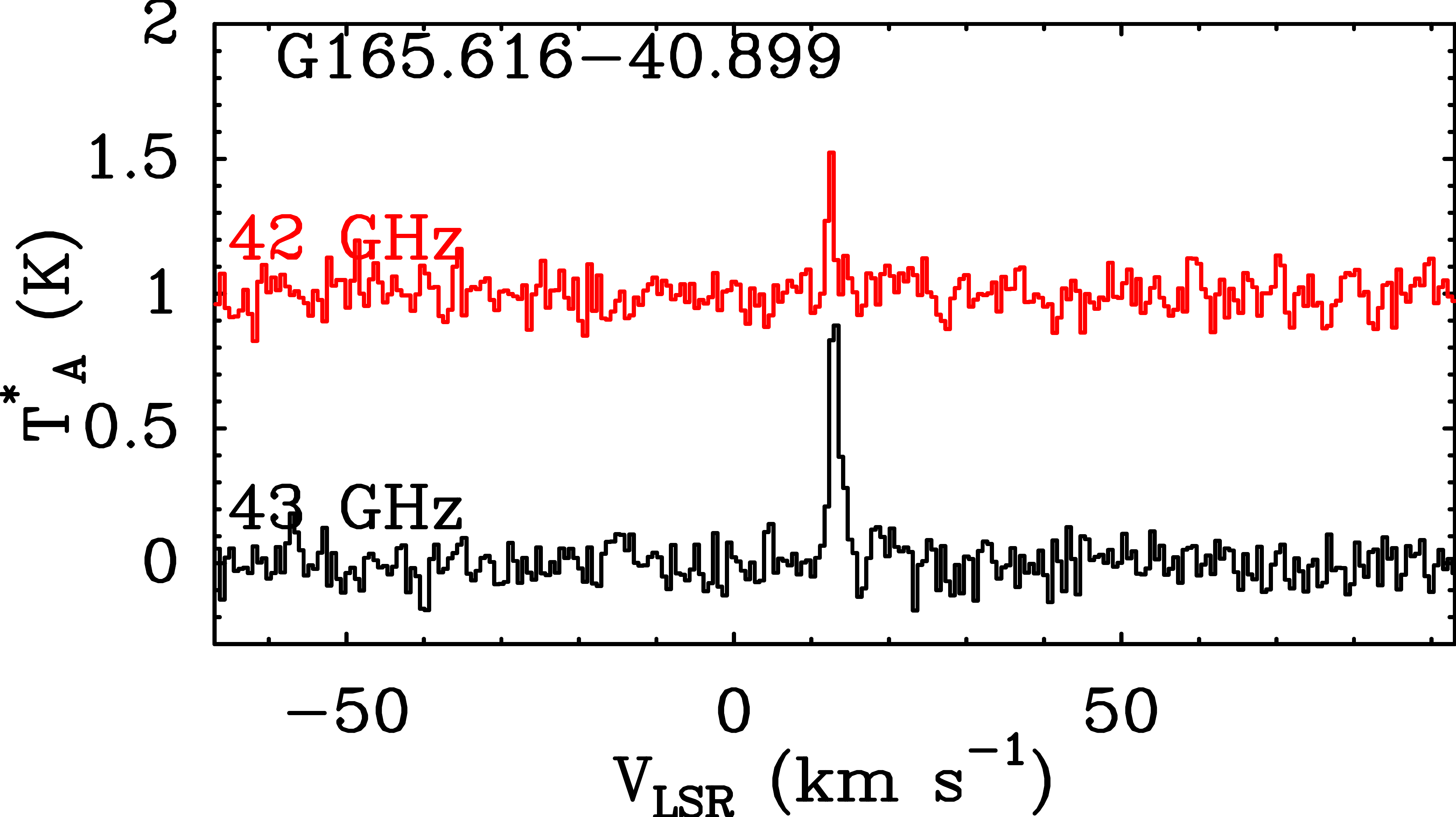}
\includegraphics[width=5.0cm]{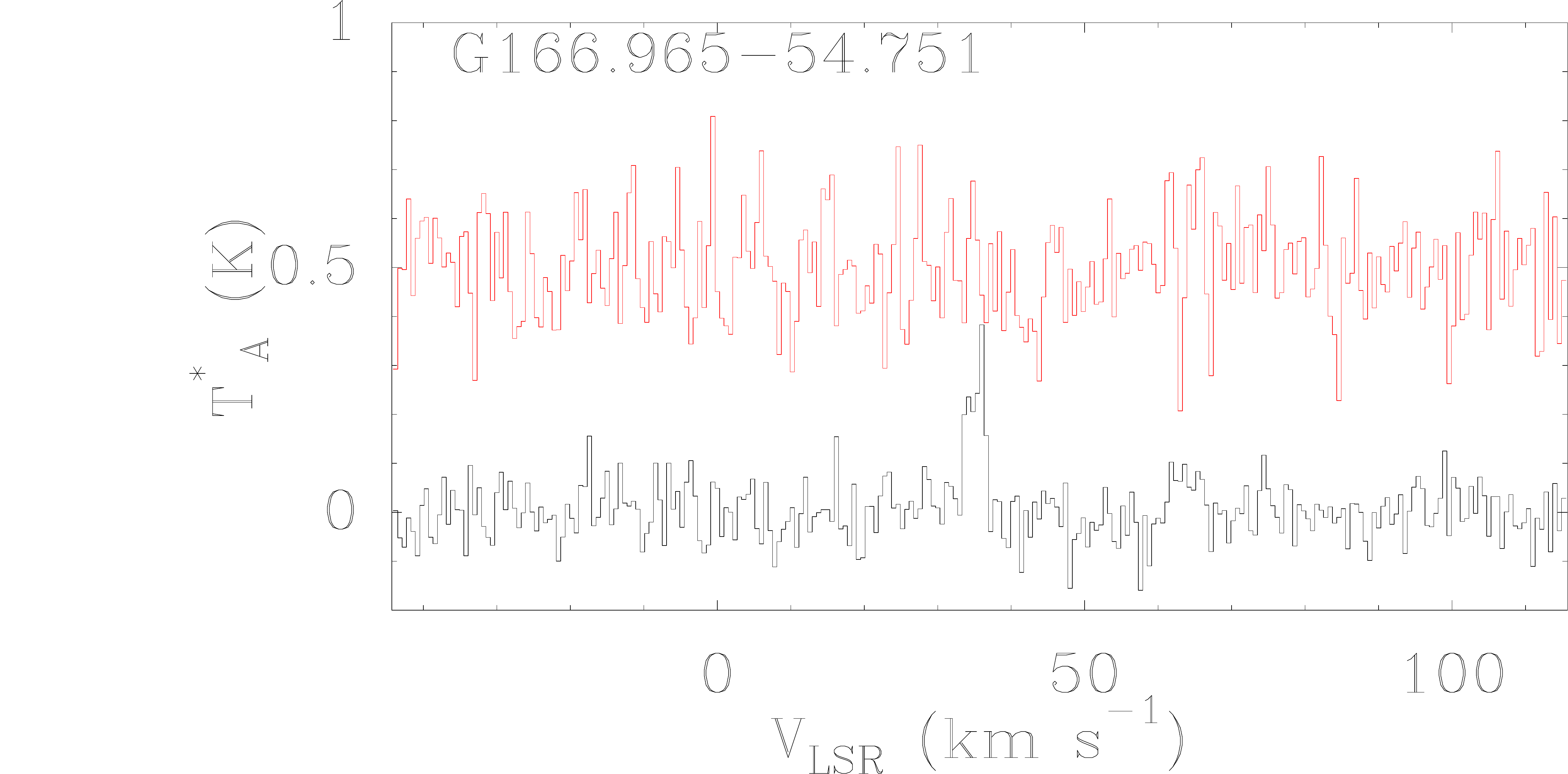}
\caption{SiO maser spectra. In case both v=1 and v=2 masers are detected, the v=2 (42 GHz)
maser lines are denoted by red histograms. Some low flux density 42 GHz maser lines are scaled up,
with the scale factors labeled in the plot.  \label{fig-A3}}
\end{figure*}   
               
\begin{figure*}
\includegraphics[width=5.0cm]{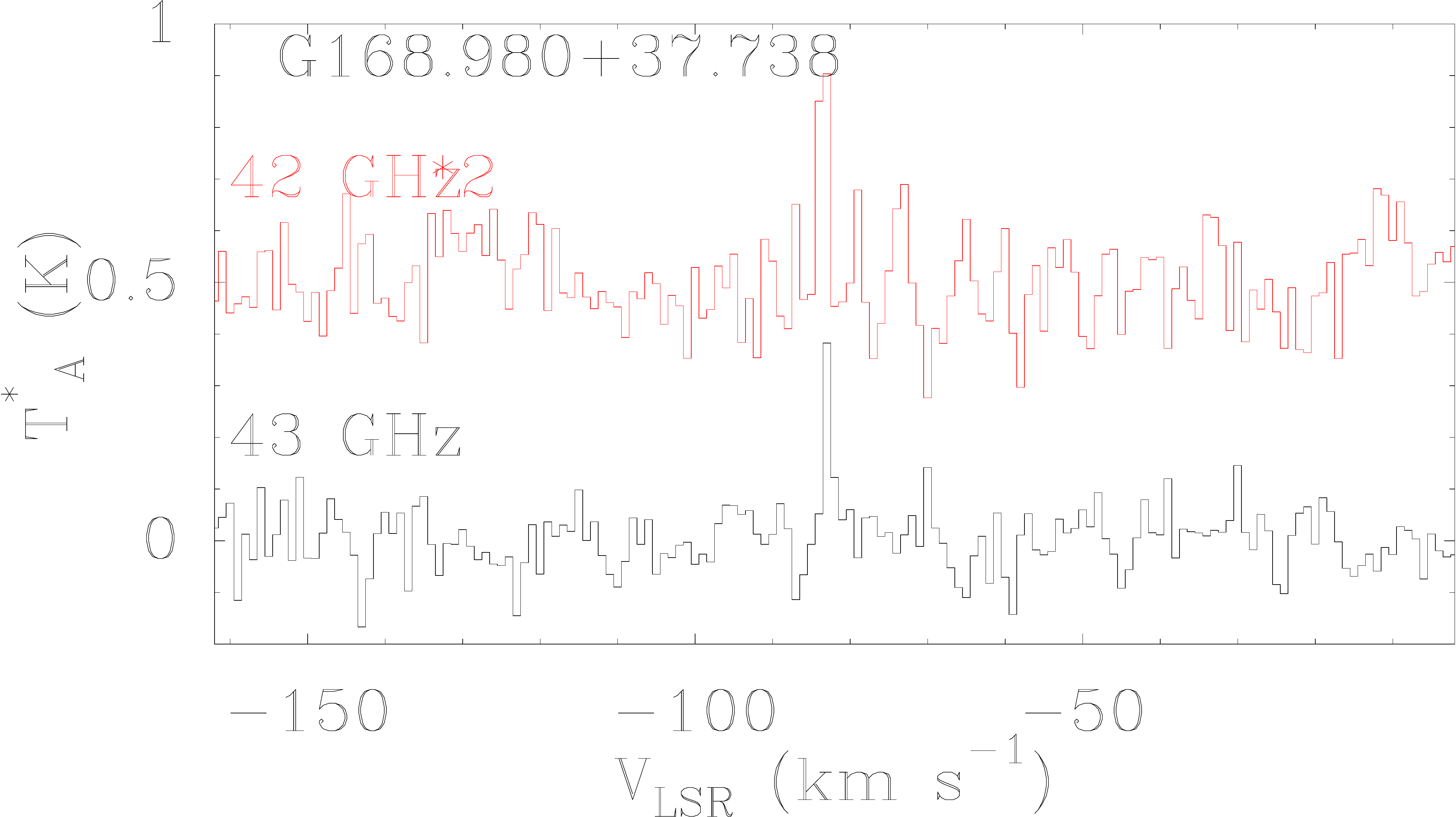}
\includegraphics[width=5.0cm]{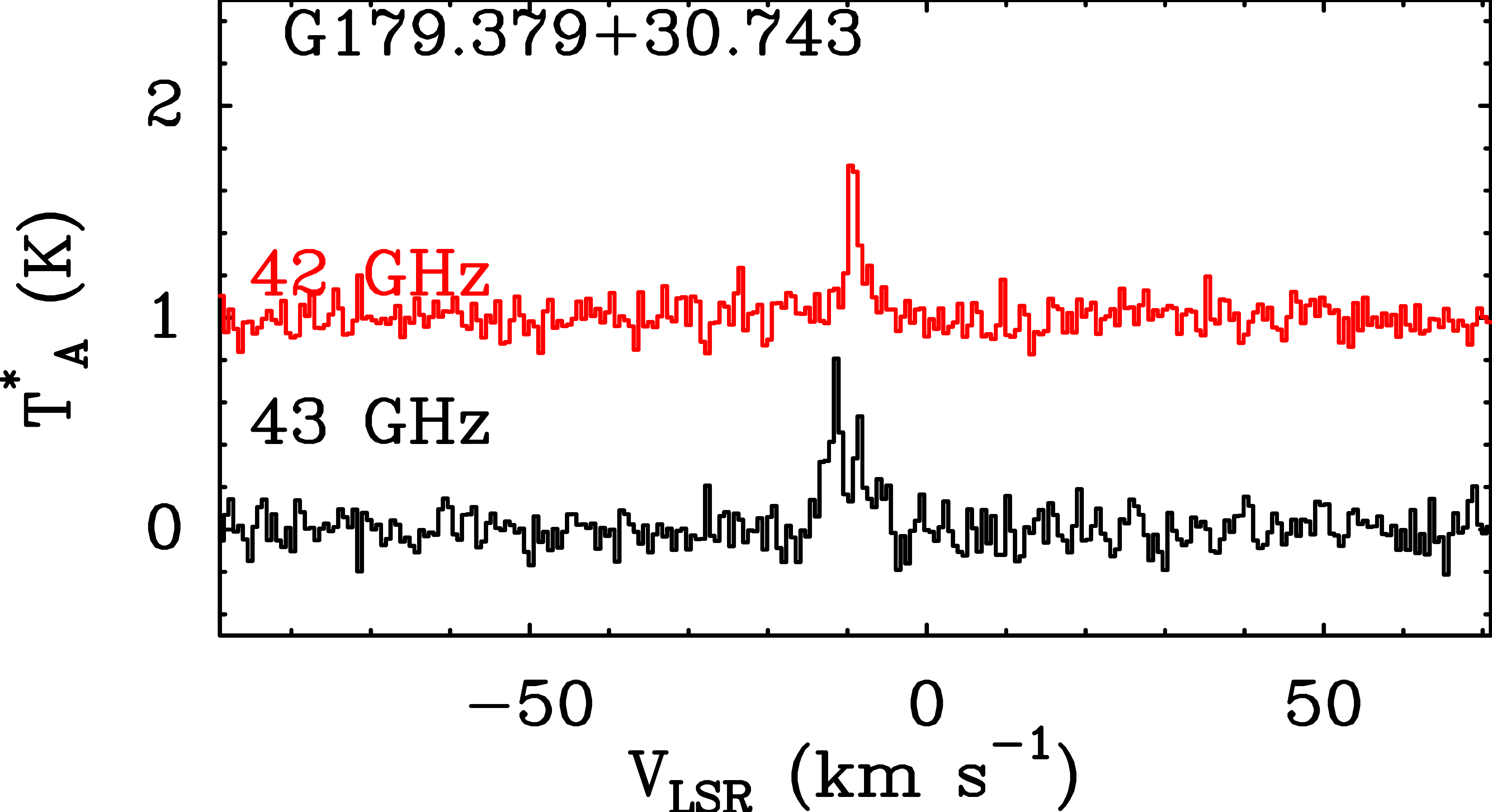}
\includegraphics[width=5.0cm]{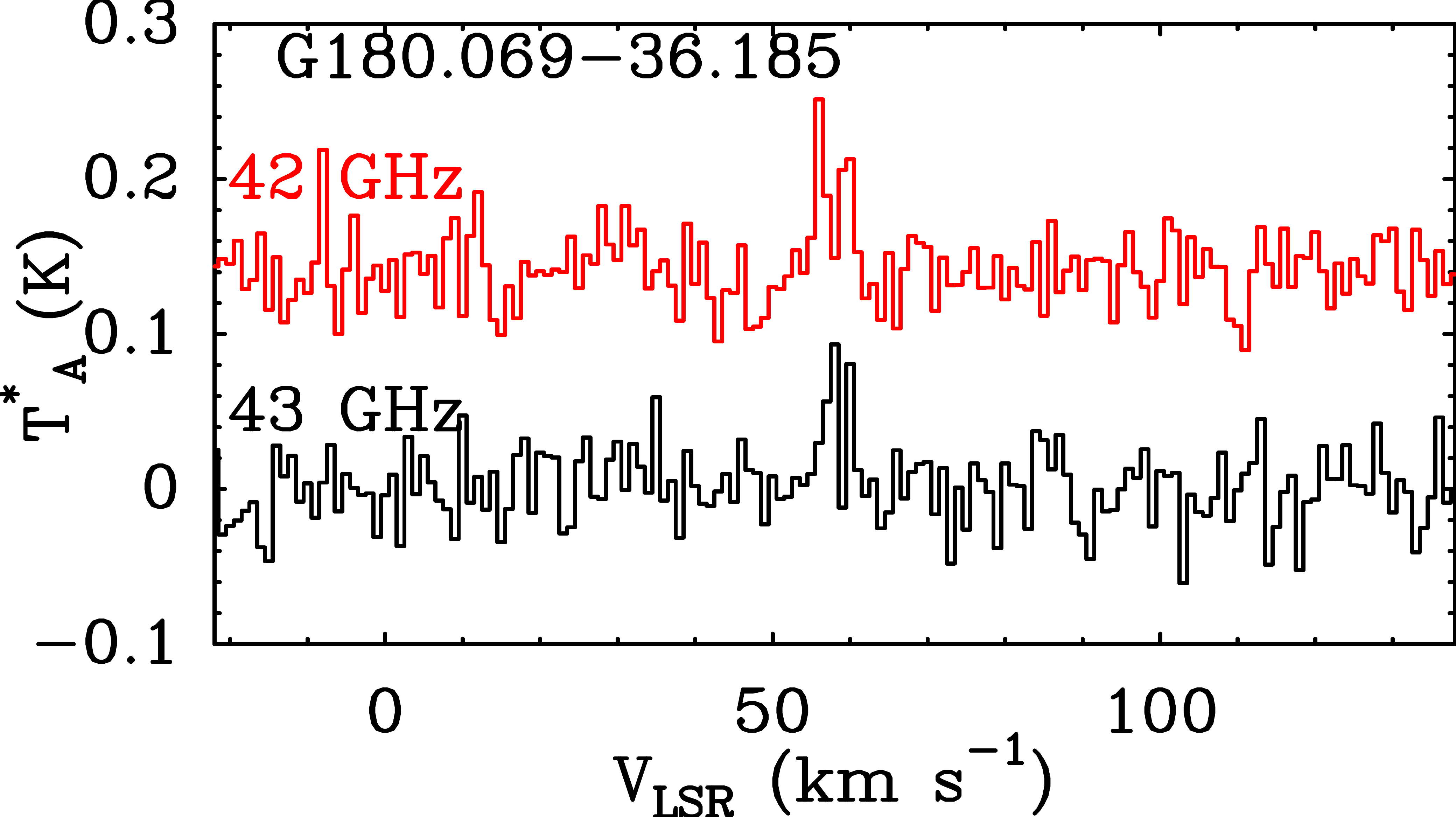}\\
\includegraphics[width=5.0cm]{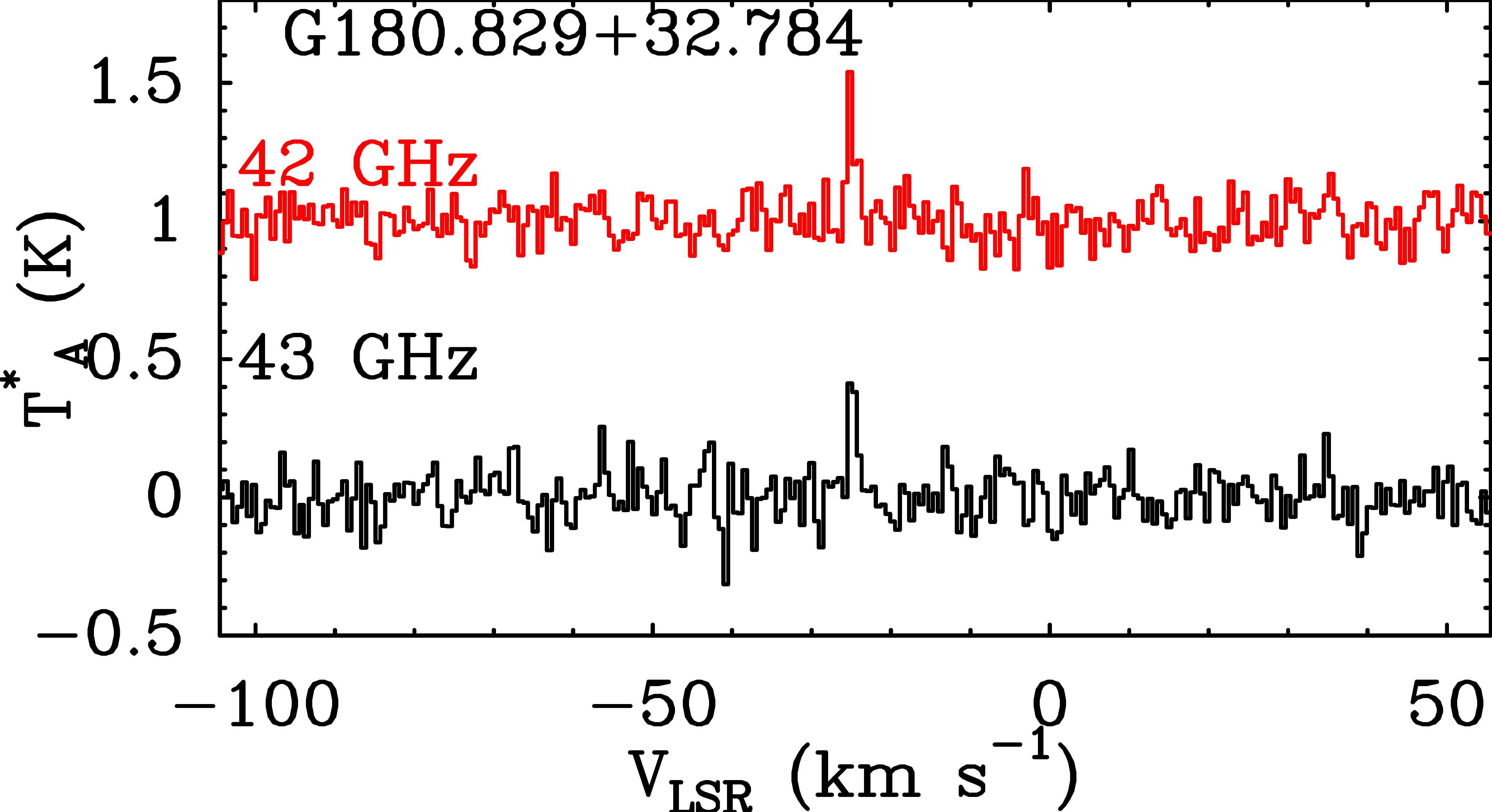}
\includegraphics[width=5.0cm]{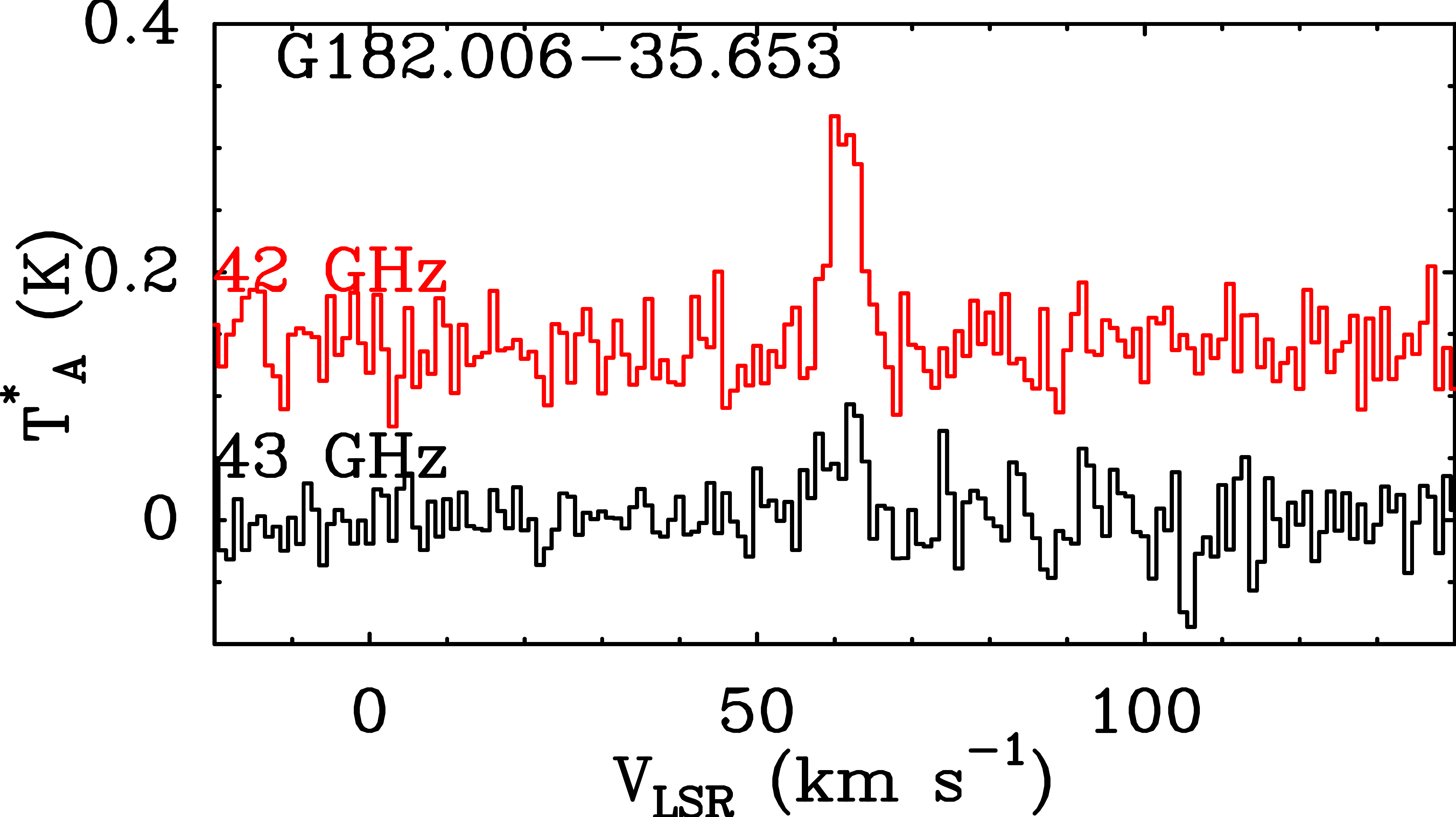}
\includegraphics[width=5.0cm]{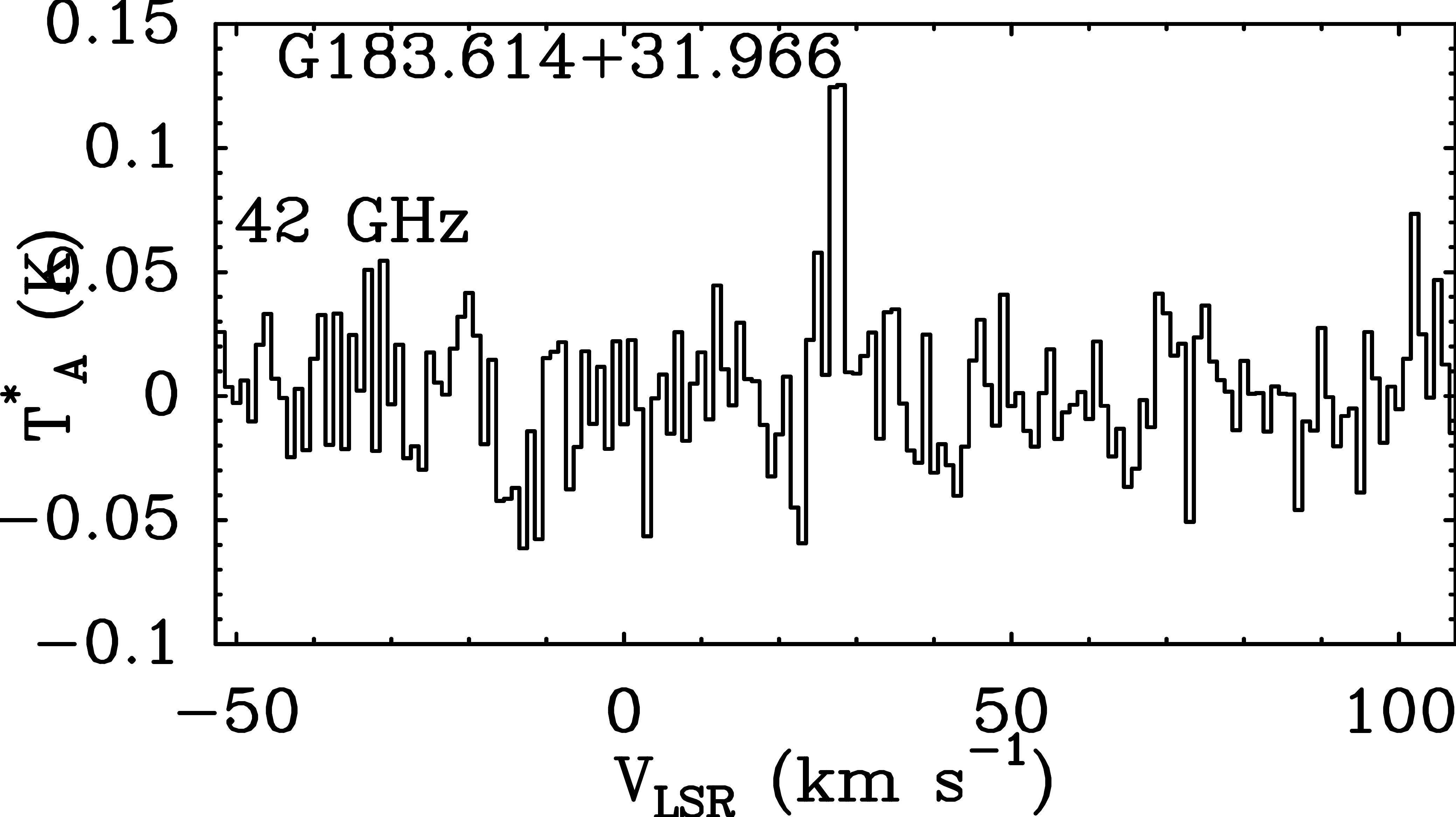}\\
\includegraphics[width=5.0cm]{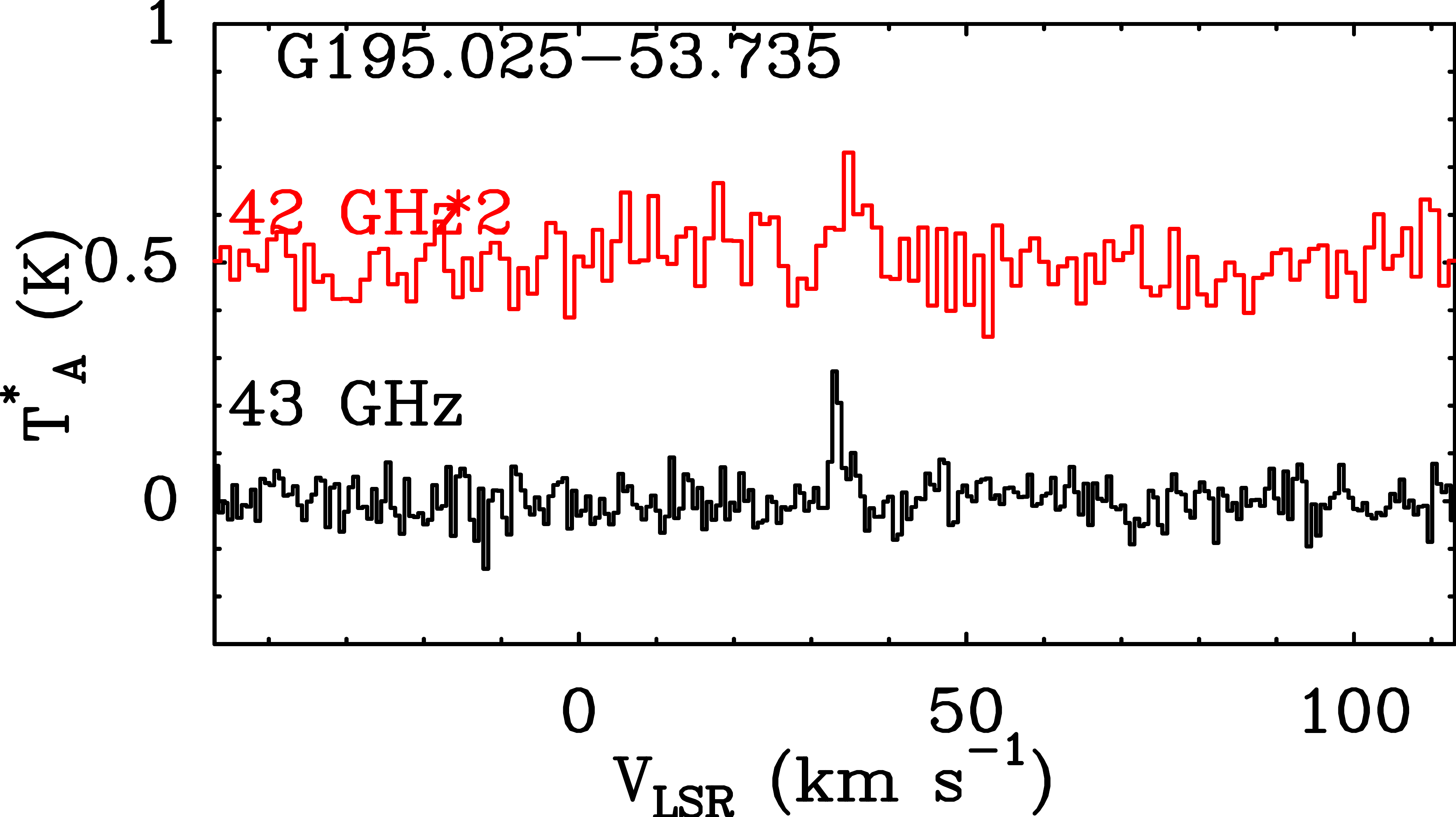}
\includegraphics[width=5.0cm]{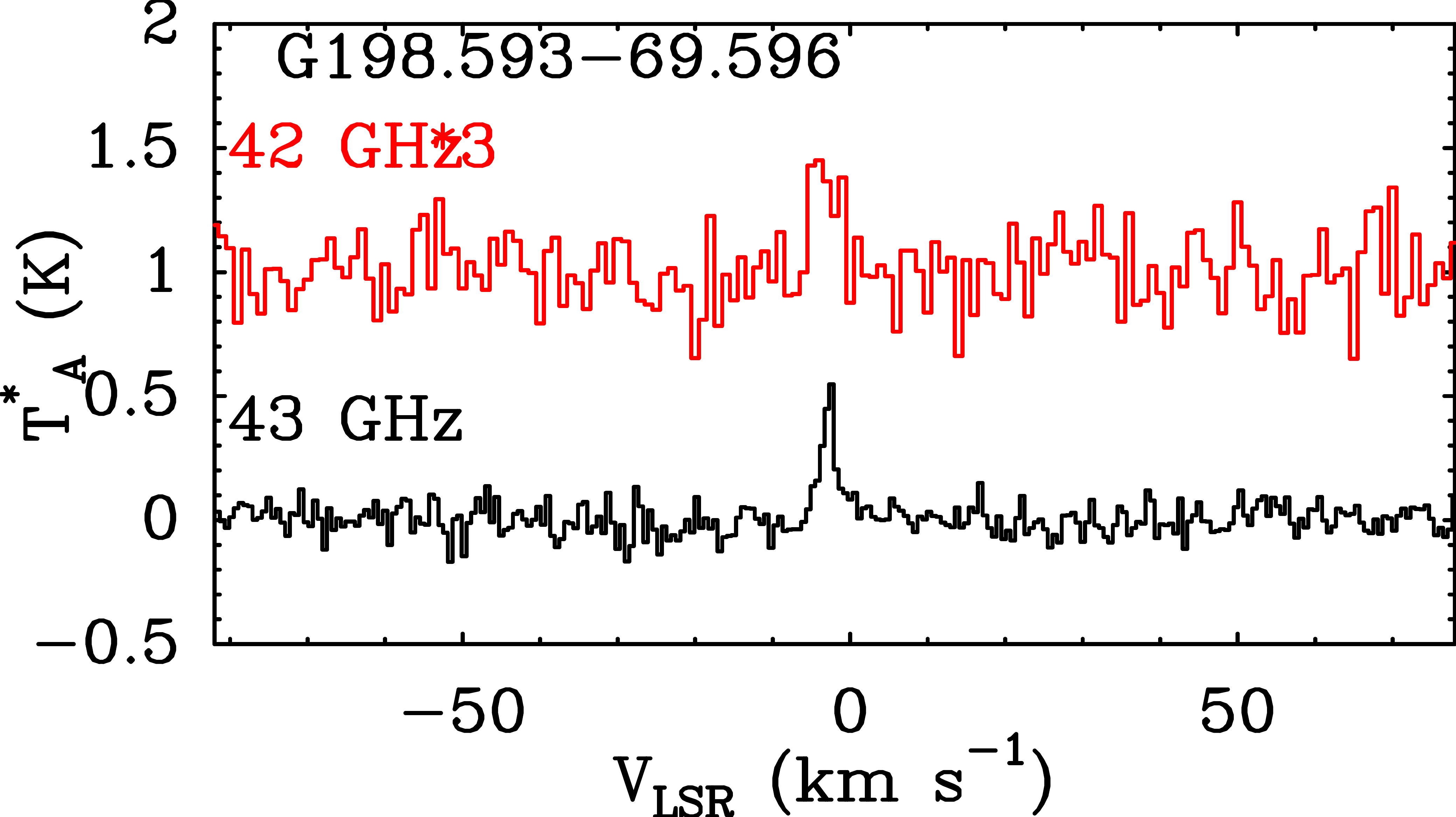}
\includegraphics[width=5.0cm]{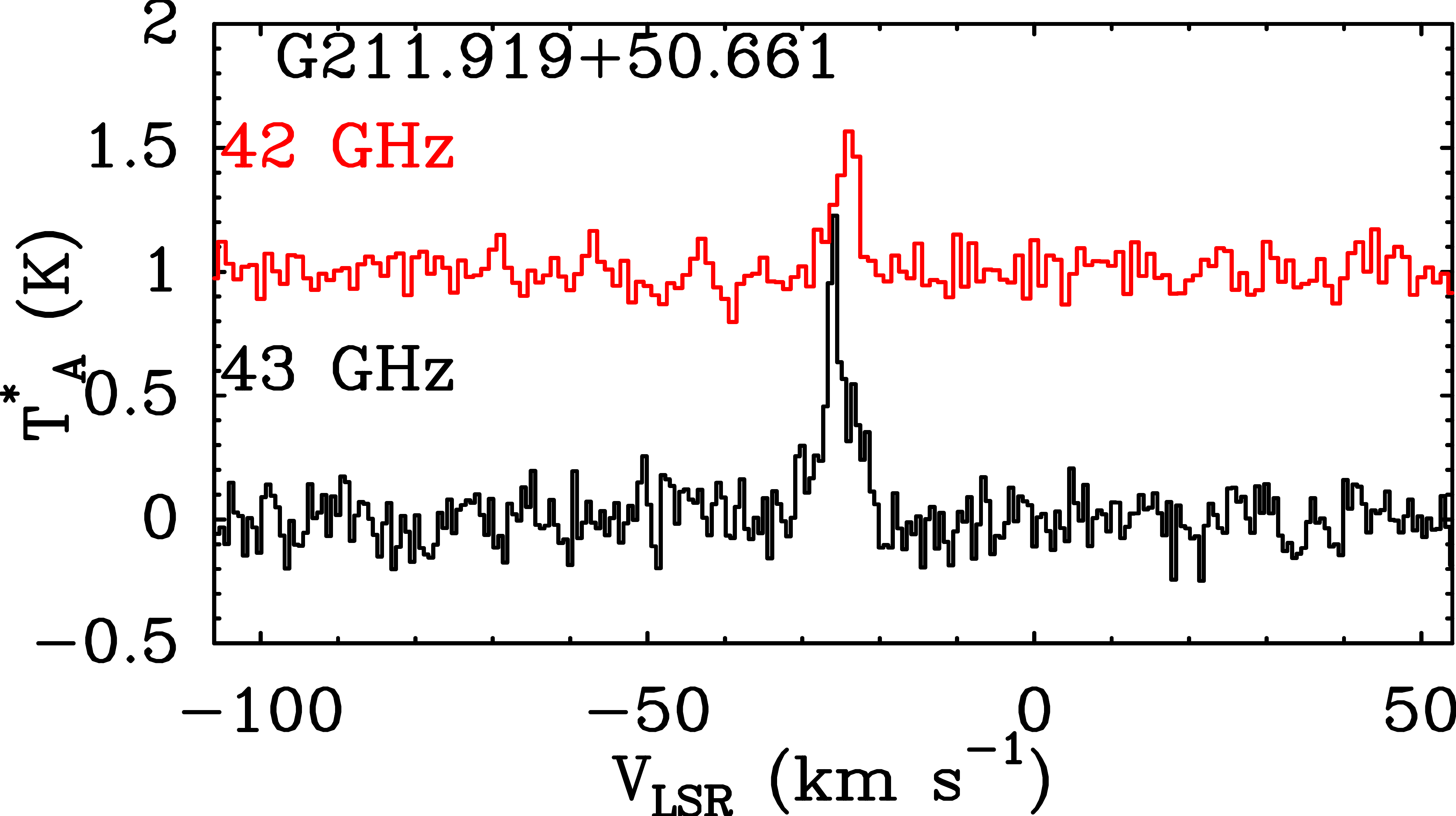}\\
\includegraphics[width=5.0cm]{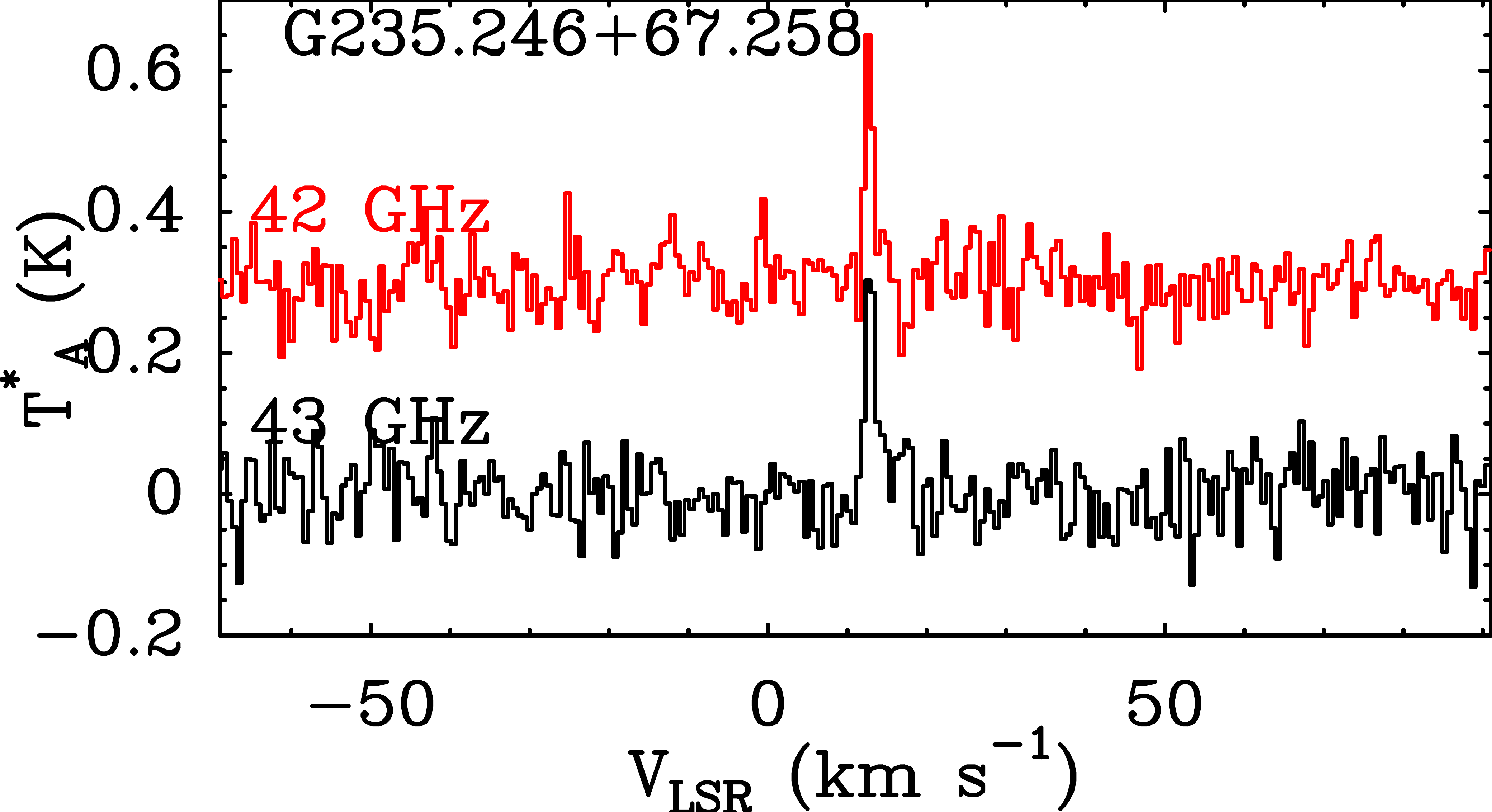}
\includegraphics[width=5.0cm]{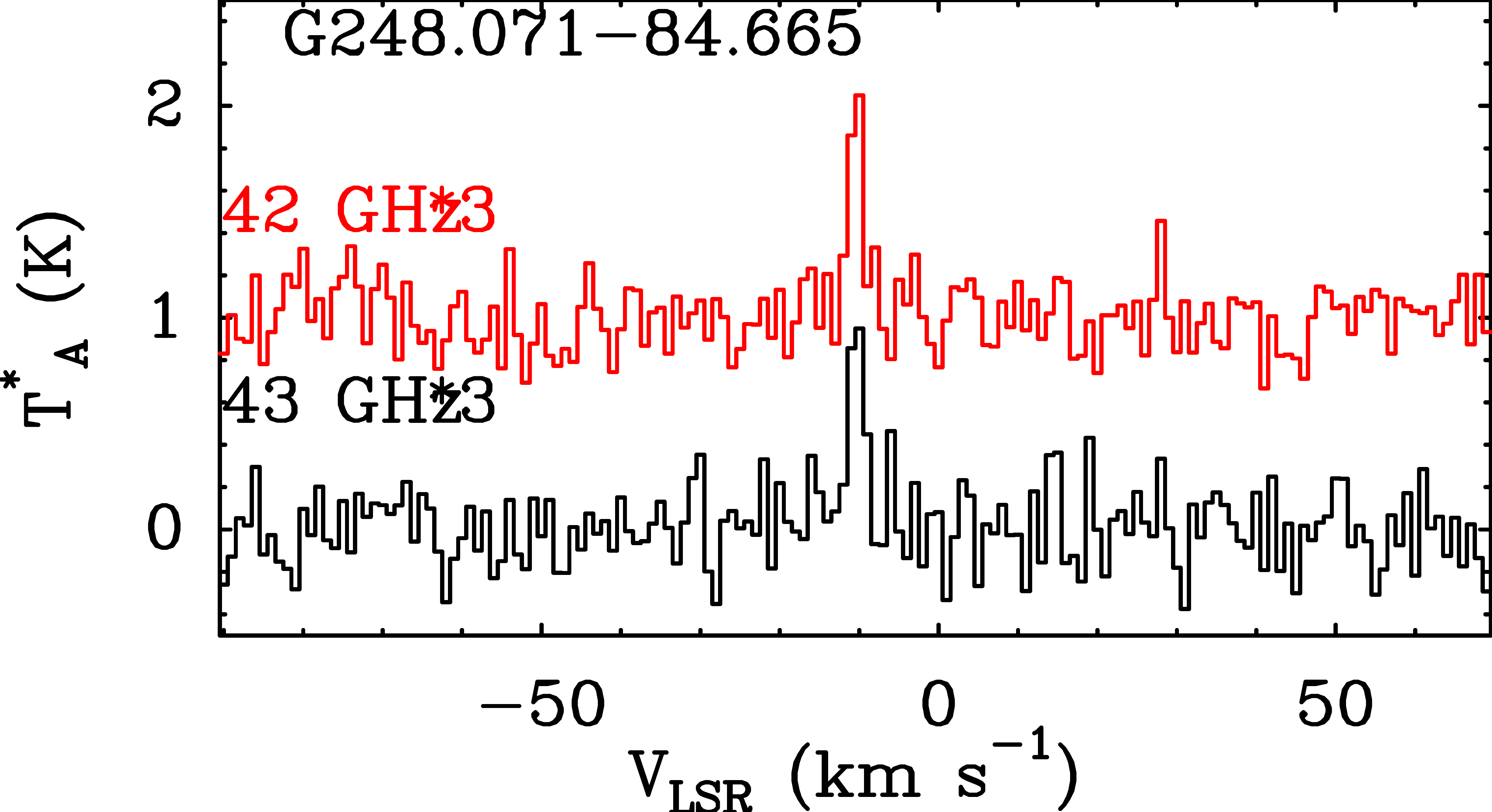}
\includegraphics[width=5.0cm]{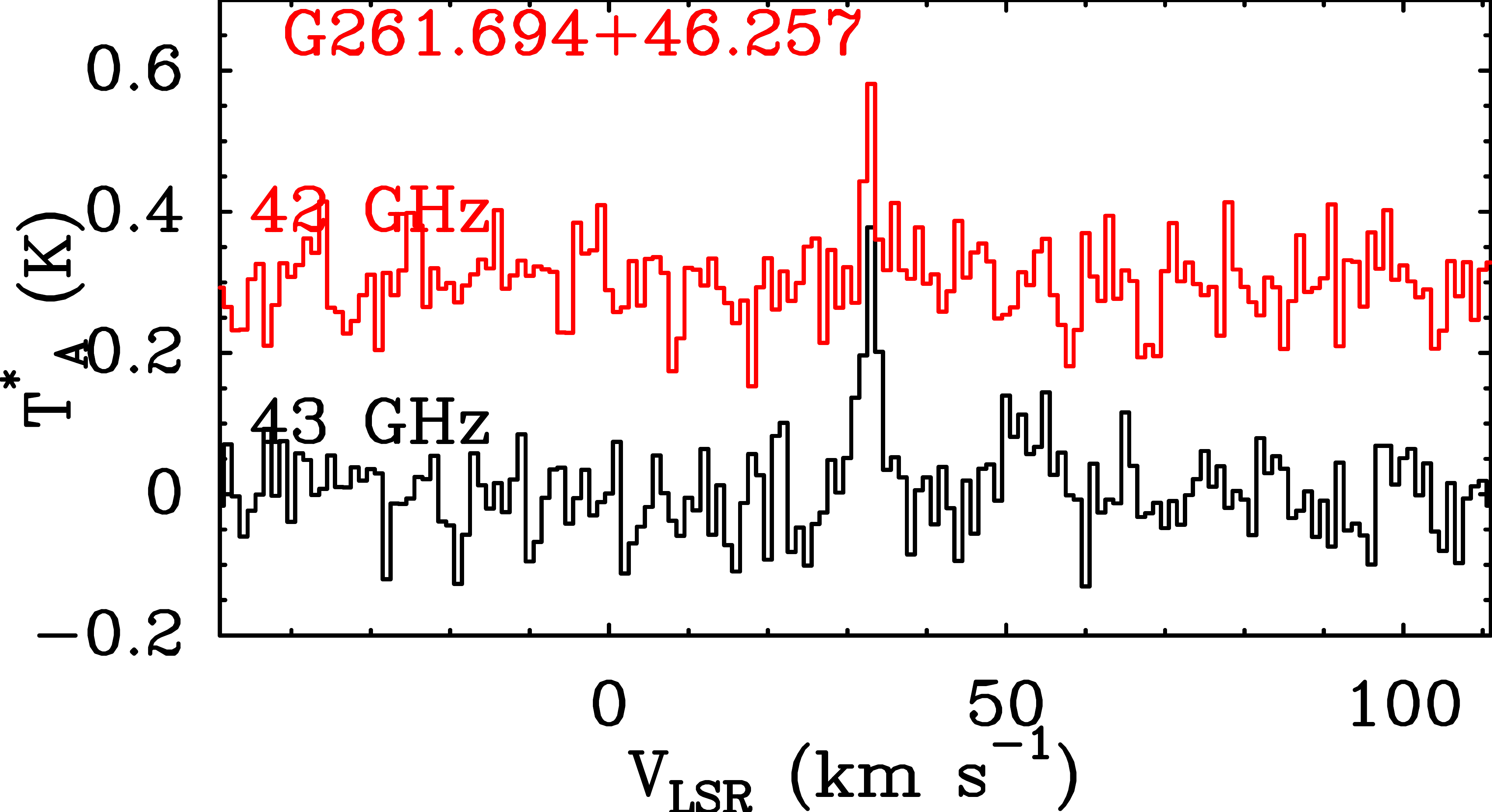}\\
\includegraphics[width=5.0cm]{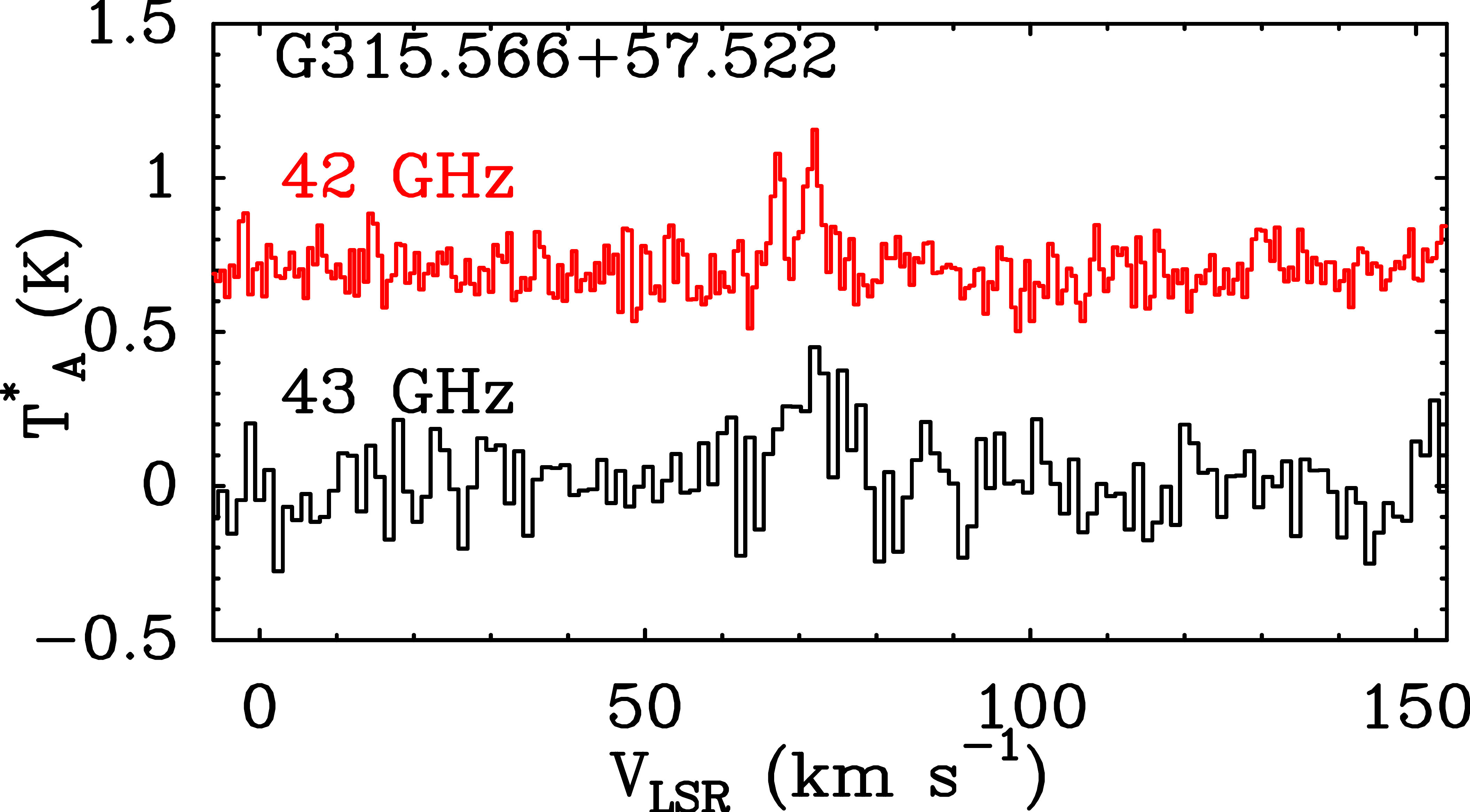}
\includegraphics[width=5.0cm]{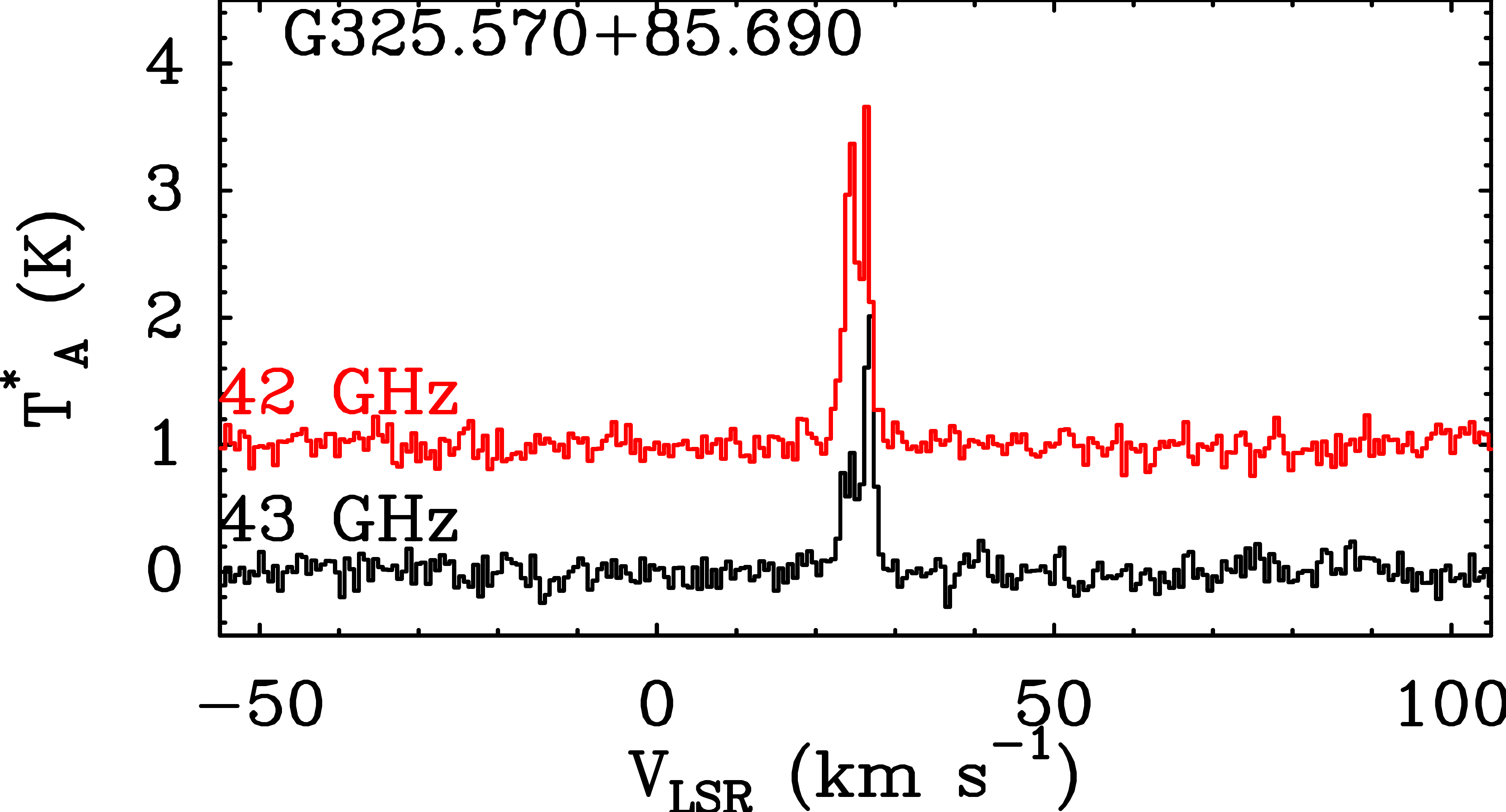}
\includegraphics[width=5.0cm]{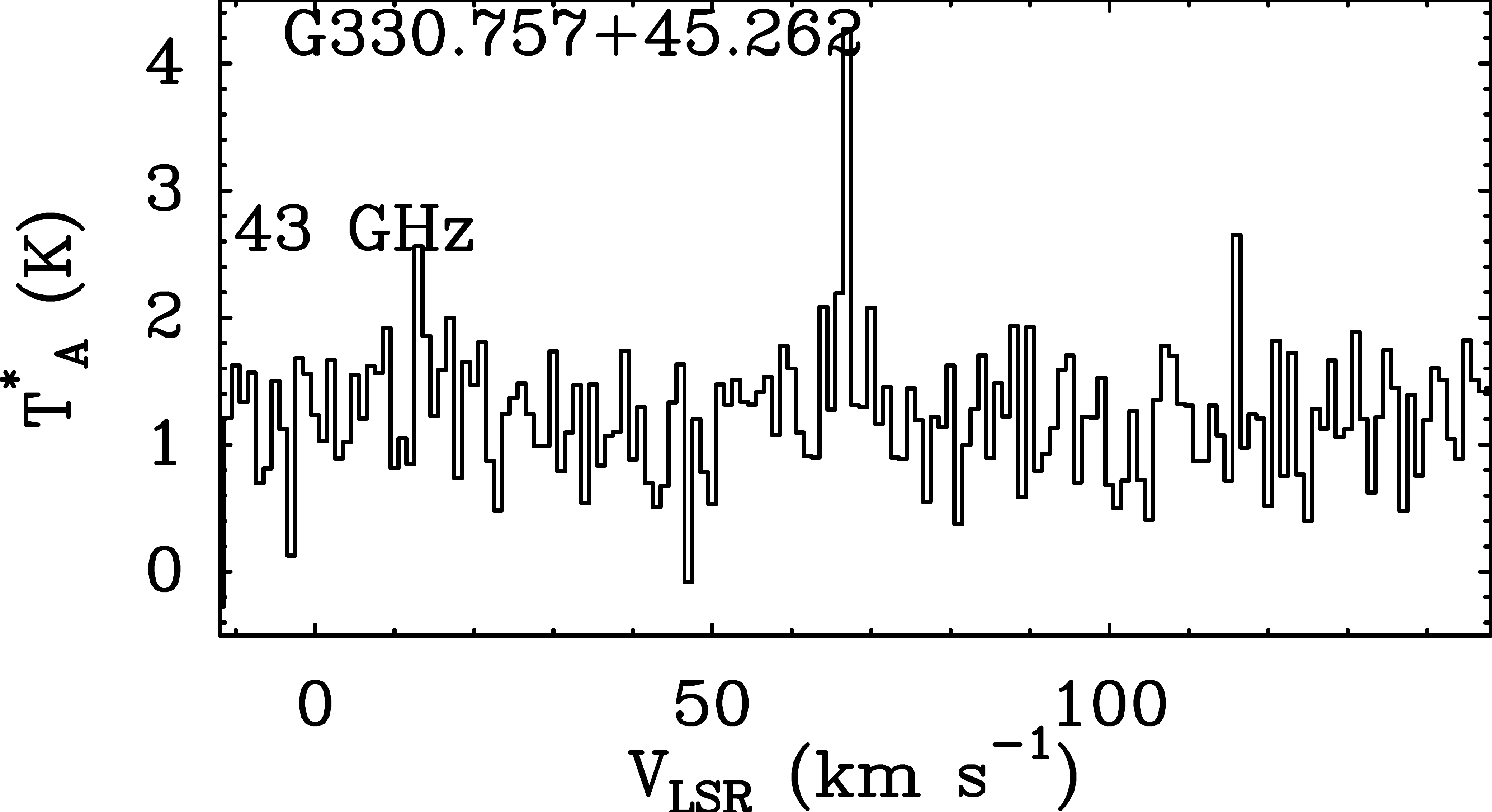}\\
\includegraphics[width=5.0cm]{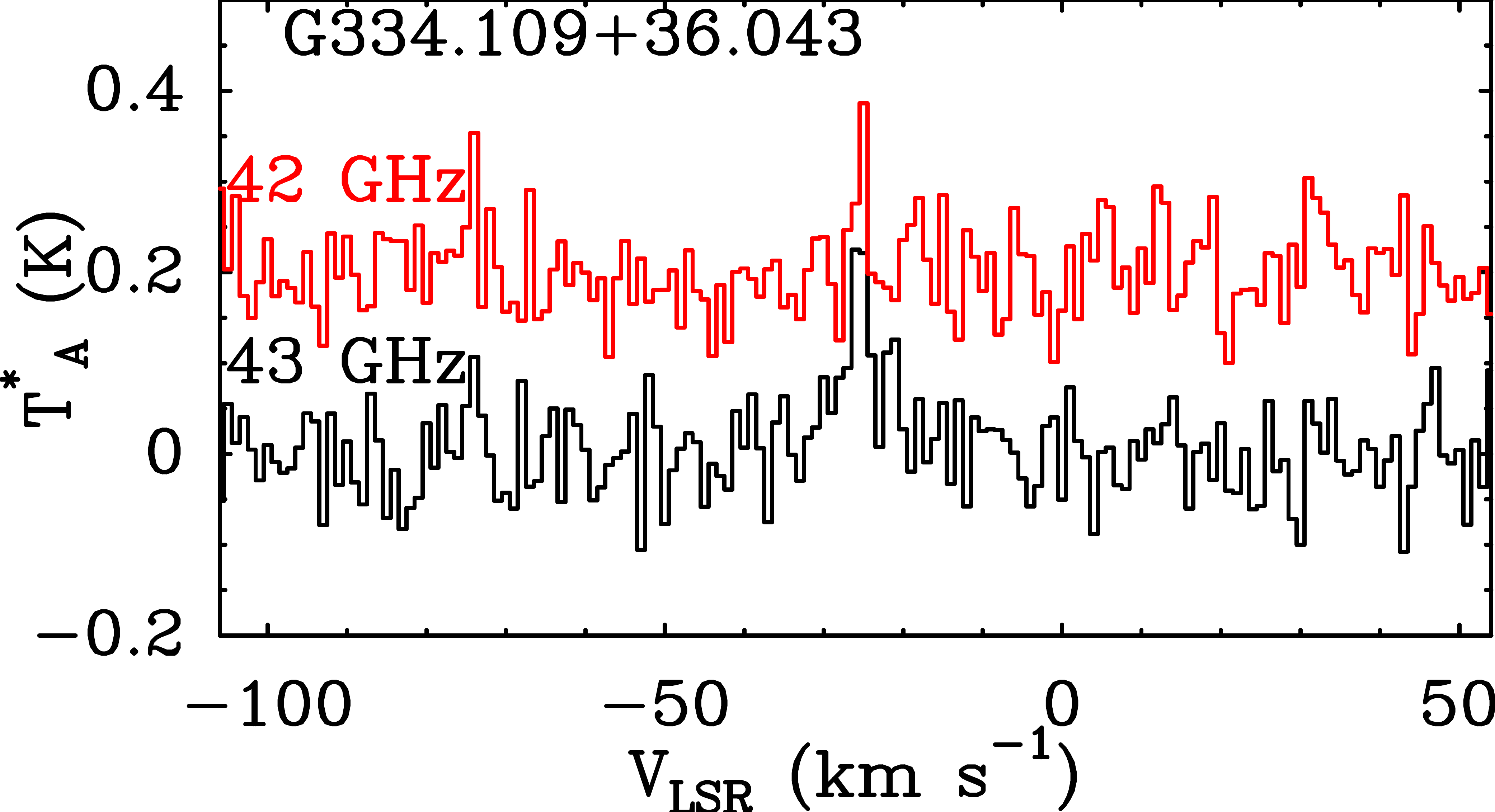}
\includegraphics[width=5.0cm]{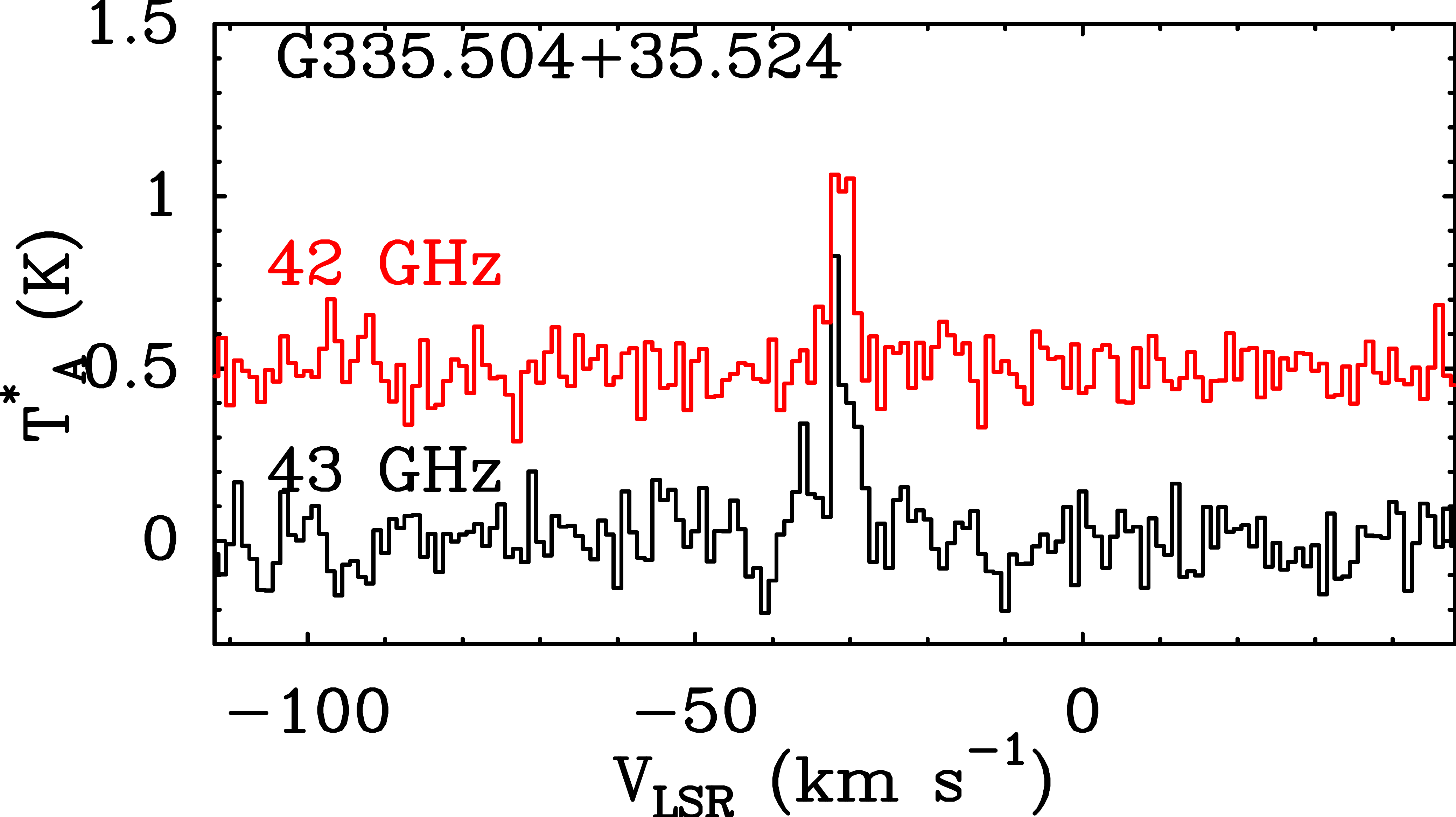}
\includegraphics[width=5.0cm]{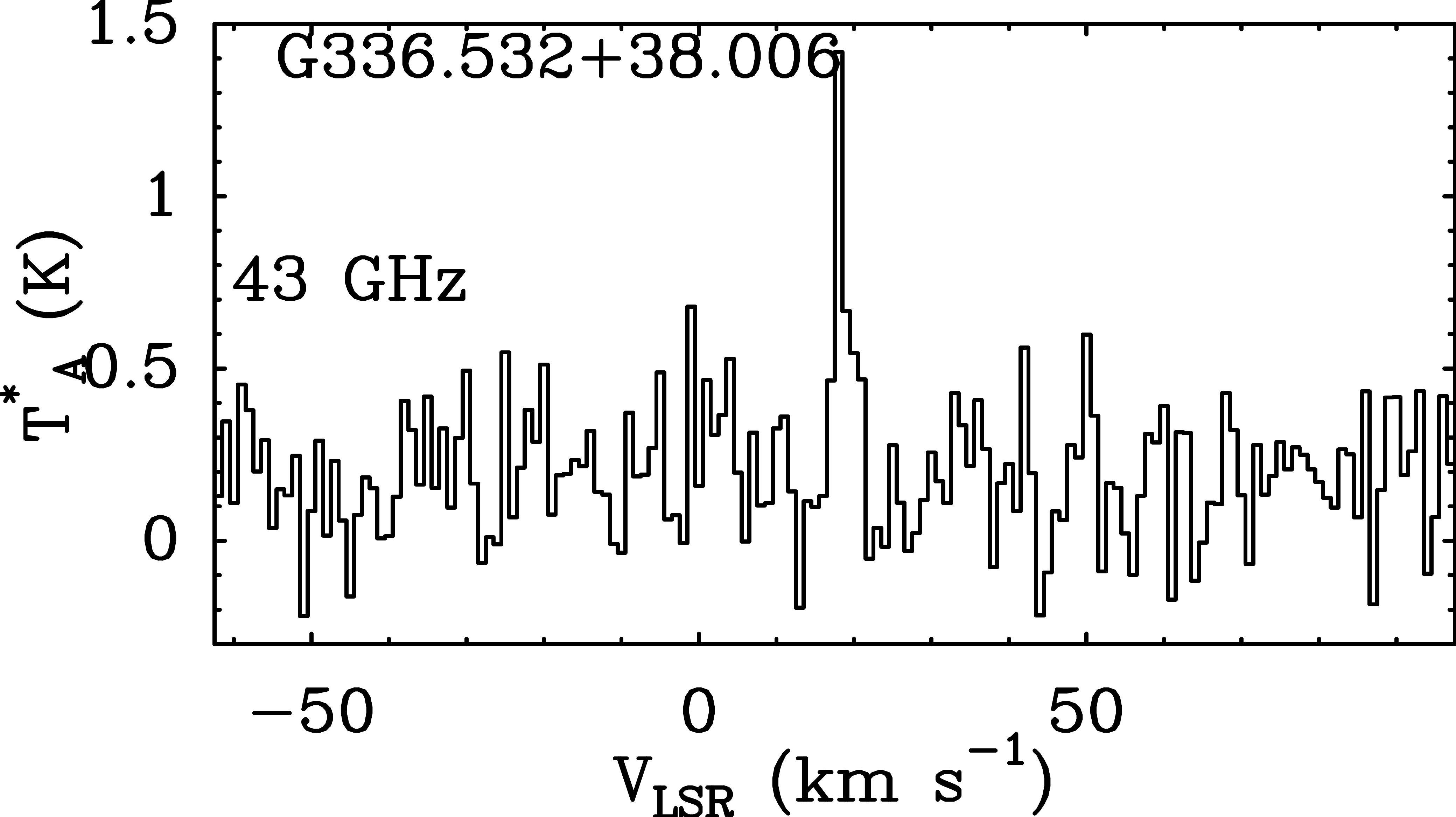}\\
\includegraphics[width=5.0cm]{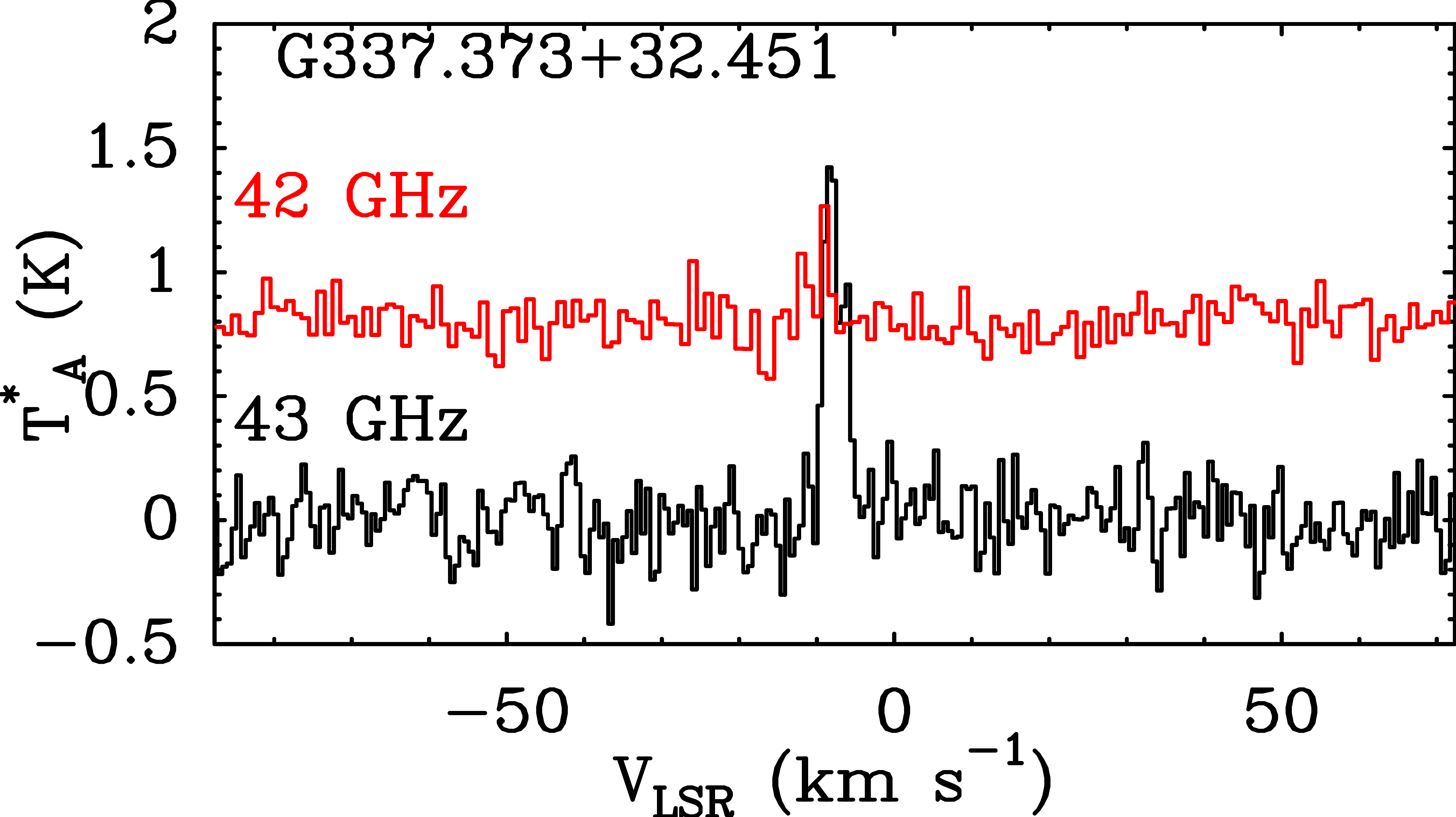}
\includegraphics[width=5.0cm]{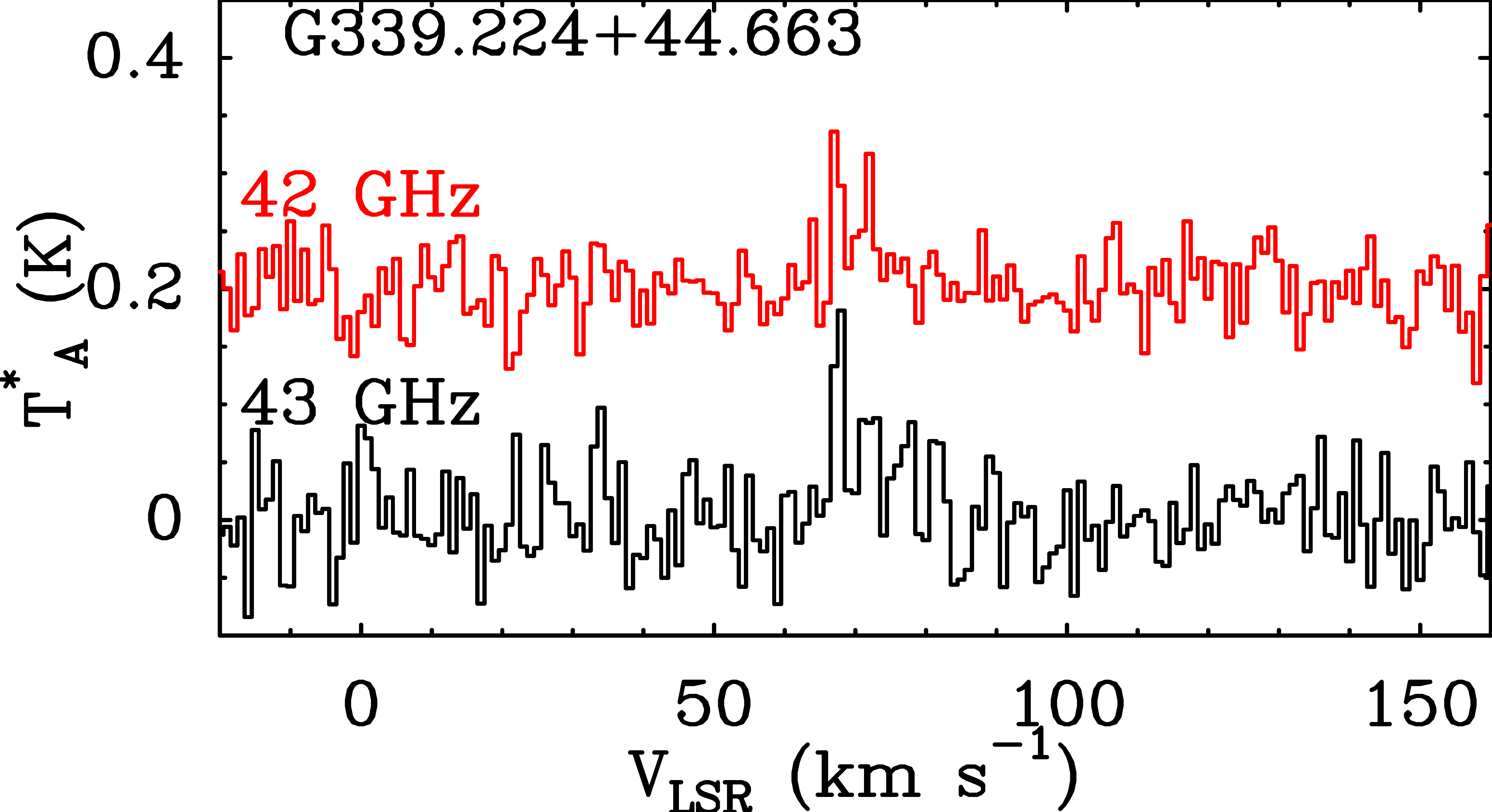}
\includegraphics[width=5.0cm]{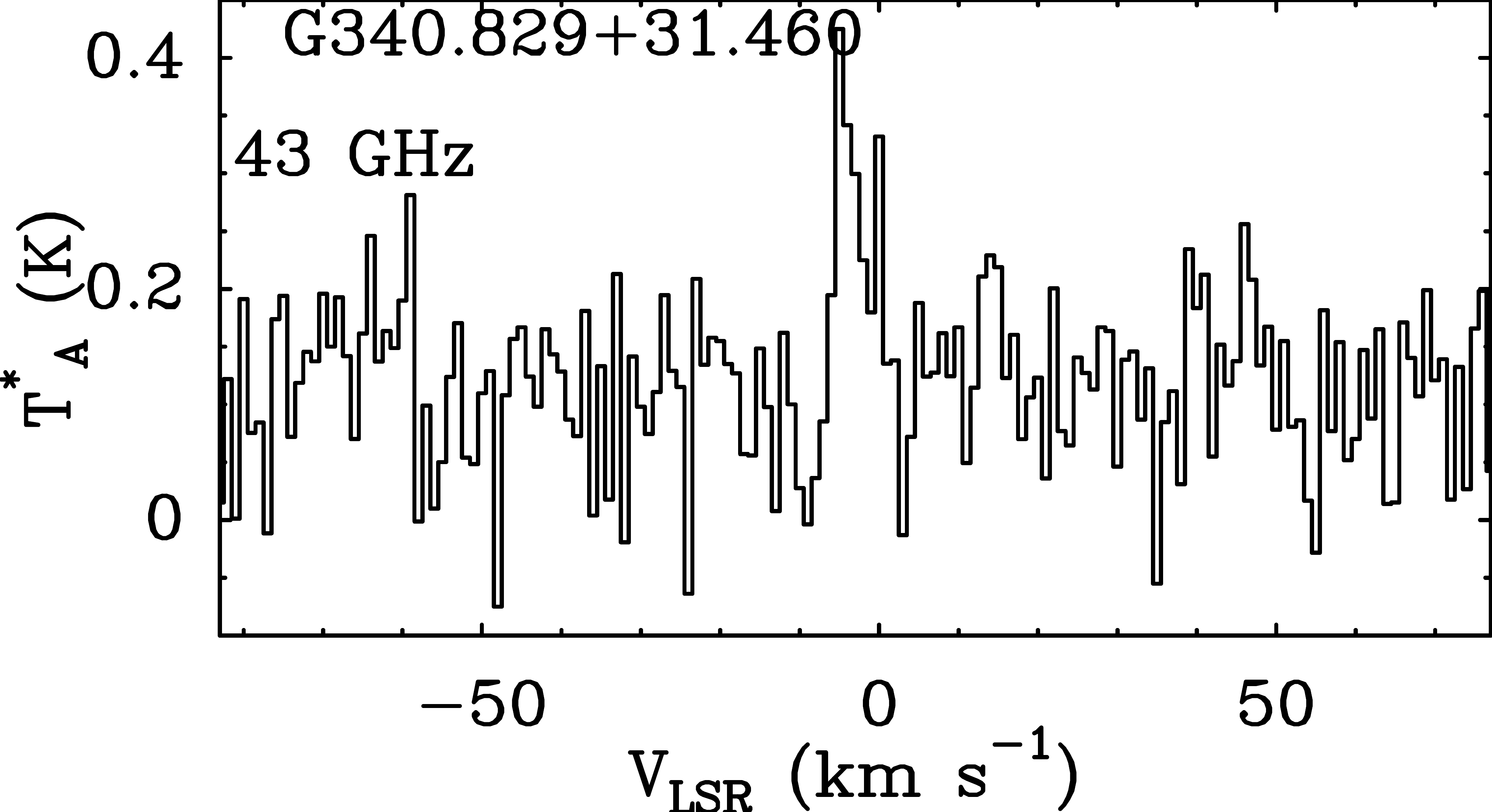}\\
\includegraphics[width=5.0cm]{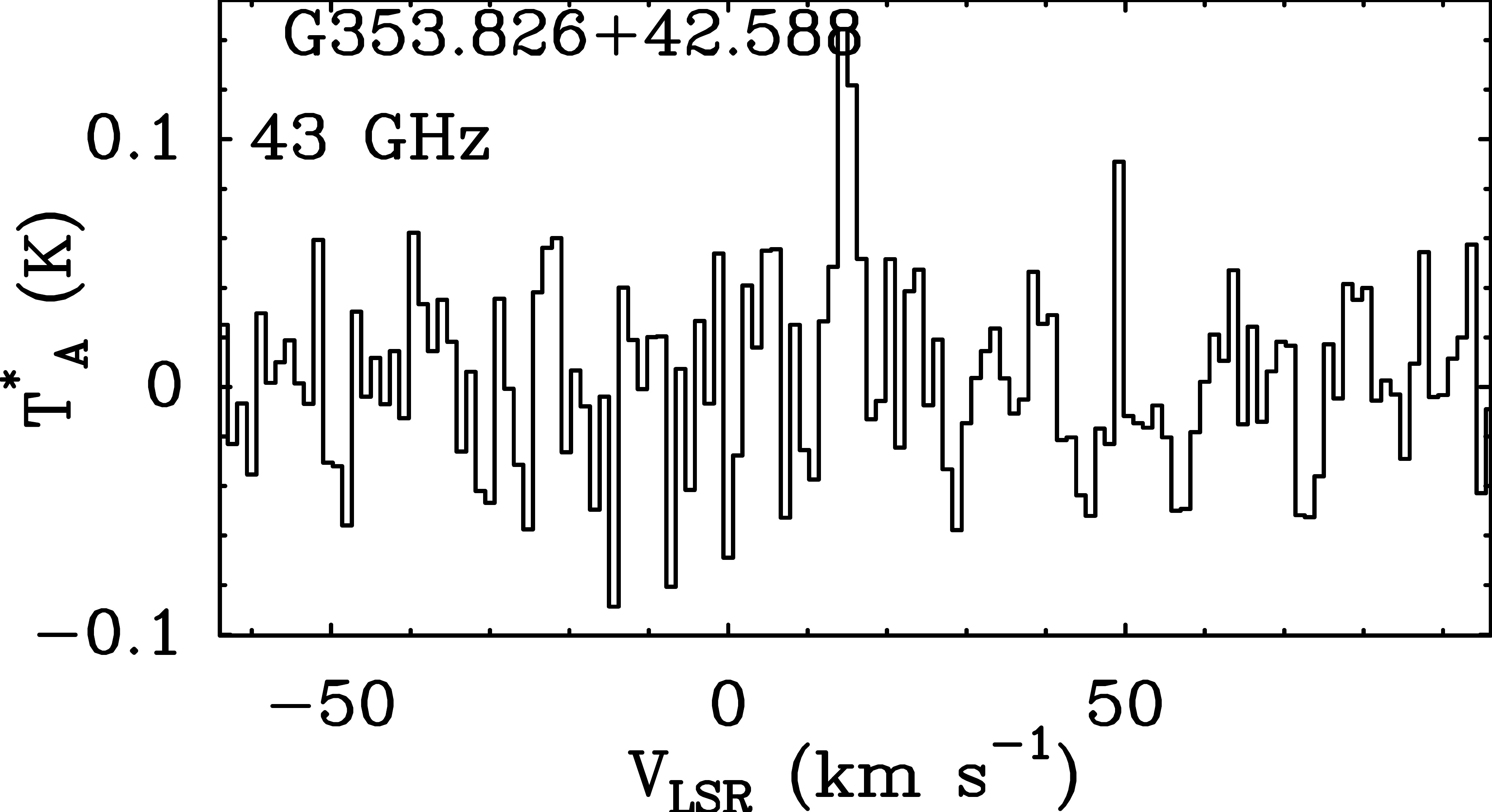}
\includegraphics[width=5.0cm]{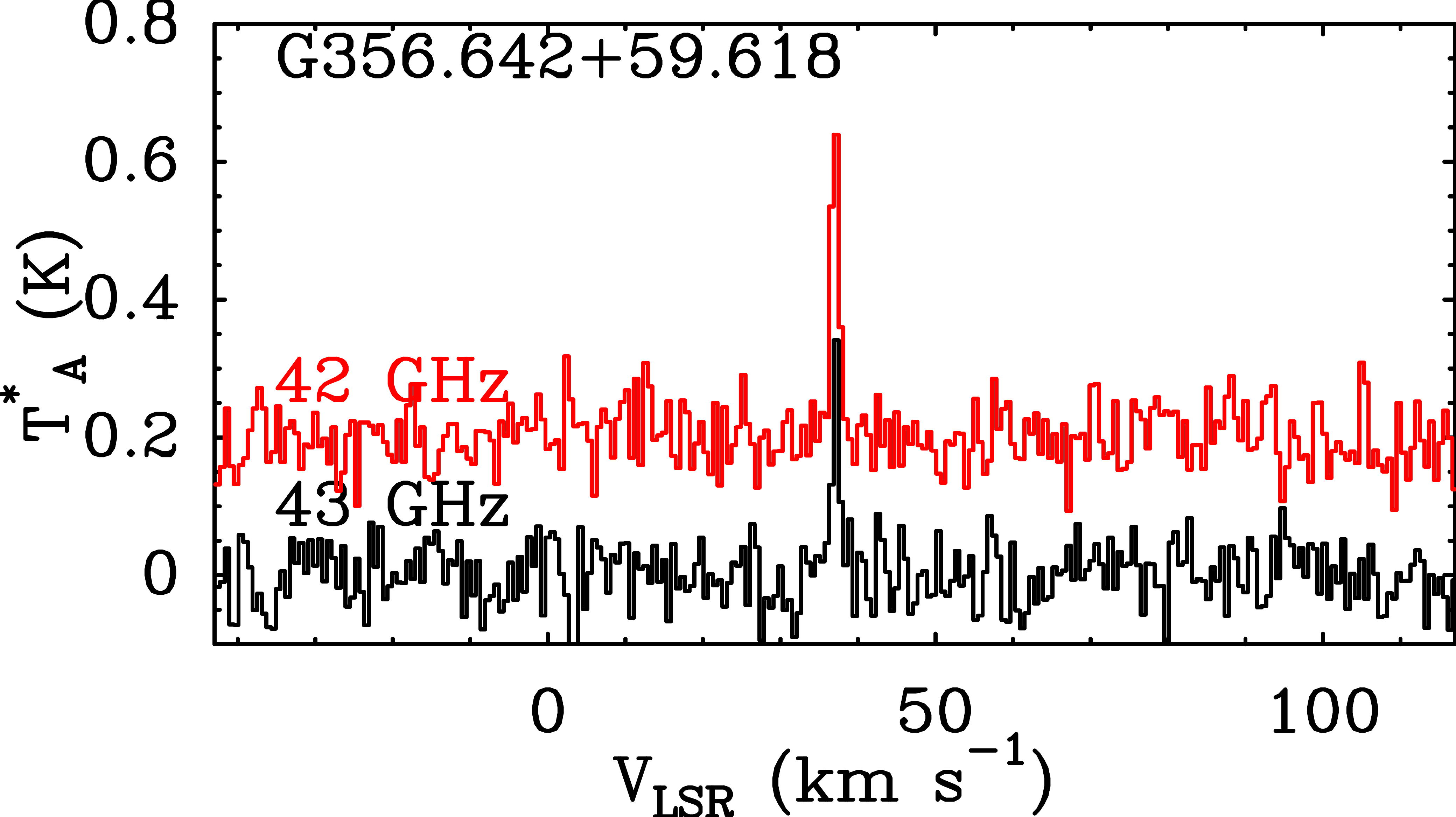}\\
Fig.~\ref{fig-A3} continue
\end{figure*}   

\begin{figure*}
\includegraphics[width=16cm]{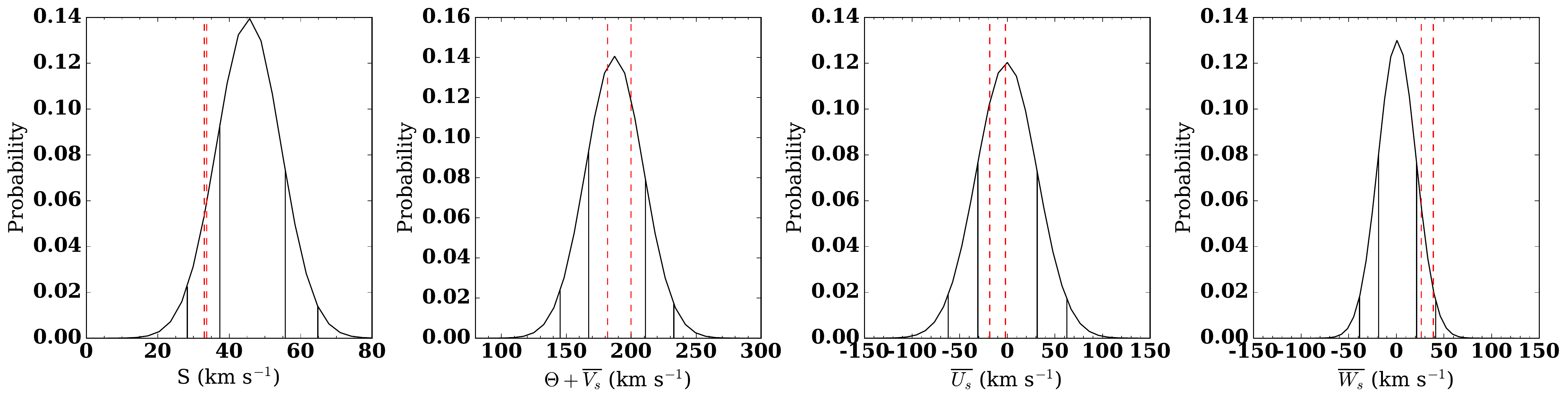} 
\includegraphics[width=16cm]{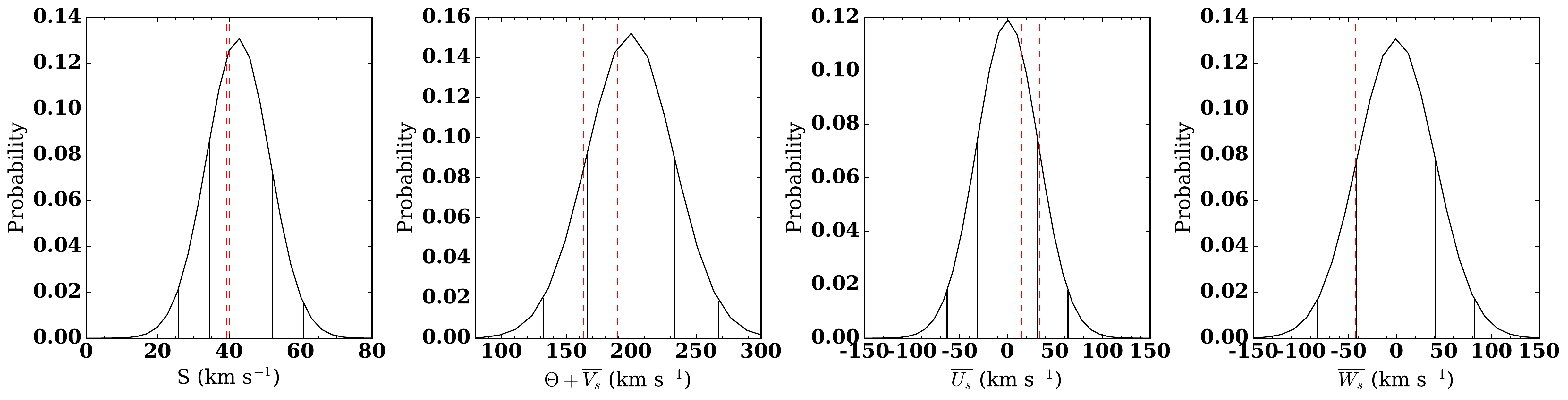} 
\includegraphics[width=16cm]{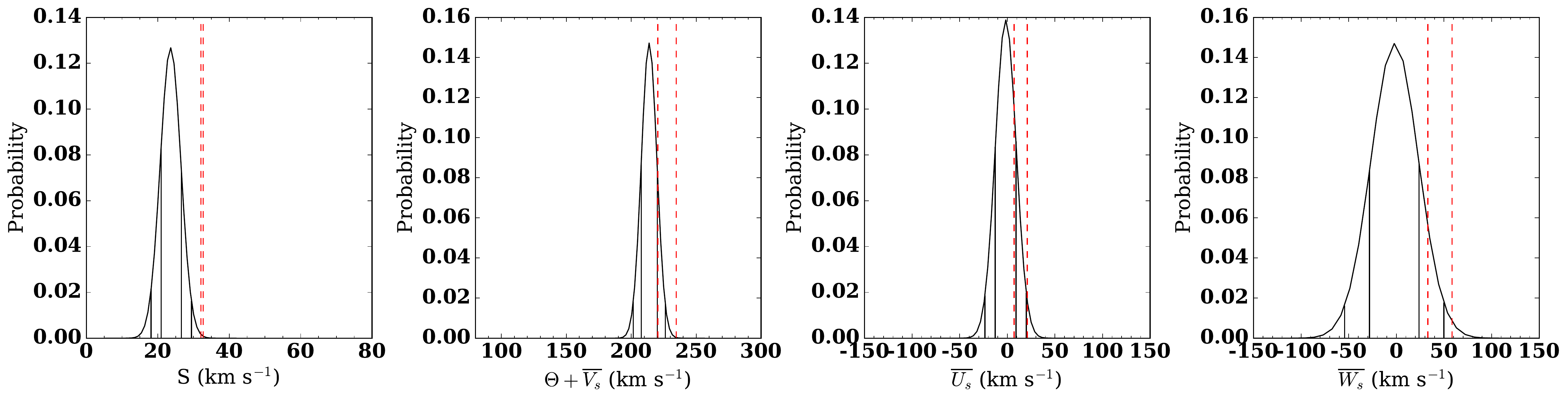} 
\caption{Probability distributions for (\textit{S}, $\overline{U_s}$,
$\Theta$+$\overline{V_s}$, $\overline{W_s}$) estimated via simulation for null
hypothesis testing. \textit{Top}, \textit{middle} and \textit{bottom} panels are results
of group A, B and C, respectively. Solid black lines
denote 1$\sigma$ (68\%) and 2$\sigma$ (95\%) boundaries of simulations. Red
dashed lines denote 1$\sigma$ range of (\textit{S}, $\overline{U_s}$,
$\Theta$+$\overline{V_s}$, $\overline{W_s}$) estimated using real data.
\label{fig-A4}}
\end{figure*}

\bsp	
\label{lastpage}

\newpage
\begin{thebibliography}{67}
\expandafter\ifx\csname natexlab\endcsname\relax\def\natexlab#1{#1}\fi

\bibitem[{{Belokurov} {et~al.}(2014){Belokurov}, {Koposov}, {Evans},
  {Pe{\~n}arrubia}, {Irwin}, {Smith}, {Lewis}, {Gieles}, {Wilkinson},
  {Gilmore}, {Olszewski}, \& {Niederste-Ostholt}}]{2014MNRAS.437..116B}
{Belokurov}, V., {Koposov}, S.~E., {Evans}, N.~W., {et~al.} 2014, \mnras, 437,
  116

\bibitem[{{Belokurov} {et~al.}(2006){Belokurov}, {Zucker}, {Evans}, {Gilmore},
  {Vidrih}, {Bramich}, {Newberg}, {Wyse}, {Irwin}, {Fellhauer}, {Hewett},
  {Walton}, {Wilkinson}, {Cole}, {Yanny}, {Rockosi}, {Beers}, {Bell},
  {Brinkmann}, {Ivezi{\'c}}, \& {Lupton}}]{2006ApJ...642L.137B}
{Belokurov}, V., {Zucker}, D.~B., {Evans}, N.~W., {et~al.} 2006, \apjl, 642,
  L137

\bibitem[{{Benson} {et~al.}(1990){Benson}, {Little-Marenin}, {Woods},
  {Attridge}, {Blais}, {Rudolph}, {Rubiera}, \& {Keefe}}]{1990ApJS...74..911B}
{Benson}, P.~J., {Little-Marenin}, I.~R., {Woods}, T.~C., {et~al.} 1990, \apjs,
  74, 911

\bibitem[{{Bland-Hawthorn} \& {Gerhard}(2016)}]{2016ARA+A..54..529B}
{Bland-Hawthorn}, J. \& {Gerhard}, O. 2016, \araa, 54, 529

\bibitem[{{Carrell} {et~al.}(2012){Carrell}, {Wilhelm}, \&
  {Chen}}]{2012AJ....144...18C}
{Carrell}, K., {Wilhelm}, R., \& {Chen}, Y. 2012, \aj, 144, 18

\bibitem[{{Cho} \& {Kim}(2012)}]{2012AJ....144..129C}
{Cho}, S.-H. \& {Kim}, J. 2012, \aj, 144, 129

\bibitem[{{Correnti} {et~al.}(2010){Correnti}, {Bellazzini}, {Ibata},
  {Ferraro}, \& {Varghese}}]{2010ApJ...721..329C}
{Correnti}, M., {Bellazzini}, M., {Ibata}, R.~A., {Ferraro}, F.~R., \&
  {Varghese}, A. 2010, \apj, 721, 329

\bibitem[{{Cutri} \& {et al.}(2012)}]{2012yCat.2311....0C}
{Cutri}, R.~M. \& {et al.} 2012, VizieR Online Data Catalog, 2311

\bibitem[{{Cutri} {et~al.}(2003){Cutri}, {Skrutskie}, {van Dyk}, {Beichman},
  {Carpenter}, {Chester}, {Cambresy}, {Evans}, {Fowler}, {Gizis}, {Howard},
  {Huchra}, {Jarrett}, {Kopan}, {Kirkpatrick}, {Light}, {Marsh}, {McCallon},
  {Schneider}, {Stiening}, {Sykes}, {Weinberg}, {Wheaton}, {Wheelock}, \&
  {Zacarias}}]{2003tmc..book.....C}
{Cutri}, R.~M., {Skrutskie}, M.~F., {van Dyk}, S., {et~al.} 2003, {2MASS All
  Sky Catalog of point sources.}

\bibitem[{{Deguchi}(2007)}]{2007IAUS..242..200D}
{Deguchi}, S. 2007, in IAU Symposium, Vol. 242, Astrophysical Masers and their
  Environments, ed. J.~M. {Chapman} \& W.~A. {Baan}, 200--207

\bibitem[{{Deguchi} {et~al.}(2004){Deguchi}, {Fujii}, {Glass}, {Imai}, {Ita},
  {Izumiura}, {Kameya}, {Miyazaki}, {Nakada}, \&
  {Nakashima}}]{2004PASJ...56..765D}
{Deguchi}, S., {Fujii}, T., {Glass}, I.~S., {et~al.} 2004, \pasj, 56, 765

\bibitem[{{Deguchi} {et~al.}(2007){Deguchi}, {Fujii}, {Ita}, {Imai},
  {Izumiura}, {Kameya}, {Matsunaga}, {Miyazaki}, {Mizutani}, {Nakada},
  {Nakashima}, \& {Winnberg}}]{2007PASJ...59..559D}
{Deguchi}, S., {Fujii}, T., {Ita}, Y., {et~al.} 2007, \pasj, 59, 559

\bibitem[{{Deguchi} {et~al.}(2012){Deguchi}, {Sakamoto}, \&
  {Hasegawa}}]{2012PASJ...64....4D}
{Deguchi}, S., {Sakamoto}, T., \& {Hasegawa}, T. 2012, \pasj, 64, 4

\bibitem[{{Deguchi} {et~al.}(2010){Deguchi}, {Shimoikura}, \&
  {Koike}}]{2010PASJ...62..525D}
{Deguchi}, S., {Shimoikura}, T., \& {Koike}, K. 2010, \pasj, 62, 525

\bibitem[{{Dehnen} \& {Binney}(1998)}]{1998MNRAS.298..387D}
{Dehnen}, W. \& {Binney}, J.~J. 1998, \mnras, 298, 387

\bibitem[{{Demers} \& {Battinelli}(2007)}]{2007A+A...473..143D}
{Demers}, S. \& {Battinelli}, P. 2007, \aap, 473, 143

\bibitem[{{Drake} {et~al.}(2013){Drake}, {Catelan}, {Djorgovski}, {Torrealba},
  {Graham}, {Mahabal}, {Prieto}, {Donalek}, {Williams}, {Larson},
  {Christensen}, \& {Beshore}}]{2013ApJ...765..154D}
{Drake}, A.~J., {Catelan}, M., {Djorgovski}, S.~G., {et~al.} 2013, \apj, 765,
  154

\bibitem[{{Drimmel} {et~al.}(2003){Drimmel}, {Cabrera-Lavers}, \&
  {L{\'o}pez-Corredoira}}]{2003A+A...409..205D}
{Drimmel}, R., {Cabrera-Lavers}, A., \& {L{\'o}pez-Corredoira}, M. 2003, \aap,
  409, 205

\bibitem[{{Fitzpatrick}(1999)}]{1999PASP..111...63F}
{Fitzpatrick}, E.~L. 1999, \pasp, 111, 63

\bibitem[{{Haikala} {et~al.}(1994){Haikala}, {Nyman}, \&
  {Forsstroem}}]{1994A+AS..103..107H}
{Haikala}, L.~K., {Nyman}, L.-A., \& {Forsstroem}, V. 1994, \aaps, 103

\bibitem[{{Hall} {et~al.}(1990){Hall}, {Allen}, {Troup}, {Wark}, \&
  {Wright}}]{1990MNRAS.243..480H}
{Hall}, P.~J., {Allen}, D.~A., {Troup}, E.~R., {Wark}, R.~M., \& {Wright},
  A.~E. 1990, \mnras, 243, 480

\bibitem[{{Helou} \& {Walker}(1988)}]{1988iras....7.....H}
{Helou}, G. \& {Walker}, D.~W., eds. 1988, {Infrared astronomical satellite
  (IRAS) catalogs and atlases. Volume 7: The small scale structure catalog},
  Vol.~7, 1--265

\bibitem[{{Honma} {et~al.}(2012){Honma}, {Nagayama}, {Ando}, {Bushimata},
  {Choi}, {Handa}, {Hirota}, {Imai}, {Jike}, {Kim}, {Kameya}, {Kawaguchi},
  {Kobayashi}, {Kurayama}, {Kuji}, {Matsumoto}, {Manabe}, {Miyaji}, {Motogi},
  {Nakagawa}, {Nakanishi}, {Niinuma}, {Oh}, {Omodaka}, {Oyama}, {Sakai},
  {Sato}, {Sato}, {Shibata}, {Shiozaki}, {Sunada}, {Tamura}, {Ueno}, \&
  {Yamauchi}}]{2012PASJ...64..136H}
{Honma}, M., {Nagayama}, T., {Ando}, K., {et~al.} 2012, \pasj, 64

\bibitem[{{Huxor} \& {Grebel}(2015)}]{2015MNRAS.453.2653H}
{Huxor}, A.~P. \& {Grebel}, E.~K. 2015, \mnras, 453, 2653

\bibitem[{{Ibata} {et~al.}(2001){Ibata}, {Lewis}, {Irwin}, {Totten}, \&
  {Quinn}}]{2001ApJ...551..294I}
{Ibata}, R., {Lewis}, G.~F., {Irwin}, M., {Totten}, E., \& {Quinn}, T. 2001,
  \apj, 551, 294

\bibitem[{{Indermuehle} \& {McIntosh}(2014)}]{2014MNRAS.441.3226I}
{Indermuehle}, B.~T. \& {McIntosh}, G.~C. 2014, \mnras, 441, 3226

\bibitem[{{Ishihara} {et~al.}(2010){Ishihara}, {Onaka}, {Kataza}, {Salama},
  {Alfageme}, {Cassatella}, {Cox}, {Garc{\'{\i}}a-Lario}, {Stephenson},
  {Cohen}, {Fujishiro}, {Fujiwara}, {Hasegawa}, {Ita}, {Kim}, {Matsuhara},
  {Murakami}, {M{\"u}ller}, {Nakagawa}, {Ohyama}, {Oyabu}, {Pyo}, {Sakon},
  {Shibai}, {Takita}, {Tanab{\'e}}, {Uemizu}, {Ueno}, {Usui}, {Wada},
  {Watarai}, {Yamamura}, \& {Yamauchi}}]{2010A+A...514A...1I}
{Ishihara}, D., {Onaka}, T., {Kataza}, H., {et~al.} 2010, \aap, 514, A1

\bibitem[{{Ita} {et~al.}(2001){Ita}, {Deguchi}, {Fujii}, {Kameya}, {Miyoshi},
  {Nakada}, {Nakashima}, \& {Parthasarathy}}]{2001A+A...376..112I}
{Ita}, Y., {Deguchi}, S., {Fujii}, T., {et~al.} 2001, \aap, 376, 112

\bibitem[{{Izumiura} {et~al.}(1995){Izumiura}, {Catchpole}, {Deguchi},
  {Hashimoto}, {Nakada}, {Onaka}, {Ono}, {Sekiguchi}, {Ukita}, \&
  {Yamamura}}]{1995ApJS...98..271I}
{Izumiura}, H., {Catchpole}, R., {Deguchi}, S., {et~al.} 1995, \apjs, 98, 271

\bibitem[{{Izumiura} {et~al.}(1998){Izumiura}, {Deguchi}, \&
  {Fujii}}]{1998ApJ...494L..89I}
{Izumiura}, H., {Deguchi}, S., \& {Fujii}, T. 1998, \apjl, 494, L89

\bibitem[{{Izumiura} {et~al.}(1994){Izumiura}, {Deguchi}, {Hashimoto},
  {Nakada}, {Onaka}, {Ono}, {Ukita}, \& {Yamamura}}]{1994ApJ...437..419I}
{Izumiura}, H., {Deguchi}, S., {Hashimoto}, O., {et~al.} 1994, \apj, 437, 419

\bibitem[{{Jiang} {et~al.}(1995){Jiang}, {Deguchi}, {Izumiura}, {Nakada}, \&
  {Yamamura}}]{1995PASJ...47..815J}
{Jiang}, B.~W., {Deguchi}, S., {Izumiura}, H., {Nakada}, Y., \& {Yamamura}, I.
  1995, \pasj, 47, 815

\bibitem[{{Kim} {et~al.}(2010){Kim}, {Cho}, {Oh}, \&
  {Byun}}]{2010ApJS..188..209K}
{Kim}, J., {Cho}, S.-H., {Oh}, C.~S., \& {Byun}, D.-Y. 2010, \apjs, 188, 209

\bibitem[{{Kirsanova} {et~al.}(2017){Kirsanova}, {Sobolev}, \&
  {Thomasson}}]{2017arXiv170502197K}
{Kirsanova}, M.~S., {Sobolev}, A.~M., \& {Thomasson}, M. 2017, ArXiv:1705.02197

\bibitem[{{Koposov} {et~al.}(2012){Koposov}, {Belokurov}, {Evans}, {Gilmore},
  {Gieles}, {Irwin}, {Lewis}, {Niederste-Ostholt}, {Pe{\~n}arrubia}, {Smith},
  {Bizyaev}, {Malanushenko}, {Malanushenko}, {Schneider}, \&
  {Wyse}}]{2012ApJ...750...80K}
{Koposov}, S.~E., {Belokurov}, V., {Evans}, N.~W., {et~al.} 2012, \apj, 750, 80

\bibitem[{{Kundu} {et~al.}(2002){Kundu}, {Majewski}, {Rhee}, {Rocha-Pinto},
  {Polak}, {Slesnick}, {Kunkel}, {Johnston}, {Patterson}, {Geisler}, {Gieren},
  {Seguel}, {Smith}, {Palma}, {Arenas}, {Crane}, \&
  {Hummels}}]{2002ApJ...576L.125K}
{Kundu}, A., {Majewski}, S.~R., {Rhee}, J., {et~al.} 2002, \apjl, 576, L125

\bibitem[{{Kwon} \& {Suh}(2012)}]{2012JKAS...45..139K}
{Kwon}, Y.-J. \& {Suh}, K.-W. 2012, Journal of Korean Astronomical Society, 45,
  139

\bibitem[{{Law} \& {Majewski}(2010{\natexlab{a}})}]{2010ApJ...718.1128L}
{Law}, D.~R. \& {Majewski}, S.~R. 2010{\natexlab{a}}, \apj, 718, 1128

\bibitem[{{Law} \& {Majewski}(2010{\natexlab{b}})}]{2010ApJ...714..229L}
{Law}, D.~R. \& {Majewski}, S.~R. 2010{\natexlab{b}}, \apj, 714, 229

\bibitem[{{Li} {et~al.}(2010){Li}, {An}, {Shen}, \&
  {Miyazaki}}]{2010ApJ...720L..56L}
{Li}, J., {An}, T., {Shen}, Z.-Q., \& {Miyazaki}, A. 2010, \apjl, 720, L56

\bibitem[{{Lian} {et~al.}(2014){Lian}, {Zhu}, {Kong}, \&
  {He}}]{2014A+A...564A..84L}
{Lian}, J., {Zhu}, Q., {Kong}, X., \& {He}, J. 2014, \aap, 564, A84

\bibitem[{{Little-Marenin} \& {Little}(1990)}]{1990AJ.....99.1173L}
{Little-Marenin}, I.~R. \& {Little}, S.~J. 1990, \aj, 99, 1173

\bibitem[{{Lynden-Bell} \& {Lynden-Bell}(1995)}]{1995MNRAS.275..429L}
{Lynden-Bell}, D. \& {Lynden-Bell}, R.~M. 1995, \mnras, 275, 429

\bibitem[{{Majewski}(2004)}]{2004PASA...21..197M}
{Majewski}, S.~R. 2004, \pasa, 21, 197

\bibitem[{{Majewski} {et~al.}(2003){Majewski}, {Skrutskie}, {Weinberg}, \&
  {Ostheimer}}]{2003ApJ...599.1082M}
{Majewski}, S.~R., {Skrutskie}, M.~F., {Weinberg}, M.~D., \& {Ostheimer}, J.~C.
  2003, \apj, 599, 1082

\bibitem[{{Matsunaga} {et~al.}(2005){Matsunaga}, {Deguchi}, {Ita}, {Tanabe}, \&
  {Nakada}}]{2005PASJ...57L...1M}
{Matsunaga}, N., {Deguchi}, S., {Ita}, Y., {Tanabe}, T., \& {Nakada}, Y. 2005,
  \pasj, 57, L1

\bibitem[{{McMillan} \& {Binney}(2010)}]{2010MNRAS.402..934M}
{McMillan}, P.~J. \& {Binney}, J.~J. 2010, \mnras, 402, 934

\bibitem[{{Nakashima} {et~al.}(2000){Nakashima}, {Jiang}, {Deguchi},
  {Sadakane}, \& {Nakada}}]{2000PASJ...52..275N}
{Nakashima}, J.-i., {Jiang}, B.~W., {Deguchi}, S., {Sadakane}, K., \& {Nakada},
  Y. 2000, \pasj, 52, 275

\bibitem[{{Nikutta} {et~al.}(2014){Nikutta}, {Hunt-Walker}, {Nenkova},
  {Ivezi{\'c}}, \& {Elitzur}}]{2014MNRAS.442.3361N}
{Nikutta}, R., {Hunt-Walker}, N., {Nenkova}, M., {Ivezi{\'c}}, {\v Z}., \&
  {Elitzur}, M. 2014, \mnras, 442, 3361

\bibitem[{{Pasetto} {et~al.}(2012){Pasetto}, {Grebel}, {Zwitter}, {Chiosi},
  {Bertelli}, {Bienayme}, {Seabroke}, {Bland-Hawthorn}, {Boeche}, {Gibson},
  {Gilmore}, {Munari}, {Navarro}, {Parker}, {Reid}, {Silviero}, \&
  {Steinmetz}}]{2012A+A...547A..70P}
{Pasetto}, S., {Grebel}, E.~K., {Zwitter}, T., {et~al.} 2012, \aap, 547, A70

\bibitem[{{Reid} {et~al.}(2014){Reid}, {Menten}, {Brunthaler}, {Zheng}, {Dame},
  {Xu}, {Wu}, {Zhang}, {Sanna}, {Sato}, {Hachisuka}, {Choi}, {Immer},
  {Moscadelli}, {Rygl}, \& {Bartkiewicz}}]{2014ApJ...783..130R}
{Reid}, M.~J., {Menten}, K.~M., {Brunthaler}, A., {et~al.} 2014, \apj, 783, 130

\bibitem[{{Ruhland} {et~al.}(2011){Ruhland}, {Bell}, {Rix}, \&
  {Xue}}]{2011ApJ...731..119R}
{Ruhland}, C., {Bell}, E.~F., {Rix}, H.-W., \& {Xue}, X.-X. 2011, \apj, 731,
  119

\bibitem[{{Samus'} {et~al.}(2003){Samus'}, {Goranskii}, {Durlevich}, {Zharova},
  {Kazarovets}, {Kireeva}, {Pastukhova}, {Williams}, \&
  {Hazen}}]{2003AstL...29..468S}
{Samus'}, N.~N., {Goranskii}, V.~P., {Durlevich}, O.~V., {et~al.} 2003,
  Astronomy Letters, 29, 468

\bibitem[{{Shi} {et~al.}(2012){Shi}, {Chen}, {Carrell}, \&
  {Zhao}}]{2012ApJ...751..130S}
{Shi}, W.~B., {Chen}, Y.~Q., {Carrell}, K., \& {Zhao}, G. 2012, \apj, 751, 130

\bibitem[{{Shiki} \& {Deguchi}(1997)}]{1997ApJ...478..206S}
{Shiki}, S. \& {Deguchi}, S. 1997, \apj, 478, 206

\bibitem[{{Sjouwerman} {et~al.}(2002){Sjouwerman}, {Lindqvist}, {van
  Langevelde}, \& {Diamond}}]{2002A+A...391..967S}
{Sjouwerman}, L.~O., {Lindqvist}, M., {van Langevelde}, H.~J., \& {Diamond},
  P.~J. 2002, \aap, 391, 967

\bibitem[{{Slater} {et~al.}(2013){Slater}, {Bell}, {Schlafly}, {Juri{\'c}},
  {Martin}, {Rix}, {Bernard}, {Burgett}, {Chambers}, {Finkbeiner}, {Goldman},
  {Kaiser}, {Magnier}, {Morganson}, {Price}, \& {Tonry}}]{2013ApJ...762....6S}
{Slater}, C.~T., {Bell}, E.~F., {Schlafly}, E.~F., {et~al.} 2013, \apj, 762, 6

\bibitem[{{Suh} \& {Kwon}(2009)}]{2009JKAS...42...81S}
{Suh}, K.-W. \& {Kwon}, Y.-J. 2009, Journal of Korean Astronomical Society, 42,
  81

\bibitem[{{Suh} \& {Kwon}(2011)}]{2011MNRAS.417.3047S}
{Suh}, K.-W. \& {Kwon}, Y.-J. 2011, \mnras, 417, 3047

\bibitem[{{Tian} {et~al.}(2015){Tian}, {Liu}, {Carlin}, {Zhao}, {Chen}, {Wu},
  {Li}, {Hou}, \& {Zhang}}]{2015ApJ...809..145T}
{Tian}, H.-J., {Liu}, C., {Carlin}, J.~L., {et~al.} 2015, \apj, 809, 145

\bibitem[{{van der Veen} \& {Habing}(1988)}]{1988A+A...194..125V}
{van der Veen}, W.~E.~C.~J. \& {Habing}, H.~J. 1988, \aap, 194, 125

\bibitem[{{Vivas} \& {Zinn}(2006)}]{2006AJ....132..714V}
{Vivas}, A.~K. \& {Zinn}, R. 2006, \aj, 132, 714

\bibitem[{{Watson}(2006)}]{2006SASS...25...47W}
{Watson}, C.~L. 2006, Society for Astronomical Sciences Annual Symposium, 25,
  47

\bibitem[{{Whitelock} {et~al.}(2008){Whitelock}, {Feast}, \& {van
  Leeuwen}}]{2008MNRAS.386..313W}
{Whitelock}, P.~A., {Feast}, M.~W., \& {van Leeuwen}, F. 2008, \mnras, 386, 313

\bibitem[{{Wright} {et~al.}(2010){Wright}, {Eisenhardt}, {Mainzer}, {Ressler},
  {Cutri}, {Jarrett}, {Kirkpatrick}, {Padgett}, {McMillan}, {Skrutskie},
  {Stanford}, {Cohen}, {Walker}, {Mather}, {Leisawitz}, {Gautier}, {McLean},
  {Benford}, {Lonsdale}, {Blain}, {Mendez}, {Irace}, {Duval}, {Liu}, {Royer},
  {Heinrichsen}, {Howard}, {Shannon}, {Kendall}, {Walsh}, {Larsen}, {Cardon},
  {Schick}, {Schwalm}, {Abid}, {Fabinsky}, {Naes}, \&
  {Tsai}}]{2010AJ....140.1868W}
{Wright}, E.~L., {Eisenhardt}, P.~R.~M., {Mainzer}, A.~K., {et~al.} 2010, \aj,
  140, 1868

\bibitem[{{Xin} \& {Zheng}(2013)}]{2013RAA....13..849X}
{Xin}, X.-S. \& {Zheng}, X.-W. 2013, Research in Astronomy and Astrophysics,
  13, 849

\bibitem[{{Yuan} {et~al.}(2013){Yuan}, {Liu}, \& {Xiang}}]{2013MNRAS.430.2188Y}
{Yuan}, H.~B., {Liu}, X.~W., \& {Xiang}, M.~S. 2013, \mnras, 430, 2188

\end{thebibliography}
\end{document}